\documentclass[11pt, letter]{article}
\pdfoutput=1

\usepackage{jheppub}
\usepackage{natbib}

\usepackage[utf8]{inputenc}

\usepackage{multirow}
\usepackage{mathtools}
\usepackage{dsfont}
\usepackage{mathdots}
\usepackage{amsmath}
\usepackage{amssymb}
\usepackage{amsthm}
\usepackage{mathrsfs}
\usepackage{slashed}
\usepackage{tabularx}
\usepackage{graphicx}
\usepackage{caption,subcaption}
\usepackage{float}
\usepackage{placeins}
\usepackage{bm}

\usepackage{tikz}
\usetikzlibrary{arrows,shapes}
\usetikzlibrary{trees}
\usetikzlibrary{patterns}
\usetikzlibrary{matrix,arrows} 				%
\usetikzlibrary{positioning}				
\usetikzlibrary{calc,through}				
\usetikzlibrary{decorations.pathreplacing}  
\usepackage{pgffor}							

\usetikzlibrary{decorations.pathmorphing}	
\usetikzlibrary{decorations.markings}

\def\be{\begin{equation}}
\def\ee{\end{equation}}
\def\bea{\begin{eqnarray}}
\def\eea{\end{eqnarray}}
\def\ba{\begin{array}}
\def\ea{\end{array}}
\def\bd{\begin{displaymath}}
\def\ed{\end{displaymath}}

\def\>{\rangle} 
\def\<{\langle} 
\def\Dsl{D \hskip-.6em \raise1pt\hbox{$ / $ } }
\def\to{\rightarrow}

\title{Off-shell Thermodynamics and Kinetics of Holographic CFTs Dual to Charged AdS Black Holes}

\author[a]{Debabrata Sahu}
\author[a,b]{Chandrasekhar Bhamidipati}
\affiliation[a]{Department of Physics, School of Basic Sciences\\ 
			Indian Institute of Technology Bhubaneswar \\ Bhubaneswar, Odisha, 752050, India}
\affiliation[b]{ Institute of Physics of the Czech Academy of Sciences \& CEICO\\ 
			Na Slovance 2, 18221 Prague, Czech Republic}

\emailAdd{deba51phys@gmail.com}
\emailAdd{chandrasekhar@iitbbs.ac.in}

\abstract{
We study the thermodynamics and phase structure of holographic conformal field theories dual to spherically symmetric charged AdS black holes using an off-shell free energy. We consider three ensembles of the dual CFT with fixed:  $(\tilde Q,{\cal V},C)$, $(\tilde \Phi,{\cal V},C)$, and $(\tilde Q,{\cal V},\mu)$ and present their corresponding phase diagrams. For the fixed $(\tilde Q,{\cal V},C)$ and $(\tilde \Phi,{\cal V},C)$ ensembles, we study the transitions between competing states using a stochastic description on the various phases given by off-shell free energy. This is described by an ensemble dependent Fokker-Planck equation, allowing us to compute the first-passage-time distribution, including the mean first passage time and its fluctuations over a range of temperatures.  We also examine how the phase structure and the associated kinetics depend on the electric charge $\tilde Q$ and the central charge $C$.
}

\begin{document}  

\maketitle

\newpage


\section{Introduction}
\label{sec:Introduction}

Hawking’s derivation of blackbody radiation from gravitational collapse \cite{Hawking:1975vcx}
established that black holes are thermal systems, sharpening the interplay between gravity,
thermodynamics and statistical physics. Soon after, phase structure began to be explored in a
variety of settings: Hut analyzed transitions for charged black holes \cite{Hut:1977ChargedBH},
and Davies identified second-order behavior in Kerr--Newman solutions associated with divergent
heat capacity \cite{Davies:1978Thermodynamics}. In asymptotically AdS spacetimes, Hawking and
Page found a first-order transition between thermal AdS and a large AdS black hole at a critical
temperature \cite{Hawking:1982dh}. For charged AdS black holes, the analogous small/large
Reissner--Nordstrom AdS transition was studied in detail by Chamblin \emph{et al.}
\cite{Chamblin:1999tk,Chamblin:1999hg}. A further development is to enlarge the
thermodynamic phase space by treating the cosmological constant as a variable and
identifying the black hole mass with enthalpy \cite{Kastor:2009wy}. In this extended framework,
Dolan emphasised the close parallel between charged AdS black holes and the van der Waals
liquid--gas system \cite{Dolan:2010ha,Dolan:2011xt}, and Kubiz\v{n}\'ak and Mann showed that in
the fixed-charge ensemble the critical behavior of RNAdS black holes matches that of the van der
Waals fluid~\cite{Kubiznak:2012wp}.\\

\noindent
The AdS/CFT correspondence \cite{Maldacena:1997re,Gubser:1998bc,Witten:1998qj} provides a
field-theoretic interpretation of these gravitational phase transitions. In particular, the
Hawking-Page transition is naturally mapped to confinement/deconfinement in the dual gauge
theory \cite{Witten:1998zw}. This perspective has motivated a broad literature on the extended thermodynamics of AdS black holes, such as, van der
Waals-type behavior and related phenomena, including geometric and microscropic view points~\cite{Cai:2013qga,Wei:2019Repulsive,Wei:2019Microstructure,Xu:2020SchwarzschildAdS,Ghosh:2020kba}. As the literature on the subject is vast, the for the complete set of references, see the recent review~\cite{Mann:2025xrb}.\\

\noindent
For our purposes it is useful to work off-shell, in terms of a free energy landscape. In
statistical physics, macroscopic phases arise from many microscopic configurations weighted by
Boltzmann factors, and the resulting free energy defines a landscape over the space of macroscopic
variables or order parameters \cite{Goldenfeld:1992RG,Frauenfelder:1991EnergyLandscapes,Frauenfelder:1994Biomolecules}.
In the black hole setting, an off-shell free energy is constructed by temporarily treating the
variables (and temperature) entering the on-shell free energy as independent parameters; imposing
the extremization condition recovers the on-shell thermodynamics. Stable and metastable phases are
then identified with extrema of the off-shell free energy and the connecting phases as fluctuating states, This approach has been applied widely to AdS black holes and related systems
\cite{Chaikin:1995CondensedMatter,Kubo:1965StatMech,Bragg:1934MeanField,Bragg:1935MeanField,Dey:2007JHEP09,Dey:2007JHEP04,Nayak:2008Prayas,Banerjee:2011IJMPA,Banerjee:2010ng,Wei:2022dzw,Yerra:2022coh},
with a recent extension to BPS black holes \cite{Sahu:2025ghv}.\\

\noindent
The same off-shell formalism also supports a kinetic description. Thermally induced switching between
competing phases can be modeled as a stochastic process on the off-shell free energy landscape,
governed by a probabilistic Fokker--Planck equation
\cite{Zwanzig:2001Nonequilibrium,Lee:2003DiffusionFPT,Lee:2003NonMarkovian,Wang:2015LandscapeFlux,Bryngelson:1989RandomEnergy}.
Thermal fluctuations enable barrier crossing, and solving the Fokker-Planck equation with suitable
boundary conditions yields the stationary probability distribution and first-passage observables,
including the mean first passage time and its fluctuations. Landscape-based kinetics has been
implemented for Hawking-Page transitions and for van der Waals--type transitions in RNAdS black
holes \cite{Li:2020khm,Li:2020nsy}, and has proved useful for linking transition rates to landscape
topography. By contrast, a comparable kinetic analysis formulated directly for the dual CFT is still
absent. This gap in literature needs to be filled and is interesting in its own right too, due to the presence of the central charge in the dual CFT and novel phase structures, which are somewhat different from bulk thermodynamics, especially because the boundary  thermodynamic variables and their interpretations are very different. \\

\noindent
The extended phase structure of AdS black holes is tied to the negative cosmological constant
$\Lambda$. Promoting $\Lambda$ to a thermodynamic variable generates new phase behavior
\cite{Kastor:2009wy}. A further refinement allows variations of Newton’s constant $G_N$
\cite{Kastor:2010gq,Kastor:2014dra,Karch:2015rpa,Sarkar:2020DifferentialEntropy,Visser:2021eqk,Cong:2021CentralCharge},
which is natural as a coupling constant across theory space and may run under quantum corrections;
it also serves as a useful bookkeeping parameter in organizing the first law. For Einstein--Maxwell
theory with $\Lambda<0$, the extended first law and generalized Smarr relation take the form
\cite{Kastor:2009wy,Visser:2021eqk}
\begin{align} 
 	d M &= \frac{\kappa}{8\pi G_N} dA + \Phi dQ +\frac{\Theta}{8\pi G_N} d \Lambda - \left ( M   - \Phi Q \right)\frac{dG_N}{G_N}\,, \label{eq:extendedfirstlaw11}\\
 		M &=\frac{d-1}{d-2} \frac{\kappa A}{8 \pi G_N} + \Phi Q - \frac{1}{d-2} \frac{\Theta \Lambda}{4 \pi G_N}\,. \label{eq:smarr1111}
 \end{align}
Here $M$ is the mass, $Q$ the electric charge and $\Phi$ the conjugate electric potential. The
quantity $\Theta$ is conjugate to $\Lambda$, and admits geometric characterizations, either in
terms of surface integrals of a Killing potential \cite{Kastor:2009wy} or in terms of a properly
weighted volume functional \cite{Jacobson:2018ahi}.\\

\noindent
A standard interpretation generally identifies $\Lambda$ with a positive bulk pressure
\cite{Kastor:2009wy,Dolan:2011xt,Dolan:2010ha,Cvetic:2010jb,Kubiznak:2014zwa}
\be\label{P}
P=-\frac{\Lambda}{8\pi G_N}\,, \qquad \text{with} \qquad 
\Lambda=-\frac{d(d-1)}{2 L^2}\,,
\ee
where $L$ is the AdS radius. At fixed $G_N$, the $\Theta\,d\Lambda/(8\pi G_N)$ term can be
rewritten as $V\,dP$ by identifying $V=-\Theta$, yielding
\bea
d M&=&T d S+\Phi d Q+ V d P\,,  \label{flaw}\\
M&=&\frac{d-1}{d-2}TS+\Phi Q-\frac{2}{d-2}PV\,.\label{smarr}
\eea
This formulation, however, comes with two persistent tensions. The first is the enthalpy
interpretation: in extended thermodynamics $M$ behaves as enthalpy rather than internal energy.
The second is structural: once $G_N$ is varied, the $\Lambda$ and $G_N$ terms cannot be combined
into a single $d(\Lambda/G_N)$. Indeed \eqref{eq:extendedfirstlaw11} can be reorganized as
\be \label{eq:firstlawfail}
dM = \frac{\kappa}{8\pi } d \left (\frac{A}{G_N} \right)
+ \Phi dQ
+ \frac{\Theta}{8\pi} d \left ( \frac{\Lambda}{G_N} \right)
- \left (M - \frac{\kappa A}{8\pi G_N}
- \Phi Q
-\frac{\Theta \Lambda}{8\pi G_N }  \right ) \frac{dG_N}{G_N}\,.
\ee
The final term is not removed by the Smarr relation and is not naturally captured in the
$(P,V)$ language. This already indicates that $(P,V)$ is not the optimal choice for holography, a
point that has been emphasized from several angles
\cite{Kastor:2014dra,Karch:2015rpa,Dolan:2016jjc,Johnson:2014yja,Dolan:2014cja,Zhang:2014uoa,Zhang:2015ova}, though there are recent developments~\cite{Borsboom:2026ash}.\\

\noindent
A cleaner holographic organization is obtained by trading bulk pressure for the boundary central
charge. Varying $\Lambda$ is naturally interpreted as varying the central charge $C$ (or $N$) in
the dual CFT \cite{Kastor:2009wy,Dolan:2014cja,Kastor:2014dra,Johnson:2014yja,Karch:2015rpa}. For
Einstein-gravity duals one has $C \sim L^{d-1}/G_N$, so variations of $C$ correspond to correlated
variations of $\Lambda$ and $G_N$. The conjugate chemical potential $\mu$ (``color susceptibility''
\cite{Karch:2015rpa}) then becomes part of the natural thermodynamic data. In \cite{Visser:2021eqk}
this statement was made precise by matching the extended bulk and boundary first laws. Using
$d\Lambda/\Lambda=-2\,dL/L$ and eliminating $\Theta$ via \eqref{eq:smarr1111}, one can rewrite
\eqref{eq:extendedfirstlaw11} as
\be \label{eq:extendedfirstlawnew}
dM  = \frac{\kappa   }{2\pi}   d \left (  \frac{A}{4G_N}    \right  )   + \frac{\Phi}{L} d (Q L )-          \frac{M}{d-1}  \frac{d L^{d-1}}{L^{d-1}}   + \left ( M  -   \frac{\kappa A}{8\pi G_N}     -  \Phi Q \right) \frac{d\! \left ( L^{d-1}/G_N \right) }{ L^{d-1}/G_N} \,.  
\ee
In this form the thermodynamic variations are organized directly in terms of $C\sim L^{d-1}/G_N$,
and no analogue of the residual $dG_N/G_N$ term of \eqref{eq:firstlawfail} appears. With the standard
dictionary for a CFT whose curvature scale coincides with the AdS scale
\cite{Karch:2015rpa,Visser:2021eqk},
\be \label{dictionary1}
E=M\, , \qquad \tilde \Phi = \Phi/L \, , \qquad \tilde Q= QL\,, \qquad {\cal V} \sim L^{d-1}\,, \qquad C \sim L^{d-1}/G_N\,,
\ee
and the usual definitions of temperature and entropy, \eqref{eq:extendedfirstlawnew} reduces to the
extended CFT first law
\begin{align} 
 	dE &= T dS + \tilde \Phi d \tilde Q - p d{\cal V} + \mu dC\,, \qquad \text{with} \label{cftfirstlawfirst}\\
    \mu &=\frac{1}{C}(E- TS - \tilde \Phi \tilde Q )\,, \quad \text{and} \quad  p = \frac{1}{d-1}\frac{E}{\cal V}\,. \label{mupintro}
\end{align}
Both the CFT charge and its conjugate potential are rescaled by $L$ \cite{Chamblin:1999tk}. The
relation for $\mu$ in \eqref{mupintro} is the Euler relation of the deconfined large-$N$ theory and
is dual to the generalized Smarr relation \eqref{eq:smarr1111}
\cite{Karch:2015rpa,Visser:2021eqk}. Further, the pressure $p$ satisfies the CFT equation of state in $d$
dimensions and the dictionary \eqref{dictionary1} can be generalized to include boundary curvature radius
$R\neq L$ \cite{Visser:2021eqk}. In this setting the matching between \eqref{cftfirstlawfirst} and
\eqref{eq:extendedfirstlaw11} persists, while ${\cal V}$ and $C$ become independent, controlled by
$R$ and $L$, respectively.\\

\noindent
The purpose of this paper is to analyze thermodynamics and kinetics of phase transitions in the
holographic CFT dual to charged AdS black holes in extended thermodynamics. The conjugate pairs
$(\tilde \Phi,\tilde Q)$, $(p,{\cal V})$ and $(\mu,C)$ define eight distinct (grand) canonical
ensembles. In \cite{Cong:2021jgb}, three of these ensembles were shown to exhibit nontrivial phase
behavior in the on-shell description, namely fixed $(\tilde Q,{\cal V},C)$, $(\tilde \Phi,{\cal V},C)$
and $(\tilde Q,{\cal V},\mu)$. Here we move these three cases off-shell: we construct the free energy landscapes and the corresponding off-shell phase diagrams. We also emphasize that the off-shell description is more general, as it gives a landscape of the various phases of  the system both in and around the equilibrium. For the fixed $(\tilde Q,{\cal V},C)$ and the $(\tilde \Phi,{\cal V},C)$ ensembles, we also solve the associated Fokker-Planck equations to extract first-passage observables, including mean first passage times and their fluctuations. We also study how these kinetic quantities depend on the electric charge $\tilde Q$ and the central charge $C$. \\

\noindent
Plan of the rest of the paper is as follows. In section-(\ref{sec:holographic}), we give a brief summary of the holographic CFT thermodynamics of charged AdS black holes as much is required for our purposes (following the notations of~\cite{Cong:2021jgb}), leading to the emergence of three ensembles listed in eqn. (\ref{freeenergies2}). Section-(\ref{sec:off}) is then devoted to the study the off-shell thermodynamics and phase structure in each of the ensembles given in eqn. (\ref{freeenergies2}), including the the novel $(\tilde Q,{\cal V},\mu)$ ensemble, where there is a zeroth order phase transition. Sections-(\ref{sec:FK}) and (\ref{subsec:kinetics_grand}) are then devoted to a study of kinetics using the Fokker-Planck equation in various cases, where we also study the effect of charge $\tilde Q$ and the central charge $C$. We end the paper with some remarks in section-(\ref{sec:conclusion}).
 

\section{Holographic CFT thermodynamics of charged AdS black holes}
\label{sec:holographic}

In this section we briefly summarize the holographic thermodynamics of electrically charged
AdS black holes in the extended thermodynamics, focussing on equations required for our analysis. Further, our presentation closely follows
\cite{Cong:2021jgb}, and is included only to fix conventions and notation. We collect here the thermodynamic relations that will be used as basic input for the
off-shell analysis in the subsequent sections. The spacetime dimension $d$ is kept
arbitrary, although all explicit plots and numerical results presented in this paper
correspond to $d=4$, i.e. the AdS$_5$/CFT$_4$ correspondence.

\subsection{Extended thermodynamics of charged AdS black holes}

We consider spherically symmetric Reissner--Nordström AdS black holes arising from
Einstein--Maxwell theory with a negative cosmological constant. The bulk action in
$d+1$ dimensions is
\begin{equation}
	I = \frac{1}{16\pi G_N} \int d^{d+1}x \sqrt{-g}
	\left( R - 2\Lambda - F^2 \right),
	\label{EMaction}
\end{equation}
where $F=dA$ and $\Lambda<0$. The static black hole metric takes the form
\begin{equation}
	ds^2 = -f(r)\,dt^2 + \frac{dr^2}{f(r)} + r^2 d\Omega_{d-1}^2 ,
	\label{eq:metric1}
\end{equation}
with the lapse function
\begin{equation}
	f(r) = 1 + \frac{r^2}{L^2} - \frac{m}{r^{d-2}} + \frac{q^2}{r^{2d-4}} .
	\label{eq:blackening}
\end{equation}
The mass and charge parameters are related to the ADM mass and electric charge via
\begin{equation}
	M = \frac{(d-1)\Omega_{d-1}}{16\pi G_N} m ,
	\label{eq:admmass}
\end{equation}
\begin{equation}
	Q = \frac{(d-1)\Omega_{d-1}}{8\pi G_N} \alpha\, q ,
	\qquad
	\alpha=\sqrt{\frac{2(d-2)}{d-1}} .
	\label{eq:electriccharge1}
\end{equation}
The gauge potential is chosen such that $A_t(r_h)=0$, yielding
\begin{equation}
	\Phi = \frac{1}{\alpha}\frac{q}{r_h^{d-2}} .
	\label{eq:electricpotential1}
\end{equation}
The horizon radius $r_h$ is defined as the largest positive root of $f(r_h)=0$, which
determines the mass parameter as
\begin{equation}
	m = r_h^{d-2}\left(1+\frac{r_h^2}{L^2}+\frac{q^2}{r_h^{2d-4}}\right).
\end{equation}
The black hole entropy and Hawking temperature follow as
\begin{equation}
	T = \frac{d-2}{4\pi r_h}\left(1+\frac{d}{d-2}\frac{r_h^2}{L^2}
	- \frac{q^2}{r_h^{2d-4}}\right),
	\qquad
	S = \frac{\Omega_{d-1} r_h^{d-1}}{4G_N}.
	\label{eq:entropytemp}
\end{equation}
These quantities satisfy the generalized Smarr relation
\begin{equation}
	M = \frac{d-1}{d-2}\frac{\kappa A}{8\pi G_N}
	+ \Phi Q
	- \frac{1}{d-2}\frac{\Theta\Lambda}{4\pi G_N},
	\label{eq:smarr1111}
\end{equation}
together with the extended gravitational first law
\begin{equation}
	dM = \frac{\kappa}{8\pi G_N}dA + \Phi dQ
	+ \frac{\Theta}{8\pi G_N}d\Lambda
	- (M-\Phi Q)\frac{dG_N}{G_N}.
	\label{eq:extendedfirstlaw}
\end{equation}

\subsection{CFT thermodynamics and central charge}
\label{subsec:CFT quantities}
In the holographic description, variations of the cosmological constant and Newton’s
constant are naturally mapped to variations of the CFT central charge
\cite{Karch:2015rpa,Visser:2021eqk}. For Einstein gravity the central charge is given by
\begin{equation}
	C = \frac{\Omega_{d-1} L^{d-1}}{16\pi G_N}.
	\label{eq:centralcharge}
\end{equation}
The spatial volume of the CFT is fixed independently by choosing a boundary curvature
radius $R$,
\begin{equation}
	{\cal V} = \Omega_{d-1} R^{d-1}.
	\label{vol}
\end{equation}
With this choice, the holographic dictionary relating bulk and boundary quantities reads
\begin{equation}
	S=\frac{A}{4G_N}, \qquad
	E = M\frac{L}{R}, \qquad
	T=\frac{\kappa}{2\pi}\frac{L}{R}, \qquad
	\tilde\Phi=\frac{\Phi}{R}, \qquad
	\tilde Q = QL .
	\label{eq:dictionaryentropyetc}
\end{equation}
Eliminating $\Theta$ using the Smarr relation and rewriting variations of $\Lambda$ in
terms of $L$, the bulk first law can be reorganized into
\begin{align}
	d\!\left(M\frac{L}{R}\right) &=
	\frac{\kappa}{2\pi}\frac{L}{R}d\!\left(\frac{A}{4G_N}\right)
	+ \frac{\Phi}{R}d(QL)
	- \frac{M}{d-1}\frac{L}{R}\frac{dR^{d-1}}{R^{d-1}} \notag\\
	&\quad
	+ \left(M\frac{L}{R}-\frac{\kappa A}{8\pi G_N}\frac{L}{R}
	- \frac{\Phi}{R}QL\right)
	\frac{d(L^{d-1}/G_N)}{L^{d-1}/G_N}.
	\label{eq:bulkfirstlaw2}
\end{align}
This maps precisely onto the extended CFT first law
\begin{equation}
	dE = T dS + \tilde\Phi d\tilde Q - p\, d{\cal V} + \mu\, dC .
	\label{first law}
\end{equation}
Comparison fixes the pressure and chemical potential to be
\begin{equation}
	p = \frac{1}{d-1}\frac{E}{\cal V},
	\qquad
	\mu = \frac{1}{C}(E-TS-\tilde\Phi\tilde Q).
	\label{pmu}
\end{equation}
Finally, introducing the dimensionless variables
\begin{equation}
	x=\frac{r_h}{L}, \qquad y=\frac{q}{L^{d-2}},
\end{equation}
the thermodynamic quantities of the dual CFT take the form
\begin{equation}
	S = 4\pi C x^{d-1}, \qquad
	\tilde Q = 2\alpha(d-1)Cy, \qquad
	\tilde\Phi = \frac{1}{\alpha R}\frac{y}{x^{d-2}},
	\label{SQphi}
\end{equation}
\begin{equation}
	E = \frac{d-1}{R}Cx^{d-2}\!\left(1+x^2+\frac{y^2}{x^{2d-4}}\right),
	\qquad
	T = \frac{d-2}{4\pi R}\frac{1}{x}
	\left(1+\frac{d}{d-2}x^2-\frac{y^2}{x^{2d-4}}\right),
	\label{energytempcft}
\end{equation}
\begin{equation}
	\mu = \frac{x^{d-2}}{R}
	\left(1-x^2-\frac{y^2}{x^{2d-4}}\right).
	\label{mu}
\end{equation}
Equations~\eqref{SQphi}–\eqref{mu} constitute the complete set of thermodynamic relations
used throughout this work. They provide the fundamental input for constructing the
off-shell free energy and for analyzing the phase structure and kinetic behavior of the
dual CFT in the ensembles considered below.\\

\noindent
Using the holographic dictionary, the on-shell thermodynamics of CFT dual to charged AdS black holes was analyzed systematically in \cite{Cong:2021jgb} across a range of
(grand) canonical ensembles. In addition to the temperature--entropy pair $(T,S)$, the
CFT description involves three independent conjugate pairs,
$(\tilde\Phi,\tilde Q)$, $(p,{\cal V})$ and $(\mu,C)$. This structure leads to
$2^3=8$ distinct (grand) canonical ensembles. It was shown that five of these ensembles
do not exhibit any nontrivial phase structure.Nontrivial phase transitions or critical behavior were found in the remaining three
ensembles, namely those at fixed $(\tilde Q,{\cal V},C)$, $(\tilde\Phi,{\cal V},C)$ and
$(\tilde Q,{\cal V},\mu)$. The corresponding free energies, which will be central to our
analysis, are defined as
\begin{equation}
\begin{aligned}  \label{freeenergies2}
&\text{fixed} \quad (\tilde Q,{\cal V},C): \qquad
&&F \equiv E - TS = \tilde\Phi \tilde Q + \mu C\,,\\
&\text{fixed} \quad (\tilde\Phi,{\cal V},C): \qquad
&&W \equiv E - TS - \tilde\Phi \tilde Q = \mu C\,,\\
&\text{fixed} \quad (\tilde Q,{\cal V},\mu): \qquad
&&G \equiv E - TS - \mu C = \tilde\Phi \tilde Q\,.
\end{aligned}
\end{equation}

\noindent
In the following three sections, we extend these on-shell results to the off-shell
regime. For the appropriate ensembles we construct the corresponding off-shell free energy,
present the associated off-shell phase diagram, and formulate the stochastic dynamics
via a Fokker--Planck equation with appropriate boundary conditions. This framework allows
us to compute the mean first passage times for transitions between competing phases, as
well as the associated fluctuations.

\section{Off-shell phase diagrams} \label{sec:off}
This section contains three subsections. We discuss the off-shell phase diagrams in three different ensembles, namely, the fixed $(\tilde Q,{\cal V},C)$, $(\tilde\Phi,{\cal V},C)$ and
$(\tilde Q,{\cal V},\mu)$ ensembles, in subsections -(\ref{sec:ce}), (\ref{subsec:phase_diagrams_grand}) and (\ref{subsec:phase_diagrams_Unique}), respectively.

\subsection{Fixed $(\tilde Q,{\cal V},C)$ ensemble} \label{sec:ce}

\noindent
We now construct the off-shell free energy of the CFT in the fixed
$(\tilde Q,{\cal V},C)$ ensemble, and use it to extract the phase structure. We denote the off-shell free energy as
\begin{equation}
    F = E - T \, S \, ,
    \label{Helmholtz}
\end{equation}
written in terms of the thermodynamic variables introduced in
section~\ref{subsec:CFT quantities}, except with one difference. 
The off-shell free energy is defined by promoting the temperature $T$ to an independent control
parameter, rather than fixing it by its equilibrium relation in eqn. \eqref{energytempcft}.
Equilibrium states are then recovered as stationary points of the off-shell free energy.
By construction, forcing the temperature to satisfy the on-shell relation in eqn. \eqref{energytempcft} will reduce our off-shell free energy to be constructed below, to the
on-shell Helmholtz free energy given in~\cite{Cong:2021jgb} (see their eqn.3.3).\\

\noindent
In the current canonical ensemble, one expects the free energy to be a function of $(T,\tilde Q,{\cal V},C)$, in addition to the dependence on an order parameter. Since the CFT volume ${\cal V}$ in \eqref{vol} depends only on the boundary curvature radius
$R$, fixing ${\cal V}$ is equivalent to fixing $R$. We therefore work with $R$ instead of
${\cal V}$. The remaining task is to eliminate the bulk parameters $x$ and $y$ in favour of
the CFT data $\tilde Q$ and $C$. From the definition of $\tilde Q$ in \eqref{SQphi} we obtain
\begin{equation}
    y =\frac{{\tilde{Q}}}{2 \sqrt{2} {C} \sqrt{\frac{d-2}{d-1}} (d-1)} \, ,
    \label{y:canonical}
\end{equation}
and substituting this into the definition of $\tilde{\Phi}$ gives
\begin{equation}
    x = 4^{\frac{1}{2-d}}
\left(
\frac{C\,(d-2)\,R\,\tilde{\Phi}}{\tilde{Q}}
\right)^{\frac{1}{2-d}} .
\label{x:canonical}
\end{equation}

\noindent
In the fixed-$\tilde Q$ ensemble the parameter $y$ is fixed, so the only remaining continuous
bulk degree of freedom is $x$. Further, at fixed $(C,\tilde Q,R)$, the only free variable left is $\tilde \Phi$. Equation \eqref{x:canonical}, then motivates the possible choice of order parameter in the boundary CFT to be:
\begin{equation}
\tilde{\delta}\equiv \tilde{\Phi}^{\,1/(2-d)} \, .
\end{equation}
With this choice, we thus look for an
off-shell potential of the form $F(\tilde{\delta},T,\tilde Q,{\cal V},C)$, with $T$ as the control parameter and $\tilde{\delta}$ as the order parameter, with the  saddle points of $F$
expected to reproduce the competing equilibrium phases.\\

\noindent
Inserting \eqref{x:canonical} and \eqref{y:canonical} into \eqref{Helmholtz}, and rewriting the
result in terms of $\tilde{\delta}=\tilde{\Phi}^{1/(2-d)}$, yields the off-shell free energy in the
canonical ensemble,
\begin{equation}
\begin{aligned}
F(\tilde{\delta},T,\tilde Q,R,C)
&= \frac{C(d-1)}{R}
\left(
4^{\frac{1}{2-d}}
\left(
\frac{C(d-2)R\,\tilde{\delta}^{\,2-d}}{\tilde Q}
\right)^{\frac{1}{2-d}}
\right)^{d-2}
\\[4pt]
&\quad\times
\Bigg[
\frac{\tilde Q^{2}}{8C^{2}(d-2)(d-1)}
\left(
4^{\frac{1}{2-d}}
\left(
\frac{C(d-2)R\,\tilde{\delta}^{\,2-d}}{\tilde Q}
\right)^{\frac{1}{2-d}}
\right)^{4-2d}
\\[4pt]
&\qquad\qquad
+\,16^{\frac{1}{2-d}}
\left(
\frac{C(d-2)R\,\tilde{\delta}^{\,2-d}}{\tilde Q}
\right)^{-\frac{2}{d-2}}
+1
\Bigg]
\\[6pt]
&\quad
-4\pi C T
\left(
4^{\frac{1}{2-d}}
\left(
\frac{C(d-2)R\,\tilde{\delta}^{\,2-d}}{\tilde Q}
\right)^{\frac{1}{2-d}}
\right)^{d-1}.
\end{aligned}
\label{offFECan}
\end{equation}

\noindent
From here on, we set $d=4$, for which $\tilde{\delta}=\sqrt{1/\tilde{\Phi}}$, and the off-shell free
energy reduces to
\begin{equation}
F = \frac{1}{8}
\left[
\frac{3\,\tilde{\delta}^{4}\,\tilde Q^{2}}{8\,C\,R^{3}}
-\frac{\sqrt{2}\,\pi\,C\,T}{
\left(\dfrac{C\,R}{\tilde{\delta}^{2}\tilde Q}\right)^{3/2}
}
+\frac{4\,\tilde Q}{\tilde{\delta}^{2}}
+\frac{3\,\tilde{\delta}^{2}\,\tilde Q}{R^{2}}
\right].
\label{offshellFECanonical4}
\end{equation}

\begin{figure}
  \centering
  \includegraphics[width=7cm]{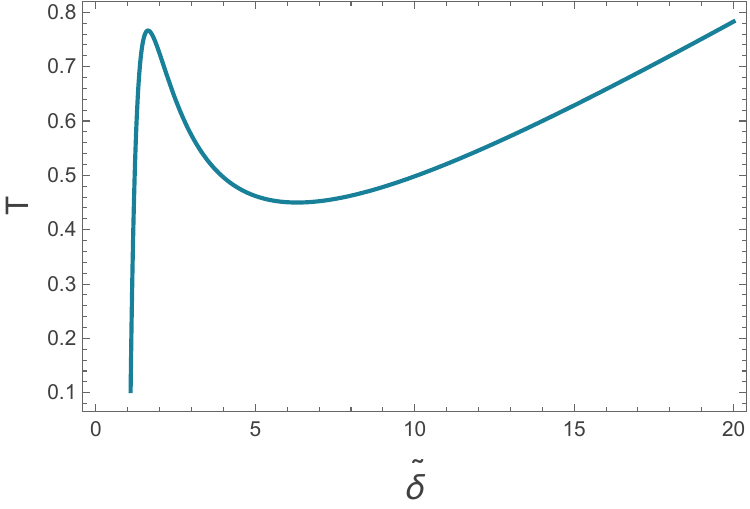}
   \hspace{0.2cm}
  \includegraphics[width=7cm]{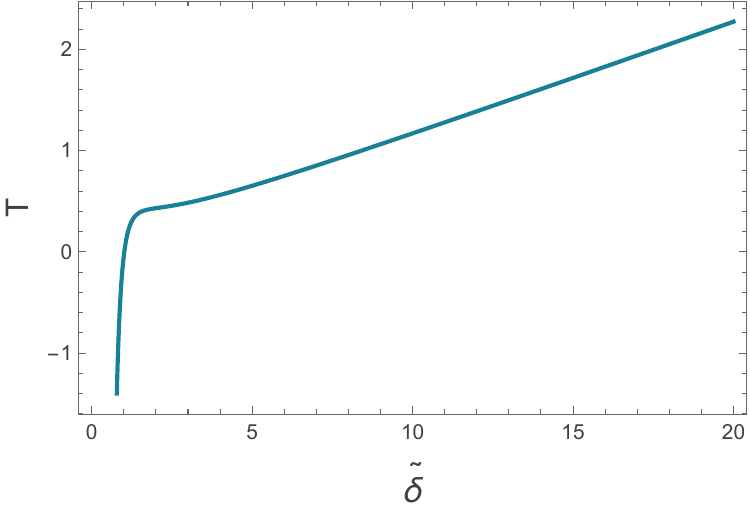}
  \caption{CFT temperature $T$ as a function of the order parameter $\tilde{\delta}$.
    Left: $C/\tilde Q > (C/\tilde Q)_{\text{crit}}$. Right: $C/\tilde Q < (C/\tilde Q)_{\text{crit}}$. }
   \label{fig_temp_can}
\end{figure}

\noindent
Before proceeding, we make a consistency check on our construction. Extremizing the off-shell free energy \eqref{offshellFECanonical4} with respect to $\tilde{\delta}$, i.e., imposing $\partial F/\partial\tilde{\delta}=0$ gives the expression for equilibrium temperature of the CFT as
\begin{equation}
T = \frac{
\tilde Q^{2}
\left(
\dfrac{C R}{\tilde{\delta}^{2}\tilde Q}
\right)^{5/2}
\left(
-16 C R^{3}
+12 C \tilde{\delta}^{4} R
+3 \tilde{\delta}^{6} \tilde Q
\right)
}{
6\sqrt{2}\,\pi\, C^{3} R^{4}
}.
\label{T4canonical}
\end{equation}
The temperature curve following from eqn. \eqref{T4canonical} is seen to exhibit a critical point in figure-(\ref{fig_temp_can}), which can be determined using standard method for finding inflection point as
\[
\frac{\partial T}{\partial \tilde{\delta}}
= 0
= \frac{\partial^{2} T}{\partial \tilde{\delta}^{2}} \, ,
\]
yielding
\begin{equation}
    \tilde{Q}_{crit} = \frac{4 {C_{}crit}}{3 \sqrt{5}},~~~~~~~~~~~~\tilde{\delta} = \sqrt{2} \sqrt[4]{5} \sqrt{{R}} \, .
    \label{Qdelcrit}
\end{equation}
Inserting \eqref{Qdelcrit} into \eqref{T4canonical} gives
$T_{crit} = \frac{4 \sqrt{3}}{5 \pi  {R}}$ and $(C/\tilde Q)_{\text{crit}} = 3\sqrt{5}/4$, in exact agreement
with \cite{Cong:2021jgb}.\\


\noindent
Figure~\ref{fig_temp_can} shows $T$ as a function of $\tilde{\delta}$ in two representative regimes:
$C/\tilde Q > (C/\tilde Q)_{\rm crit}$ (left panel) and
$C/\tilde Q < (C/\tilde Q)_{\rm crit}$ (right panel).
For $C/\tilde Q < (C/\tilde Q)_{\rm crit}$ the temperature is monotonic in $\tilde{\delta}$.
For $C/\tilde Q > (C/\tilde Q)_{\rm crit}$ it develops a local minimum and maximum, signalling the
multi-branch structure relevant for the phase transition.
For $C/\tilde Q > (C/\tilde Q)_{\rm crit}$, the extrema are fixed by
\begin{equation}
    \frac{\partial T}{\partial \tilde{\delta}} = 0\,,
    \label{Tderi}
\end{equation}
which determines $\tilde{\delta}_{\min}$ and $\tilde{\delta}_{\max}$, and hence
$T_{\min}$ and $T_{\max}$ from \eqref{T4canonical}. For generic $(R,C,\tilde Q)$ the explicit expressions
are unwieldy; we determine them numerically. For $C/\tilde Q > (C/\tilde Q)_{\rm crit}$ in the window $T_{\min}<T<T_{\max}$ three branches coexist,
corresponding to low-, intermediate-, and high-entropy configurations, with the intermediate branch unstable.
This is the standard first-order small/large-entropy structure in the canonical ensemble.\\

\begin{figure}[!htbp]
    \centering
    \includegraphics[width=.7\linewidth]{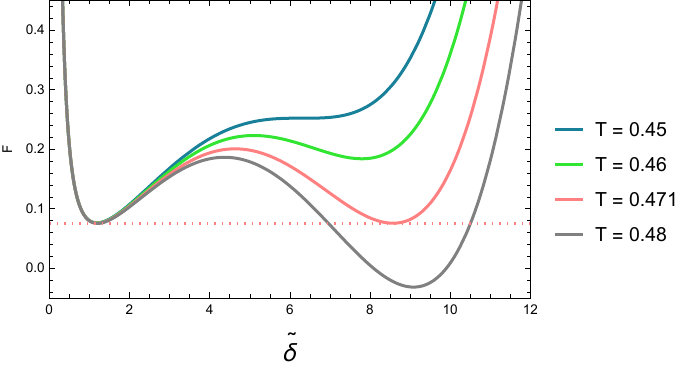}
    \caption{Off-shell free energy $F$ versus $\tilde{\delta}$ for $R=1$, $C=1$, $\tilde Q=0.1$.
    The temperature increases from top to bottom:
    $T=0.45,\,0.46,\,0.471,\,0.48$ (blue, green, red, grey).
    The red curve corresponds to $T=T_{\rm tran}=0.471$, where the low- and high-entropy minima are degenerate.
    For these parameters $T_{\min}=0.44997$ and $T_{\max}=0.766396$.}
    \label{figphaseDcan1}
\end{figure}

\noindent
We now turn to the off-shell phase diagrams. Figure~\ref{figphaseDcan1} shows $F$ as a function of $\tilde{\delta}$
for $C/\tilde Q>(C/\tilde Q)_{\rm crit}$.

\FloatBarrier

\noindent
For $T_{\min}<T<T_{\max}$ the potential admits three stationary points. Since smaller $\tilde{\delta}$
corresponds to smaller $x$, \eqref{SQphi} implies the left minimum is a low-entropy phase, the right minimum a
high-entropy phase, and the intervening maximum the unstable intermediate branch. The transition temperature
$T_{\rm tran}$ is obtained by solving for the value of $T$ at which the two stable minima have equal free energy.
For $T_{\min}<T<T_{\rm tran}$ the low-entropy phase dominates; for $T_{\rm tran}<T<T_{\max}$ the high-entropy phase
dominates. Outside the window $[T_{\min},T_{\max}]$ the potential has no stationary point and no equilibrium
configuration. Away from the stationary points, the curve interpolates between fluctuating configurations, which
is precisely the off-shell content absent in the on-shell description.\\

\noindent
The equilibrium values of the order parameter $\tilde{\delta}$ follow from the
off-shell free energy~\eqref{offshellFECanonical4} by imposing the stationarity
condition
$\partial F/\partial\tilde{\delta}=0$.
This yields $\tilde{\delta}$ as a function of the thermodynamic parameters
$(T,R,C,\tilde{Q})$.
Since closed-form expressions are not available, the solutions are obtained
numerically.
We denote the resulting values of the order parameter corresponding to the
low-, intermediate-, and high-entropy CFT states by
$\tilde{\delta}_l$, $\tilde{\delta}_m$, and $\tilde{\delta}_h$, respectively.\\

\begin{figure}[!htbp]
    \centering
    \includegraphics[width=.7\linewidth]{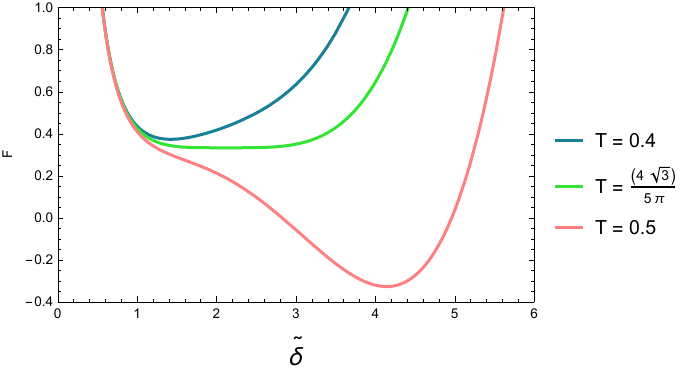}
    \caption{Off-shell free energy $F$ versus $\tilde{\delta}$ at $C/\tilde Q=(C/\tilde Q)_{\rm crit}$.
    Parameters: $R=1$, $C=1$, $\tilde Q=\frac{4}{3\sqrt{5}}$.
    Temperatures (top to bottom): $T=0.4$, $T=\frac{4\sqrt{3}}{5\pi}$, $T=0.5$ (blue, green, red).
    The green curve corresponds to $T=T_{\rm crit}=\frac{4\sqrt{3}}{5\pi}$, where the transition becomes second order.}
    \label{figphaseDcan2}
\end{figure}

\noindent
At the critical ratio $C/\tilde Q=(C/\tilde Q)_{\rm crit}$ the phase diagram changes to the degenerate case shown
in Fig.~\ref{figphaseDcan2}.

\begin{figure}[!htbp]
    \centering
l    \includegraphics[width=.7\linewidth]{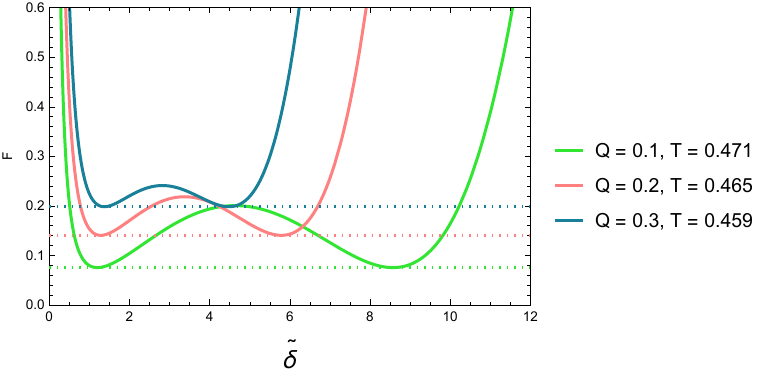}
    \caption{Off-shell free energy $F$ versus $\tilde{\delta}$. For all curves
   $R=1$, $C=1$. For green, red and blue curves $\tilde Q = 0.1, T = T_{trans} = 0.471 $, $\tilde Q = 0.2, T = T_{trans} = 0.465 $ and $\tilde Q = 0.3, T = T_{trans} = 0.459$ respectively.
   }
    \label{figphaseDcandiffQ}
\end{figure}
\FloatBarrier

\noindent
At $C/\tilde Q=(C/\tilde Q)_{\rm crit}$ there is a single stationary point for each $T$. For $T<T_{\rm crit}$ it lies on
the low-entropy side, while for $T>T_{\rm crit}$ it moves to the high-entropy side. The crossover at $T=T_{\rm crit}$
is continuous, i.e.\ second order. We have also checked that for $C/\tilde Q<(C/\tilde Q)_{\rm crit}$ the qualitative
structure remains single-branched (no phase transition), and we therefore do not display it separately.\\

\noindent
Comparative plots for phase diagrams at different $\tilde{Q}$ has been given in Fig \ref{figphaseDcandiffQ}. We can see that with increasing $\tilde{Q}$, both the transition temperature and the free energy where the transition happens decreases.

\subsection{Fixed $(\tilde \Phi,{\cal V},C)$ ensemble}
\label{subsec:phase_diagrams_grand}
We now construct the off-shell free energy of the holographic CFT in the fixed
$(\tilde \Phi,{\cal V},C)$ ensemble, starting from the expression
\begin{equation}
   W \equiv E - TS - \tilde\Phi \tilde Q 
    \label{Gibbs}
\end{equation}
written in terms of the thermodynamic variables introduced in
section~\ref{subsec:CFT quantities}. For fixed $(\tilde \Phi,{\cal V},C)$ ensemble case, 
$\tilde{Q}$ turns out to be the correct order parameter. Hence, we seek off-shell free energy of the form $W(Q,T,\tilde \Phi,{\cal V},C)$, where now, apart from temperature $T$, $\tilde\Phi$ is also treated as a control parameter 
and not apriori fixed to their on-shell values given in~\ref{subsec:CFT quantities}. Equilibrium states are then recovered as stationary points of the off-shell free energy.
By construction, imposing the on-shell values of the control parameters reduces the off-shell expression to usual Gibbs free energy.\\

\noindent
We first use the definition of $\tilde Q$ and $\tilde \Phi$ in eq~\ref{SQphi} to eliminate variables $x$ and $y$ as
\begin{equation}
    x =
4^{\frac{1}{2-d}}
\left(
\frac{C\,(d-2)\,R\,\tilde{\Phi}}{\tilde{Q}}
\right)^{\frac{1}{2-d}},~~~~~~~~~y = \frac{\sqrt{\frac{d-2}{d-1}}\,\tilde{Q}}
{2\sqrt{2}\,C\,(d-2)} \, .
\label{xygrand}
\end{equation}
Plugging eq~\ref{xygrand} in~\ref{Gibbs}, and treating $T$ and $\tilde \Phi$ as free parameters, gives rise to our desired off-shell free energy as
\begin{equation}
\begin{aligned}
W(Q,T,\tilde \Phi,{\cal V},C) = &\frac{C(d-1)}{R}
\Bigg[
\frac{\tilde Q^{2}}{8C^{2}(d-2)(d-1)}
\left(
4^{\frac{1}{2-d}}
\left(
\frac{C(d-2)R\tilde{\Phi}}{\tilde Q}
\right)^{\frac{1}{2-d}}
\right)^{4-2d}
\\[6pt]
&\qquad
+\,16^{\frac{1}{2-d}}
\left(
\frac{C(d-2)R\tilde{\Phi}}{\tilde Q}
\right)^{-\frac{2}{d-2}}
+ 1
\Bigg]
\left(
4^{\frac{1}{2-d}}
\left(
\frac{C(d-2)R\tilde{\Phi}}{\tilde Q}
\right)^{\frac{1}{2-d}}
\right)^{d-2}
\\[8pt]
&\quad
- 4\pi C T
\left(
4^{\frac{1}{2-d}}
\left(
\frac{C(d-2)R\tilde{\Phi}}{\tilde Q}
\right)^{\frac{1}{2-d}}
\right)^{d-1}
- \tilde Q \tilde{\Phi} \, .
\end{aligned}
\end{equation}
Once again, specializing to the case $d = 4$ for brevity, the off-shell free energy is
\begin{equation}
W = \frac{1}{8}
\left[
\frac{3\,\tilde Q^{2}}{8\,C\,R^{3}\tilde{\Phi}^{2}}
-
\frac{\sqrt{2}\,\pi\,C\,T}
{\left(\dfrac{C R \tilde{\Phi}}{\tilde Q}\right)^{3/2}}
+
\frac{3\,\tilde Q}{R^{2}\tilde{\Phi}}
-
4\,\tilde Q\,\tilde{\Phi}
\right].
\label{offshellgrand4}
\end{equation}
Now, the equilibrium temperature of the CFT can be obtained by extremizing the
off-shell free energy \eqref{offshellgrand4} with respect to $\tilde{Q}$.
Imposing $\partial W/\partial\tilde{Q}=0$ gives

\begin{equation}
T = \frac{
C \tilde{\Phi}
\left[
4 C R \tilde{\Phi}
\left(3 - 4 R^{2} \tilde{\Phi}^{2}\right)
+ 3 \tilde Q
\right]
}
{
6 \sqrt{2}\,\pi\, \tilde Q^{2}
\left(
\dfrac{C R \tilde{\Phi}}{\tilde Q}
\right)^{3/2}
}.\label{Tempgrand4}
\end{equation}
\begin{figure}
  \centering
  \includegraphics[width=7cm]{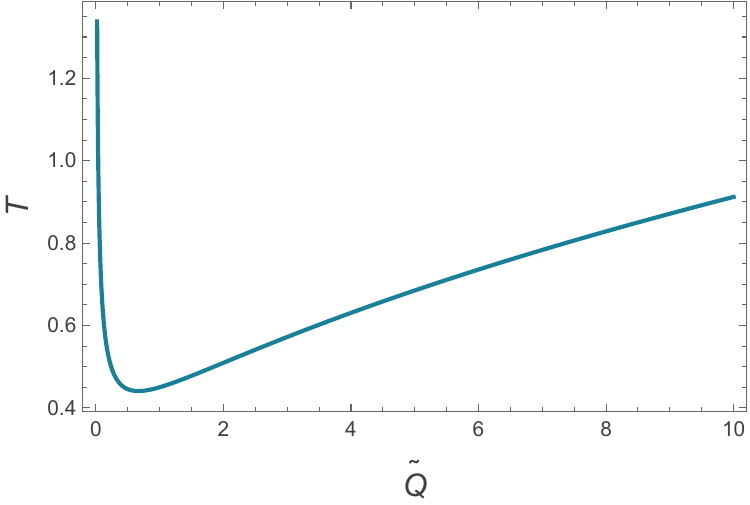}
   \hspace{0.2cm}
  \includegraphics[width=7cm]{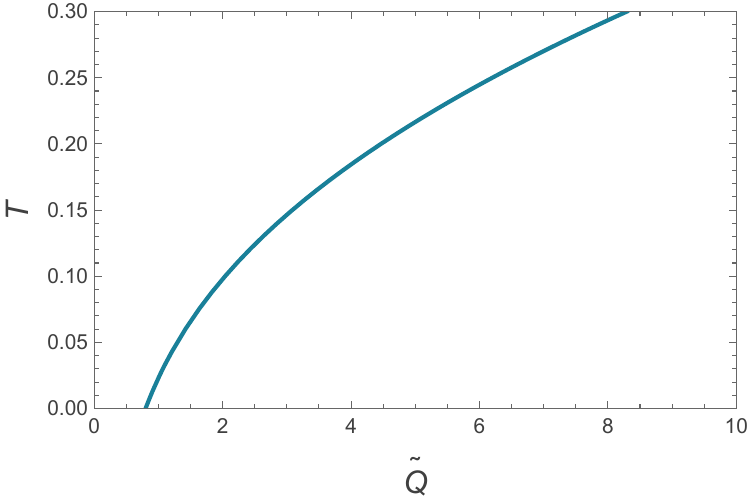}
  \caption{CFT temperature $T$ as a function of the order parameter $\tilde{Q}$.
    Left: $\tilde{\Phi} = (0.2) \tilde{\Phi}_{crit}$. Right:$\tilde{\Phi} = (1.1)\tilde{\Phi}_{crit}$ }
   \label{fig_temp_grand}
\end{figure}

\noindent
Figure-(\ref{fig_temp_grand}) displays this CFT temperature $T$ as a function of the order
parameter $\tilde Q$ in the grand canonical ensemble. We observe the existence of a
critical value of the electric potential,
$\tilde{\Phi}_{\rm crit} = \sqrt{3}/2$, which is independent of the boundary
curvature radius $R$ and the central charge $C$. For
$\tilde{\Phi} < \tilde{\Phi}_{\rm crit}$, the temperature curve develops two distinct
branches separated by a minima (left panel), whereas for $\tilde{\Phi} > \tilde{\Phi}_{\rm crit}$ only a single branch remains (right panel).\\

\begin{figure}[!htbp]
    \centering
    \includegraphics[width=.7\linewidth]{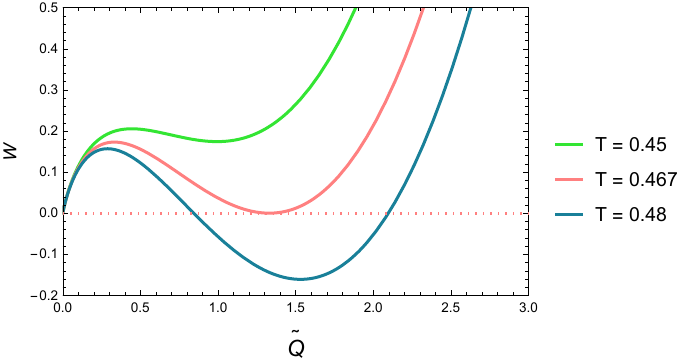}
    \caption{Off-shell free energy $W$ versus $\tilde{Q}$. For all curves
   $R=1$, $C=1, \Phi = (0.2)\Phi_{crit}$. For green, red and blue curves represents $T = 0.45, T= T_{cd} = 0.467$ and $T = 0.48$ respectively.
   }
    \label{fig_dual_HP}
\end{figure}

\noindent
We now study the off-shell phase diagrams in the grand canonical ensemble.
Figure~\ref{fig_dual_HP} shows $W$ as a function of the
order parameter $\tilde Q$ for $\tilde{\Phi} < \tilde{\Phi}_{\rm crit}$.
From the left panel of Fig.~\ref{fig_temp_grand}, it is clear that in this regime
there exists a minimum temperature $T_{\min}$, below which no CFT state is present.
For $T > T_{\min}$, two CFT states appear: a low-entropy state and a high-entropy
state, corresponding respectively to a local maximum and a local minimum of $W$. There exists a temperature $T = T_{\rm cd}$ at which a first-order phase transition
takes place between the confined phase ($W=0$) and the deconfined phase
(the large-entropy CFT state). This transition is dual to the generalized
Hawking--Page transition. The transition temperature can be obtained by
solving the simultaneous conditions $\frac{\partial W}{\partial \tilde Q}=0$,
together with $W=0$.\\

\begin{figure}[!htbp]
    \centering
    \includegraphics[width=.7\linewidth]{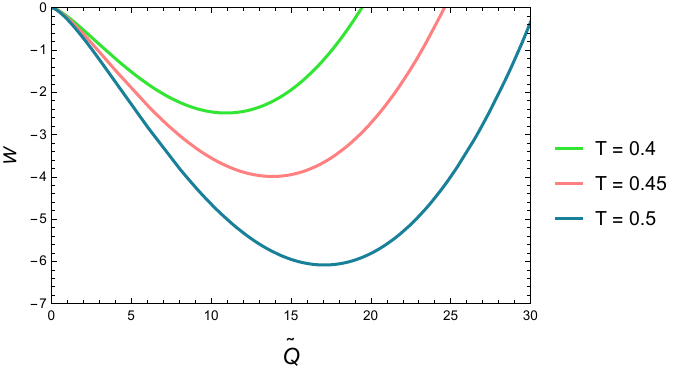}
    \caption{Off-shell free energy $W$ versus $\tilde{Q}$. For all curves
   $R=1$, $C=1, \Phi =\Phi_{crit}$. For green, red and blue curves represents $T = 0.4, T= T_{cd} = 0.45$ and $T = 0.5$ respectively.
   }
    \label{fig_phaseD_grand_2}
\end{figure}

\noindent
The phase diagrams for $\tilde{\Phi} = \tilde{\Phi}_{\rm crit}$ are shown in
Fig.~\ref{fig_phaseD_grand_2}. At the critical value of the electric potential,
the low-entropy CFT branch disappears. The large-entropy CFT state always has
negative free energy, and therefore no confinement--deconfinement type transition
occurs in this case. For $\tilde{\Phi} > \tilde{\Phi}_{\rm crit}$, the phase diagrams are qualitatively similar to Fig.~\ref{fig_phaseD_grand_2}. The only difference is that the entire free-energy curve shifts further downward (i.e.\ becomes more negative).
Since no new structure appears, we do not display those cases separately.

\subsection{$(\tilde Q,{\cal V},\mu)$ ensemble}
\label{subsec:phase_diagrams_Unique}

\begin{figure}
  \centering
  \includegraphics[width=7cm]{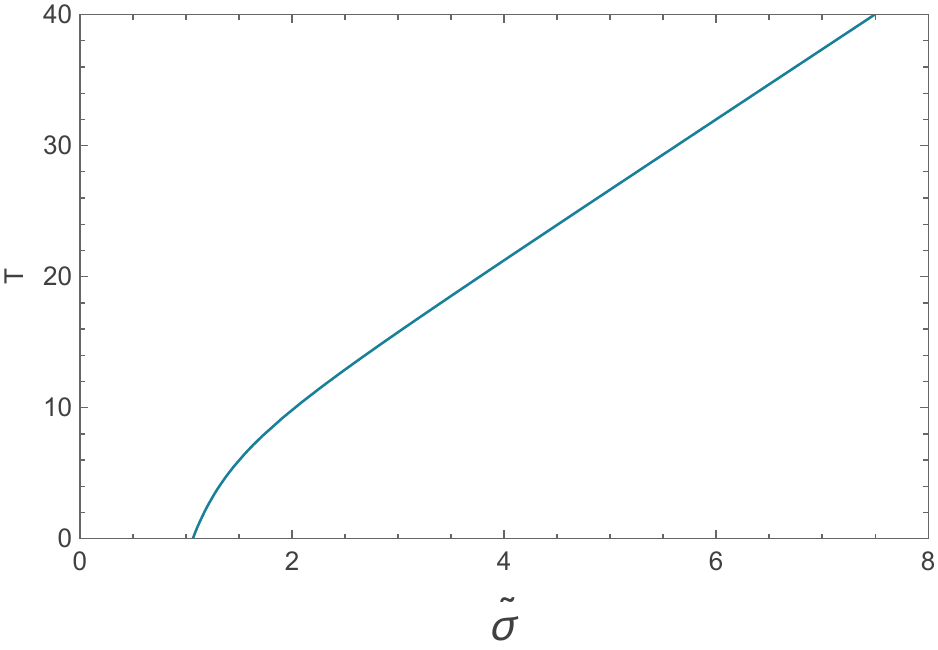}
   \hspace{0.2cm}
  \includegraphics[width=7cm]{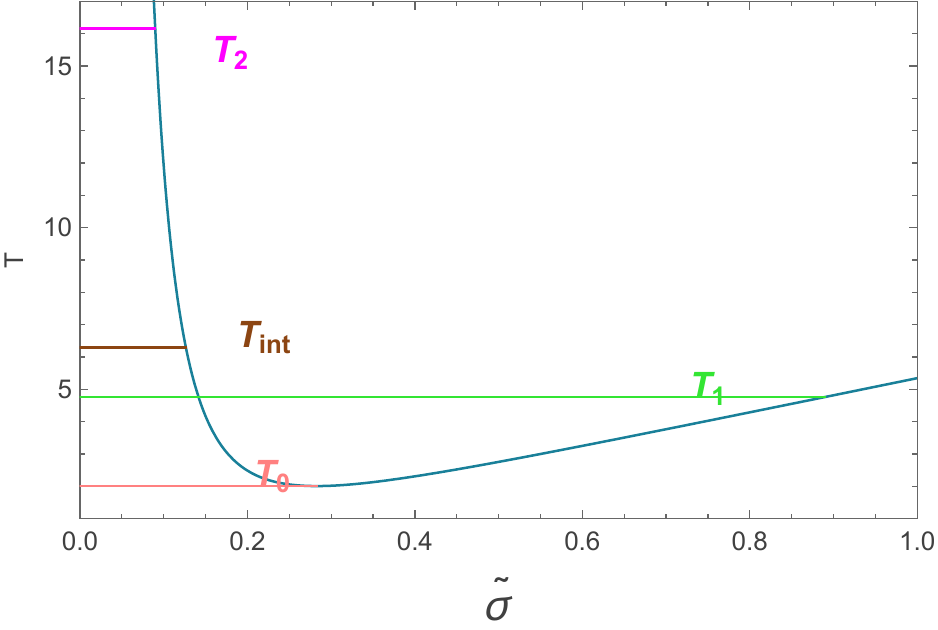}
  \caption{For both panels we fix $\tilde{Q} = 1$, $R = 0.1$. The left panel corresponds to $\mu = -60$  while the right panel corresponds to $\mu = .1$}
   \label{fig_temp_Unique}
\end{figure}

In this subsection we construct the off-shell free-energy landscape for an ensemble with fixed $(\tilde Q,{\cal V},\mu)$, first introduced in ref.~\cite{Cong:2021jgb}. In contrast to the earlier ensembles, where the central charge C was fixed (i.e., fixed number of degrees of freedom or $N^2$ in the large $N$ $SU(N)$ gauge theory), fixing the chemical potential $\mu$ means one is allowed to vary $C$, which works in a situation when one considers a family of dual CFTs (i.e., a varying $\Lambda$ and $G_N$ in the current context). This ensemble is particularly interesting since the system exhibits a zeroth-order phase transition, and the off-shell construction of such a transition in the context of black holes is quite novel. The formal expression for the free energy in this case is~\cite{Cong:2021jgb},
\begin{equation}
\label{Unique_Onshell}
G=E-TS-\mu C.
\end{equation}
Now,  to construct the off-shell free energy, we keep $\mu$ fixed as the ensemble control parameter, and treat $T$ as the external temperature of the heat bath, without imposing the on-shell relations of eqs.~\eqref{energytempcft} and \eqref{mu}. Once again, we continue to work with $R$ instead of
${\cal V}$. Thus, we expect the off-shell free energy to be a function of the variables as $G(\tilde Q, R,\mu,T)$,  together with its dependence on a yet to be chosen order parameter. To identify the order parameter we proceed by rearranging the remaining thermodynamic variables. First, from eq.~\eqref{SQphi}, the central charge can be expressed as
\begin{equation}
\label{Unique_C}
C=\frac{\tilde Q}{2\alpha(d-1)y}.
\end{equation}
and inversion of eq.~\eqref{mu} gives
\begin{equation}
\label{Unique_y}
y=x^{\frac{d-4}{2}}
\sqrt{-x^{d+2}+x^d-\mu R x^2}.
\end{equation}
Now, to identify the order parameter, one can actually
eliminate $y$ from the expressions for $\tilde Q$ and $\tilde\Phi$ in eq.~\eqref{SQphi}, to see that
\begin{equation}
\label{Unique_x}
x=
\left(
\frac{\tilde Q}{2\alpha^2 C(d-1)R\tilde\Phi}
\right)^{\frac{1}{d-2}}.
\end{equation}
Equation~\eqref{Unique_x} shows that, for fixed $\tilde Q$ and $R$, it is the combination $C\tilde\Phi$ which remains to be chosen. 
This motivates us to introduce the order parameter
\begin{equation} \label{sigma}
\tilde{\sigma}
=
\left(\frac{1}{\tilde\Phi C}\right)^{\frac{1}{d-2}}.
\end{equation}
With this set up, we now substitute eqs.~\eqref{Unique_C}, \eqref{Unique_y}, and \eqref{Unique_x} successively into eq.~\eqref{Unique_Onshell}, to obtain the desired off-shell free energy for the fixed $(\tilde Q,{\cal V},\mu)$ ensemble as 
\begin{equation}
\label{Unique_Offshell}
\begin{aligned}
G(\tilde Q, R, \mu, T, \tilde{\sigma} )=
\frac{\tilde Q\,\mathcal{A}^{-\frac{d}{2}}}{2\sqrt{2}\sqrt{d-2}\sqrt{d-1}\,R\,
\sqrt{\mathcal{B}}}
\Big[
&-\,16^{\frac{1}{2-d}}\,d\,\mu R
\left(\frac{\tilde Q\,\tilde{\sigma}^{d-2}}{(d-2)R}\right)^{\frac{2}{d-2}}
\\[4pt]
&+\,2(d-1)\,\mathcal{A}^{d}
-4\pi R T\,\mathcal{A}^{d+1}
\Big],
\end{aligned}
\end{equation}
where we have defined
\begin{equation}
\mathcal{A}
=
4^{\frac{1}{2-d}}
\left(
\frac{\tilde Q\,\tilde{\sigma}^{d-2}}{(d-2)R}
\right)^{\frac{1}{d-2}},
\end{equation}
\begin{equation}
\mathcal{B}
=
\left[
\left(
-\frac{\tilde Q\,\tilde{\sigma}^{d-2}}{8R-4dR}
\right)^{\frac{1}{d-2}}
\right]^{d}
-
16^{\frac{1}{2-d}}
\left(
\frac{\tilde Q\,\tilde{\sigma}^{d-2}}{(d-2)R}
\right)^{\frac{2}{d-2}}
\left\{
\left[
\left(
-\frac{\tilde Q\,\tilde{\sigma}^{d-2}}{8R-4dR}
\right)^{\frac{1}{d-2}}
\right]^{d}
+\mu R
\right\}.
\end{equation}
\noindent
Here, we have used the definition of $\alpha$ given in eq.~\eqref{eq:electriccharge1}.
From now on we restrict the discussion to $d=4$. In this case the off-shell free
energy takes the form
\begin{equation}
\label{Unique_Offshell_d4}
G(\tilde Q,R,\mu,T,\tilde{\sigma})=
\frac{\tilde Q\left[
-\pi R^{2}T\left(\frac{\tilde Q\tilde{\sigma}^{2}}{R}\right)^{3/2}
+3\sqrt{2}\,\tilde Q\tilde{\sigma}^{2}
-16\sqrt{2}\,\mu R^{2}
\right]}
{\sqrt{3}\,R^{2}
\sqrt{
-\frac{\tilde Q\tilde{\sigma}^{2}}{R^{3}}
\left(
\tilde Q^{2}\tilde{\sigma}^{4}
-8\tilde Q R\tilde{\sigma}^{2}
+64\mu R^{3}
\right)
}} .
\end{equation}
We must remember that in the above free energy, $\mu$ and $T$ are free parameters and can apriori take any value. The equilibrium CFT temperature can  be obtained from the off-shell
free energy \eqref{Unique_Offshell_d4}, by imposing the saddle-point condition
$\partial G/\partial\tilde{\sigma}=0$, and comes out to be
\begin{equation}
\label{Unique_T_off}
T(\tilde{\sigma})=
\frac{3\tilde Q^{2}\tilde{\sigma}^{4}+64\mu R^{3}}
{4\sqrt{2}\pi\,\tilde{\sigma}^{3}(\tilde Q R)^{3/2}} \, ,
\end{equation}
\noindent
which is shown in Fig.~\ref{fig_temp_Unique} as a function of the order parameter
$\tilde{\sigma}$. For negative
$\mu$, there is a single branch corresponding to a single CFT state. In
contrast, for positive $\mu$, two branches appear, corresponding to the
low- and high-entropy CFT phases. \\

\noindent Before presenting the off-shell phase
diagrams, we first elaborate on the various temperatures regimes shown in the
right panel of Fig.~\ref{fig_temp_Unique}, and later on discuss how the phase structure
changes around them.
We first find the value of critical temperature where the free energy vanishes. To do so we set $G = 0$ from eqn. (\ref{Unique_Offshell_d4}), to find that 
\begin{equation}
\label{Unique_T_mu0}
T_{G=0}=
\frac{\sqrt{2}\left(3\tilde Q\,\tilde{\sigma}^{2}-16\mu R^{2}\right)}
{\pi\,\tilde{\sigma}^{3}\sqrt{\tilde Q^{3}R}}.
\end{equation}
The values of the order parameter where the equilibrium temperature in eqn. (\ref{Unique_T_off}) reaches the temperature obtained in eqn. (\ref{Unique_T_mu0}), gives two distinct solutions for $\tilde{\sigma}$, corresponding to high entropy and low entropy states respectively, as 
\begin{equation}\label{sigma_crit}
\tilde{\sigma}_{\rm c1}
=
2\sqrt{
\frac{
R\left(\sqrt{1-4\mu R} + 1\right)
}{
\tilde Q
}
}\, , \qquad \qquad 
\tilde{\sigma}_{\rm c2}
=
2\sqrt{
\frac{
R\left(\sqrt{1-4\mu R} -1\right)
}{
\tilde Q
}
}.
\end{equation}
Now plugging the values of the order parameters in eqn. (\ref{sigma_crit}) into either \ref{Unique_T_off} or \ref{Unique_T_mu0} gives rise to their corresponding critical temperatures 
 \begin{equation}\label{T1T2}
T_1=
\frac{
-4\mu R+3\sqrt{1-4\mu R}+3
}{
\sqrt{2}\pi R\left(1+\sqrt{1-4\mu R}\right)^{3/2}
}\, ,~~~~~~~
T_2=
\frac{
-4\mu R-3\sqrt{1-4\mu R}+3
}{
\sqrt{2}\pi R\left(1-\sqrt{1-4\mu R}\right)^{3/2}
}\, ,
\end{equation}
which both agree with the corresponding expressions in~\cite{Cong:2021jgb} (see their eqn. (3.15)). The CFT which is a low entropy state until $T_1$, gives way to a new CFT state with low entropy after $T_1$. This new low entropy CFT state also ceases to exist at $T_1$. Since, the free energy itself makes an abrupt jump as the temperature changes beyond $T_1$, this marks the transition as zeroth order. The coincident point where $T_1=T_2$, gives a bound on the value of the chemical potential as $\mu_{\rm coin}=1/(4R)$, while for $\mu>\mu_{\rm coin}$ the temperatures become complex, and hence
no physical CFT branch exists. Thus, a nontrivial phase structure is present only
for $0<\mu<\mu_{\rm coin}$. Hence, while discussing the phase diagrams below, we focus our attention on two representative cases: $\mu \leq 0$, and  $0<\mu<\mu_{\rm coin}$. \\

\noindent 
The minimum temperature $T_0$ shown in Fig.~\ref{fig_temp_Unique}, below which no CFT states exist. $T_0$ is easily found from the stationary point of $T(\tilde{\sigma})$, i.e.,
\begin{equation*}
   \frac{\partial T(\tilde{\sigma})}{\partial\tilde{\sigma}} = 0~~~ \text{at} ~~~ \tilde{\sigma} = \frac{2\sqrt{2}\,\mu^{1/4} R^{3/4}}{Q^{1/2}} \, ,
\end{equation*}
giving us
\begin{equation}
    T_0 = \frac{2}{\pi R}(\mu R)^{1/4}\, .\\
\end{equation}

\noindent
Next we would like to know the stationary points of the off-shell free energy $G(\tilde{\sigma})$ (Eq \ref{Unique_Offshell_d4}), i.e., which is generally obtained by looking for solutions of the following simultaneous equations
\begin{equation}
    \frac{\partial G(\tilde{\sigma})}{\partial \tilde{\sigma}} = 0 = \frac{\partial^2 G(\tilde{\sigma})}{\partial \tilde{\sigma}^2} \, ,
\end{equation}
\noindent
which straightforwardly gives
 \begin{equation}
T_{\rm int}
=
\frac{\mu\left(\frac{1}{\mu R}\right)^{3/2}(12\mu R+1)}
{4\sqrt{2}\pi}~~~~~\text{at}~~~~~ \tilde{\sigma}_{int} = \frac{4R\sqrt{\mu}}{\sqrt{Q}}
\end{equation}
In this ensemble at fixed $(Q,R)$,  the various temperatures $T_0, T_1, T_2$ and $T_{\rm int}$ and their corresponding values of order parameters are all fixed once a value of $\mu$ is chosen. 
We note that  $T_{\rm int}$ (and the corresponding $\tilde{\sigma}_{int}$) marks certain stationary points around which the CFT states will change their nature from being stable to unstable, as the CFT $T$ temperature crosses $T_{\rm int}$. One should of course remember that the stationary point will be clearly visible when the equilibrium temperature $T =T_{\rm int}$, i.e., the point where the maximum and minimum coalesce. For nearby temperatures, the extrema are split apart, producing one minimum and one maximum on the opposite sides, precisely a situation corresponding to saddle-node bifurcation\footnote{More details on the fold catastrophe can be found in~\cite{arnold1992catastrophe,strogatz2018nonlinear}.}. This will clearly be seen in figure-(\ref{fig_PhaseD_Unique_near_T1}) in the sequel. 
In particular, there exists a critical value
\begin{equation}
\mu_*=\frac{16\sqrt{7}-35}{324R},
\end{equation}
such that for $\mu_*<\mu<\mu_{\rm coin}$ the transition occurs between the stable low-entropy branch and the high-entropy branch, while for $0<\mu<\mu_*$ it occurs between the unstable low-entropy branch and the stable high-entropy branch. $\mu_*$ can be calculated by equating $T_{int} = T_1$ and solving for $\mu$. 

\subsubsection{Case-1: $0<\mu<\mu_{\rm coin}$ }
\noindent
In the following we set $R=0.1,  \tilde{Q} = 1$, for which $\mu_*\simeq0.22$, and consider $\mu=0.1$, which still keeps the discussion general. In this case, the various temperatures of importance for studying the phase diagram are:
\begin{equation*}
      T_0 = 2.01, ~~T_1 = 4.77,~~ T_{int} = 6.30,~~ T_2 = 16.16
     \end{equation*}
As the temperature is a free variable and reaches close to the above temperatures, the phase diagram is expected to show different features, which we now explore. In all the figures to follow, for reference, the value of the order parameter $\tilde{\sigma} = \tilde{\sigma}_{int} $ (which does not change with temperature) corresponding to $T_{\rm int}$, is also marked. The use of  $\tilde{\sigma} $ will be seen in the phase diagrams close to $T_{\rm int}$, where it will correspond to the spinodal point.  \\

\begin{figure}[!htbp]
    \centering
    \includegraphics[width=.7\linewidth]{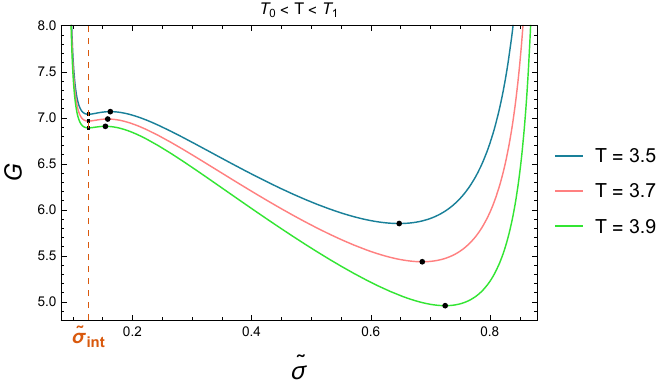}
    \caption{Off-shell free energy $G$ versus $\tilde{\sigma}$ for $R=.1$, $\tilde{Q}=1$, $\mu=0.1$.
    The temperature increases from top to bottom:
    $T=3.5,\,3.7,\,3.9$ (blue, red, green).}
    \label{fig_phaseD_Unique_T0_T1}
\end{figure}

\begin{figure}[!htbp]
    \centering
    \includegraphics[width=.7\linewidth]{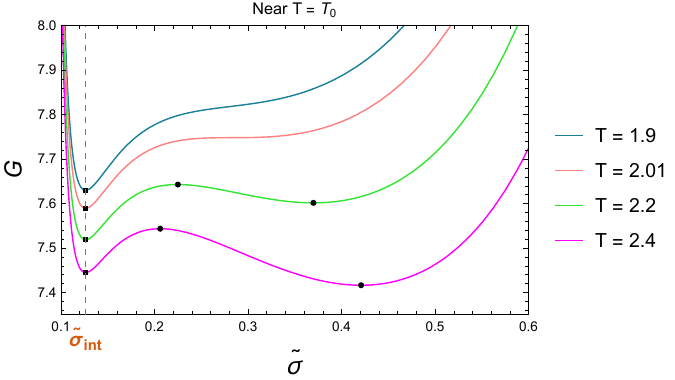}
   \caption{Off-shell free energy $G$ as a function of $\tilde{\sigma}$ for $R=0.1$, $\tilde Q=1$, and $\mu=0.1$. The blue, red, green, and magenta curves correspond to $T=1.9$, $T=T_{0}=2.01$, $T=2.2$, and $T=2.4$, respectively.}
    \label{fig_PhaseD_Unique_near_T0}
\end{figure}

\begin{figure}[!htbp]
    \centering
    \includegraphics[width=.7\linewidth]{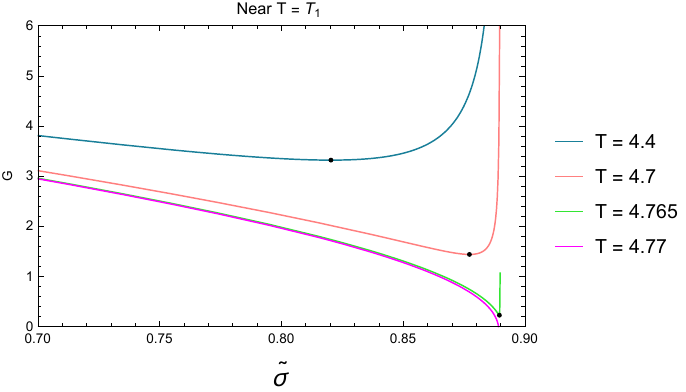}
   \caption{Off-shell free energy $G$ as a function of $\tilde{\sigma}$ for $R=0.1$, $\tilde Q=1$, and $\mu=0.1$. The blue, red, green, and magenta curves correspond to $T=4.4$, $T=4.7$, $T=4.765$, and $T = T_1 =4.77$, respectively.}
    \label{fig_PhaseD_Unique_near_T1}
\end{figure}

\noindent
\bm{$T_0<T<T_1$} :\\
\noindent
The off-shell phase diagrams for $T_0<T<T_1$ are shown in
Fig.~\ref{fig_phaseD_Unique_T0_T1}. We see two physical CFT branches 
present in this temperature range: the unstable low-entropy CFT state, which
appears as a local maximum, and the stable high-entropy CFT state, which appears
as the right local minimum. As the temperature increases, the entropy of the
high-entropy branch increases, while that of the low-entropy branch decreases. \\
\\
\noindent
\textbf{Near} \bm{$ T = T_0$} :\\
\noindent
The off-shell phase diagrams near $T=T_0$ are shown in
Fig.~\ref{fig_PhaseD_Unique_near_T0}. We see that for $T>T_0$ two CFT branches
are present: an unstable low-entropy CFT state and a stable high-entropy CFT
state, represented by the green and magenta curves. By contrast, for
$T\leq T_0$ no CFT state exists (no extrema), as illustrated by the blue and red curves. \\
\\
\noindent
\textbf{Near} \bm{$ T = T_1$} :\\
\noindent
The off-shell phase diagrams near $T=T_1$ are shown in
Fig.~\ref{fig_PhaseD_Unique_near_T1}. The unstable low-entropy CFT branch is present in this region, but is not displayed here due to the different parameter range; our focus is on the behavior of the high-entropy branch close to $T_1$.  We observe that the high-entropy CFT state terminates at $T=T_1$. For $T>T_1$, only the low-entropy CFT state persists, up to $T=T_2$. Since the free energy is discontinuous across this transition, it corresponds to a zeroth-order phase transition. For $T > T_1$, the free energy of the high entropy CFT state $G(\tilde{\sigma}_h)$ becomes complex, hence our order parameter has a natural upper bound given as $\tilde\sigma_{cut} = \tilde\sigma_h(T_1)$.\\

\begin{figure}[!htbp]
    \centering
    \includegraphics[width=.7\linewidth]{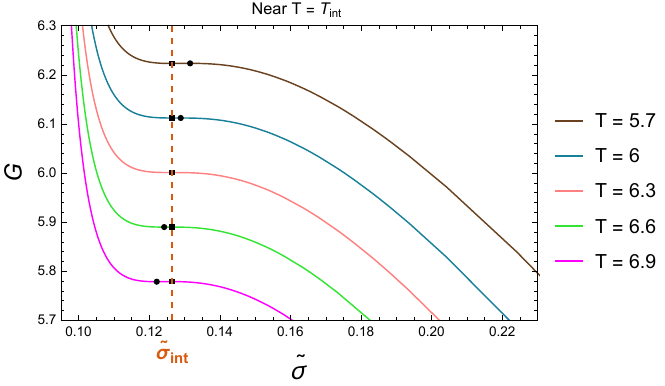}
   \caption{Off-shell free energy $G$ as a function of $\tilde{\sigma}$ for $R=0.1$, $\tilde Q=1$, and $\mu=0.1$. The brown, blue, red, green and magenta curves correspond to  $T=5.7$, $T=6$, $T = T_{int} =6.3$, $T = 6.6$ and $T= 6.9$ respectively.}
    \label{fig_PhaseD_Unique_near_Tint}
\end{figure}

\noindent
\textbf{Near} \bm{$ T = T_{int}$} :\\
\noindent
Next we present the off-shell phase diagrams near $T=T_{\rm int}$ in
Fig.~\ref{fig_PhaseD_Unique_near_Tint}. For $T<T_{\rm int}$ (teal, orange and blue
curves) there exists a low-entropy CFT branch that appears as a local maximum and is therefore unstable 
$\left(
\left.
\frac{\partial^2 G}{\partial \tilde{\sigma}^{\,2}}
\right|_{\tilde{\sigma}=\tilde{\sigma}_l}
<0
\right)$. As the temperature increases within this range, the entropy of this branch decreases.  For $T>T_{\rm int}$ (green, magenta and purple curves) the situation reverses: the low-entropy CFT branch now appears as a local minimum and becomes stable  $\left(
\left.
\frac{\partial^2 G}{\partial \tilde{\sigma}^{\,2}}
\right|_{\tilde{\sigma}=\tilde{\sigma}_l}
>0
\right)$, again with decreasing entropy as the temperature increases.  The red curve corresponds to $T=T_{\rm int}$, where the stable and unstable extrema merge, marking the point at which the low-entropy CFT branch changes its stability.\\

\begin{figure}[!htbp]
    \centering
    \includegraphics[width=.7\linewidth]{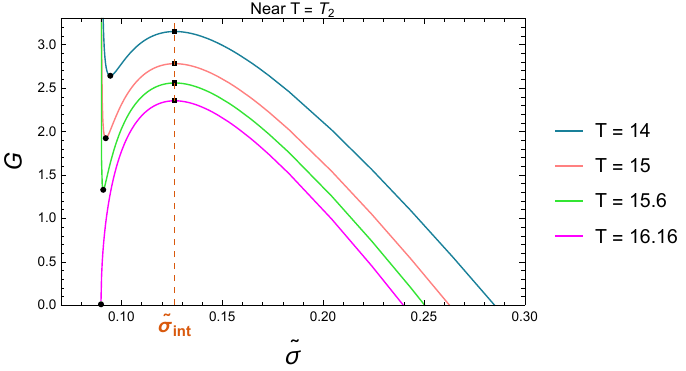}
   \caption{Off-shell free energy $G$ as a function of $\tilde{\sigma}$ for $R=0.1$, $\tilde Q=1$, and $\mu=0.1$. The blue, red, green and magenta curves correspond to $T=14$, $T=15$, $T=15.6$ and $T = T_2 = 16.16$ respectively.}
    \label{fig_PhaseD_Unique_near_T2}
\end{figure}

\noindent
\textbf{Near} \bm{$ T = T_2$} :\\
\noindent
Finally, the off-shell phase diagrams near $T=T_2$ are shown in
Fig.~\ref{fig_PhaseD_Unique_near_T2}. As expected, the stable low-entropy CFT
state terminates at $T=T_2$.   One can verify that the results presented here are consistent with on-shell results of \cite{Cong:2021jgb}, if the various external control parameters are taken to their on-shell values. \\

\begin{figure}[!htbp]
    \centering
    \includegraphics[width=.6\linewidth]{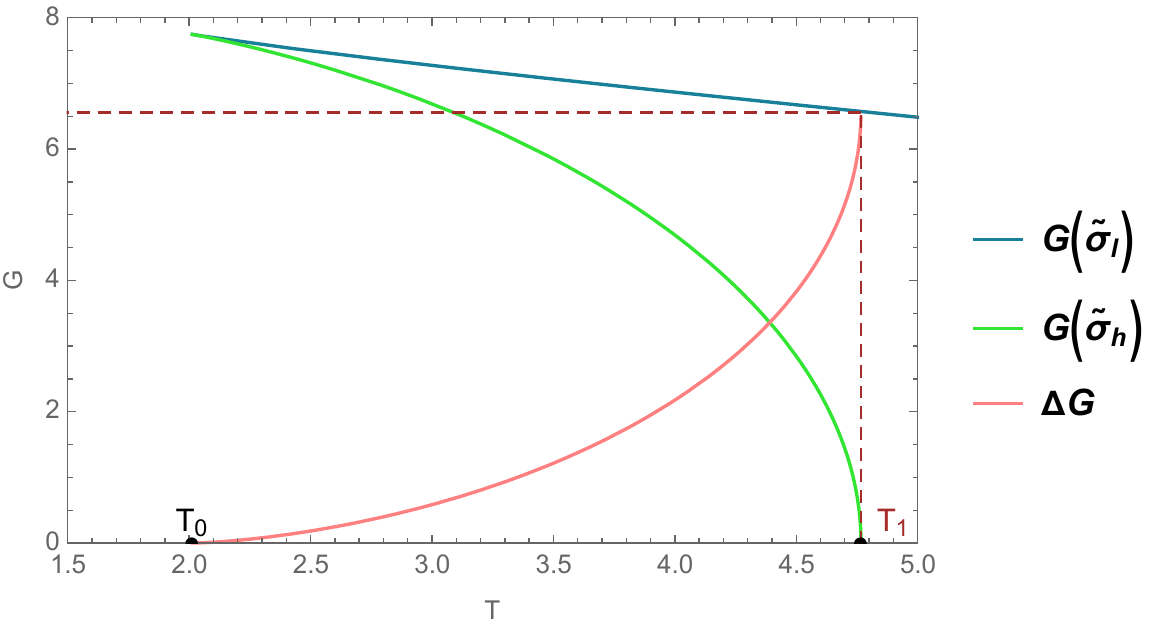}
   \caption{ Difference of free energies of low and high entropy CFT states $ \Delta G = G(\tilde{\sigma_l}) - G(\tilde{\sigma_h})$ as a function of CFT temperature $T$, plotted together with  $G(\tilde{\sigma_l})$ and $G(\tilde{\sigma_h})$. }
    \label{fig_Delta_G_Vs_T1}
\end{figure}

\noindent
Now, to make the zeroth-order phase transition more explicit, we can actually plot the free-energy difference between the low- and high-entropy CFT states in figure-(\ref{fig_Delta_G_Vs_T1}), computed from,
\[
\Delta G = G(\tilde{\sigma}_l)-G(\tilde{\sigma}_h),
\]
as a function of the CFT temperature $T$ in
Fig.~\ref{fig_Delta_G_Vs_T1}. Here, we used
eq.~\eqref{Unique_T_off} to obtain the order parameters of the low- and
high-entropy CFT states, which we denoted by $\tilde{\sigma}_l$ and
$\tilde{\sigma}_h$, respectively. Substituting these back into
eq.~\eqref{Unique_Offshell_d4}, one obtains the corresponding off-shell free
energies $G(\tilde{\sigma}_l)$ and $G(\tilde{\sigma}_h)$, which are also plotted in figure-(\ref{fig_Delta_G_Vs_T1}). Since the
resulting expressions are rather lengthy, we do not display them
explicitly here.  We find that $\Delta G=0$ at $T=T_0$ and
then increases, reaching its maximum value at $T=T_1$. The fact that
$\Delta G$ remains nonzero at $T=T_1$, i.e., the free energy is discontinuous, and one of the stable high entropy branches (figure-(\ref{fig_Delta_G_Vs_T1}, Green)) terminates at this temperature, is a clear indication that the transition is
zeroth order.

\subsubsection{Case-2: $\mu \le 0$}
\begin{figure}[!htbp]
    \centering
    \includegraphics[width=.7\linewidth]{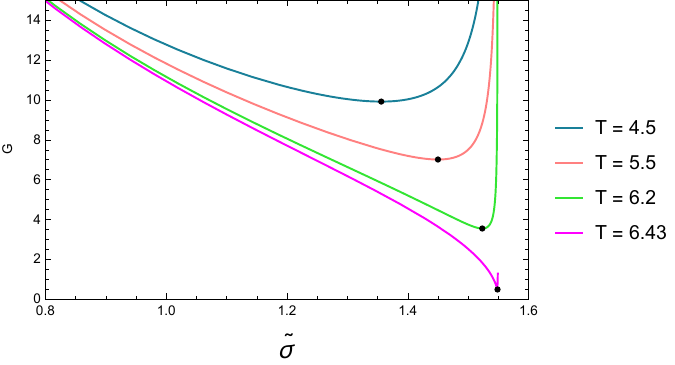}
    \caption{Off-shell free energy $G$ versus $\tilde{\sigma}$ for $R=.1$, $\tilde{Q}=1$, $\mu=-60$.
    The temperature increases from top to bottom:
    $T=4.5,\,5.5,\,6.2,\,6.43$ (blue, red, green, magenta).
    The magenta curve corresponds to critical temperature  $T=T_{1}=6.43$, where the CFT phase disappears.}
    \label{fig_phaseD_Unique_1}
\end{figure}

In this case, the off-shell free energy $G$ at different temperatures is shown in Fig (\ref{fig_phaseD_Unique_1}) for $\mu = -60$. The critical temperature $T_2$ is unphysical while $T_{1} = 6.43$ (with $ R = .1)$ . For every temperature $T<T_{1}$, we have one equilibrium point, and hence one CFT phase. At $T = T_1$, the CFT state has vanishing free energy, and for $T> T_1$ there are no CFT states. With increasing $T$, the order parameter $\tilde{\sigma}$ increases, thereby increasing the entropy, which is very much expected. The $\mu = 0$ case is physically same as the $\mu < 0$ case, and hence we do not show the phase diagrams separately.

\begin{figure}
  \centering
  \includegraphics[width=7.5cm]{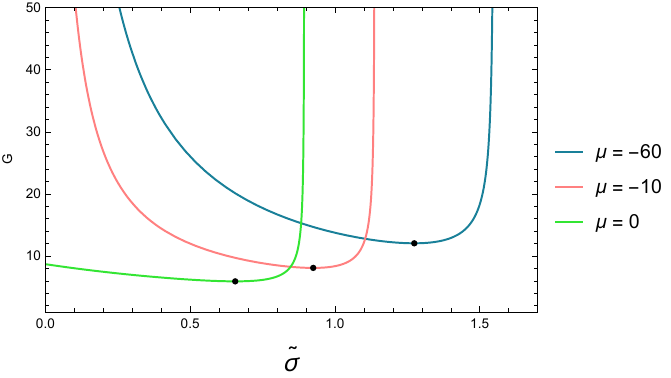}
   \hspace{0.2cm}
  \includegraphics[width=7.5cm]{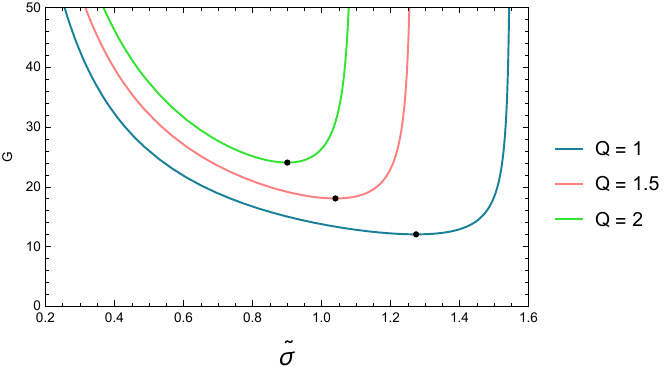}
  \caption{\textit{Left}: For all curves we set $R = .1$, $\tilde{Q} = 1$ and $T = 3.5$. The blue,red and green curve corresponds to $\mu = -60, -10$ and $0$ respectively.
\textit{Right}: For all curves we set $R = .1$, $\mu = -60$ and $T = 3.5$. The blue,red and green curve corresponds to $\tilde{Q} = 1,1.5$ and $2$ respectively. }
   \label{fig_phaseD_Unique_2}
\end{figure}
\noindent
From figure-(\ref{fig_phaseD_Unique_2}, left panel) we can see that with an increase in $\mu$, the free energy decreases, and $\tilde{\sigma}$ decreases too, and hence entropy decreases. On the other hand, figure-(\ref{fig_phaseD_Unique_2}, right panel) shows that with increasing $\tilde{Q}$, the free energy increases, but $\tilde{\sigma}$ decreases, and hence the entropy still decreases.

\section{Fokker–Planck dynamics in the fixed $(\tilde Q,{\cal V},C)$ ensemble } \label{sec:FK}

\noindent
In this section we study the transition state theory describing switching between
distinct CFT phases, treating the CFT configurations as states in an extended phase
space. As established earlier and illustrated in
Fig.~\ref{figphaseDcan1}, for $C/\tilde Q>(C/\tilde Q)_{\rm crit}$ the off-shell free energy
landscape as a function of the order parameter $\tilde{\delta}$ develops a double-well
structure in the temperature window $T_{\min}<T<T_{\max}$. The two local minima correspond
to stable low- and high-entropy CFT phases, while the local maximum represents an
intermediate-entropy configuration that is thermodynamically unstable. Our aim is to study the dynamical evolution of the system induced by thermal
fluctuations. In this framework, CFT configurations are treated as states in the
extended phase space, parametrized by the order parameter $\tilde{\delta}$.
Accordingly, the time evolution of the ensemble is described by a probability
density $\rho(\tilde{\delta},t)$, representing the probability of finding the system
in a given CFT state at time $t$. The stochastic evolution on the off-shell free
energy landscape is governed by a Fokker-Planck equation, which we set up below.\\

\noindent
Fokker--Planck dynamics in the context of black hole phase transitions has been important in recent years. In particular, the stochastic description of the
Hawking--Page transition for Schwarzschild--AdS black holes was developed in
\cite{Li:2020khm}, while the Fokker--Planck framework for van der Waals--type phase
transitions in Reissner--Nordstr\"om--AdS black holes was explored in \cite{Li:2020nsy}.
These analyses focus entirely on bulk gravitational degrees of freedom. To the best of
our knowledge, an analogous kinetic description formulated directly in terms of the
dual CFT states has not been carried out so far. In this work, we initiate such a study
by constructing and analyzing the Fokker--Planck dynamics on the off-shell phases of the holographic CFT free--energy.\\
\subsection{Forward Fokker–Planck dynamics }
\noindent
The Fokker-Planck equation for the probabilistic evolution on the holographic CFT free energy landscape is explicitly given by~\cite{Zwanzig:2001Nonequilibrium,
Lee:2003DiffusionFPT,
Lee:2003NonMarkovian,
Wang:2015LandscapeFlux,
Bryngelson:1989RandomEnergy,
Li:2020khm,
Li:2020nsy}
\begin{eqnarray}\label{ForwardFPequation}
\frac{\partial \rho(\tilde{\delta},t)}{\partial t}
= D\,\frac{\partial}{\partial \tilde{\delta}}
\left\{
e^{-\beta G(\tilde{\delta})}
\frac{\partial}{\partial \tilde{\delta}}
\left[
e^{\beta G(\tilde{\delta})}\,
\rho(\tilde{\delta},t)
\right]
\right\}\,.
\end{eqnarray}
\noindent
In the above equation, the inverse temperature is $\beta=1/kT$ and the diffusion coefficient is  $D=kT/\zeta$ with $k$ being the Boltzman constant and $\zeta$ being dissipation coefficient. Following~\cite{Li:2020khm,
Li:2020nsy} Without loss of generality, we will take $k=\zeta=1$ in the following.\\

\noindent
To solve the Fokker--Planck equation, one must specify suitable boundary conditions at the
edges of the domain in the order-parameter space. The choice of boundary condition is
dictated by the physical process one wishes to describe. For concreteness, we list below
the two standard boundary conditions imposed at $\tilde{\delta}=\tilde{\delta}_0$.
\begin{itemize}
\item \emph{Reflecting boundary condition:}
\begin{eqnarray}\label{bc1}
\left.
e^{-\beta G(\tilde{\delta})}
\frac{\partial}{\partial \tilde{\delta}}
\left[
e^{\beta G(\tilde{\delta})}\rho(\tilde{\delta},t)
\right]
\right|_{\tilde{\delta}=\tilde{\delta}_0}
=0 \; .
\end{eqnarray}
This condition ensures that there is no probability flux across the boundary, so that the
system is reflected back into the allowed region. It may be written in an equivalent and
more transparent form as
\begin{eqnarray}\label{bc11}
\left.
\beta G'(\tilde{\delta})\,\rho(\tilde{\delta},t)
+\rho'(\tilde{\delta},t)
\right|_{\tilde{\delta}=\tilde{\delta}_0}
=0 \; .
\end{eqnarray}

\item \emph{Absorbing boundary condition:}
\begin{eqnarray}\label{bc2}
\rho(\tilde{\delta}_0,t)=0 \; .
\end{eqnarray}
This condition corresponds to complete absorption at the boundary, meaning that once the
system reaches $\tilde{\delta}=\tilde{\delta}_0$ it is removed from the stochastic
dynamics.
\end{itemize}
\noindent
We now study the time evolution of the probability distribution in the canonical ensemble,
which consists of a continuous set of CFT states parametrized by the order parameter
$\tilde{\delta}\in(0,\infty)$. From the explicit form of the off-shell free energy in
\eqref{offshellFECanonical4}, it is clear that the free energy diverges in both limits
$\tilde{\delta}\rightarrow 0$ and $\tilde{\delta}\rightarrow\infty$. These divergences act
as effective barriers in the order-parameter space and confine the stochastic dynamics to
a finite region. As a result, the probability cannot escape the physical domain during time evolution.
It is therefore natural to impose reflecting boundary conditions at the boundaries of the
$\tilde{\delta}$ domain. With this choice, the probability current vanishes at the
boundaries, ensuring conservation of total probability and preserving the normalization
of the distribution throughout the evolution.\\

\noindent
Solving the forward equation~\eqref{ForwardFPequation} requires specifying an initial
probability distribution. We take the system to start from a localized configuration
around $\tilde{\delta}_i$, and choose a Gaussian profile with width $a$,
\begin{eqnarray}\label{initial}
\rho(\tilde{\delta},0)=\frac{1}{\sqrt{\pi}\,a}
\,\exp\!\left[-\frac{(\tilde{\delta}-\tilde{\delta}_i)^2}{a^2}\right]\;.
\end{eqnarray}

\noindent
Throughout our numerical analysis we fix the width of the initial Gaussian to
$a=0.1$, which provides a sufficiently localized wave packet for stable numerical
evolution.
The initial distribution is normalized, and we have explicitly verified that this
normalization is preserved during the time evolution.
We choose the initial position of the wave packet to coincide with one of the
equilibrium values of the order parameter, taking either
$\tilde{\delta}=\tilde{\delta}_l$ or $\tilde{\delta}=\tilde{\delta}_h$.
These choices correspond to initializing the system in the low-entropy or
high-entropy CFT phase, respectively.\\

\noindent
We now discuss the kinetic description of phase transitions by analyzing
first-passage processes between distinct CFT states on the underlying
off-shell free energy landscape.
In this framework, the first passage time characterizes the stochastic
time required for the system to reach the intermediate transition state,
identified with the top of the free energy barrier, for the first time.
The associated mean first passage time (MFPT) sets the characteristic
timescale governing such thermally activated transitions.
As a first step in this analysis, we examine the behavior of the free
energy barrier height as a function of the CFT temperature $T$ in figure.
\begin{figure}[!htbp]
\centering
\includegraphics[width=.6\linewidth]{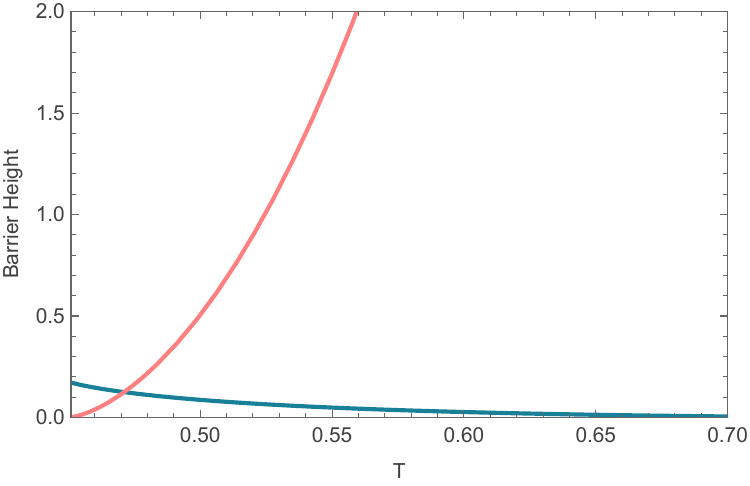}
    \caption{Barrier heights as functions of the CFT temperature $T$. The red curve corresponds to the  barrier height
$F(\tilde{\delta}_m)-F(\tilde{\delta}_h)$
between intermediate-entropy
and high-entropy CFT state,
while the blue curve corresponds to barrier height
$F(\tilde{\delta}_m)-F(\tilde{\delta}_l)$,
between intermediate-entropy
and low-entropy CFT state. For both the plot we fix $C=1$, $R=1$ and $\tilde Q=0.1$.
}
    \label{figbarrierheight1}
\end{figure}
From fig~\ref{figbarrierheight1}, it is obvious that they are all monotonic functions of the CFT temperature $T$. When temperature increases, the free energy barrier from low entropy CFT state to the intermediate entropy CFT state decreases while the free energy barrier from high entropy CFT state to the intermediate entropy CFT  state increases. This is consistent with the free energy landscape as a function of black hole radius at varying temperatures as shown in Fig \ref{figphaseDcan1}.\\

\noindent
Firstly, let us consider the first passage time of system from the low entropy CFT state to the intermediate CFT state. We denote the distribution of first passage times by $F_p(t)$ and define $\Sigma(t)$ to be the probability that the CFT state has not made a first passage by time $t$. The distributions $F_p(t)$ and $\Sigma(t)$ are related by
\begin{eqnarray}\label{FPTequation}
F_p(t)=-\frac{d\Sigma(t)}{dt}\;.
\end{eqnarray}

\noindent
According to the definition, $\Sigma(t)$ is defined as the probability of CFT state
being in the system at time $t$. So we have
\begin{eqnarray}\label{Sigmaequation}
\Sigma(t)=\int_{0}^{\tilde{\delta}_m} \rho(\tilde{\delta}, t) dr\;.
\end{eqnarray}

\noindent
It is obvious that $F_p(t)dt$ is the probability that a low entropy CFT state passes through the intermediate entropy  CFT state (free energy barrier top) for the first time in the time interval $(t, t+dt)$. In this setup, we have made the assumption that the time taken from the intermediate entropy CFT state to the high entropy CFT state is much smaller than the first passage time. Suppose there is a perfect absorber placed at the site $\tilde{\delta_m}$ (transition state at the free energy barrier top). If the  CFT state makes the first passage under the thermal fluctuation, this CFT state leaves the system. The normalization of the probability distribution will not be preserved in this case. Therefore, at very late time, the probability of the black hole still in the system becomes zero, i.e. $\Sigma(r, t)|_{t\rightarrow +\infty}=0$.\\

\noindent
By substituting Eq.(\ref{Sigmaequation}) into eq.(\ref{FPTequation}), and using the forward Fokker-Planck equation (\ref{ForwardFPequation}), one can get~\cite{Li:2020nsy}
\begin{equation}
  F_p(t)=-\frac{d}{dt}\int_{0}^{\tilde{\delta}_m} \rho(\tilde{\delta}, t) dr  =\left.-D\frac{\partial}{\partial \tilde{\delta}}\rho(\tilde{\delta},t)\right|_{\tilde{\delta}=\tilde{\delta}_m}
  \label{FPTcal}
\end{equation}
\noindent
We impose a reflecting boundary condition at $\tilde{\delta}=0$, ensuring that the probability current vanishes and no probability flux escapes the configuration space from this side. An absorbing boundary condition is imposed at $\tilde{\delta}=\tilde{\delta}_m$, which corresponds to the transition state located at the top of the free energy barrier. The absorbing boundary effectively removes the state from the system  once the state reaches the barrier, and therefore probability is no longer conserved. This choice is deliberate, as it allows a direct computation of first passage time statistics. With the initial condition and boundary conditions specified, the Fokker--Planck equation determines the full distribution of first passage times. In practice, since the off-shell free energy diverges as $\tilde{\delta}\to 0$, we implement the reflecting boundary slightly away from the singular point, at $\tilde{\delta}=0.1$. This regulator does not affect the physical results and ensures numerical stability.\\

\begin{figure}[!htbp]
    \centering
    \includegraphics[width=.6\linewidth]{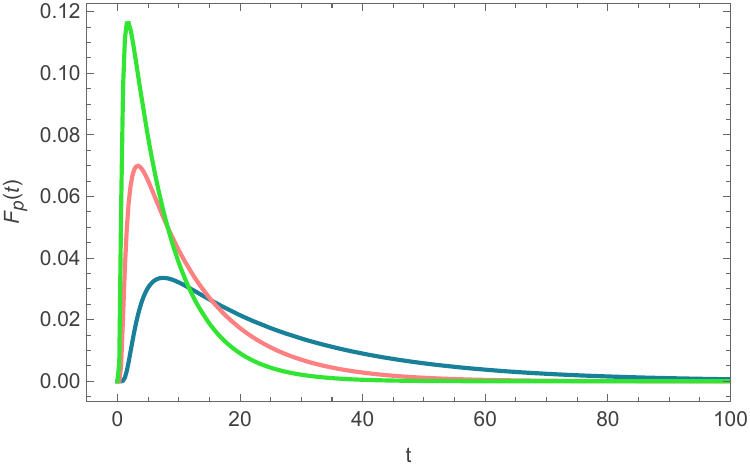}
    \caption{The distributions of first passage time $F_p(t)$ from low entropy CFT state to high entropy CFT state transition at different temperatures. The blue, red and green curves represents $T = 0.46, 0.50$ and $0.54$ respectively, and here for all curves we fix $\tilde{Q}=.1,C=1,R=1$. The initial distribution is Gaussian wave pocket located at the low entropy CFT state. }
    \label{figFpcan1}
\end{figure}

\noindent
In Fig.~\ref{figFpcan1} we show the first passage time distributions at different
temperatures as a function of time $t$. The initial probability profiles are taken to be Gaussian wave packets
centered around the low-entropy CFT state. These distributions capture the generic
features of the first passage dynamics across the free energy landscape. For all the temperatures considered, the distribution exhibits a single dominant peak,
indicating that a significant fraction of transition events occur over relatively short
timescales, before the distribution enters its asymptotic decay regime. As the
temperature is increased, the peak sharpens and shifts toward smaller times. This trend
reflects the reduction of the free energy barrier separating the low- and high-entropy
states via the intermediate configuration, as illustrated in
Fig.~\ref{figbarrierheight1} as CFT temperature $T$ increases. A lower barrier enhances the probability of thermally
activated crossings, thereby accelerating the transition process. The long-time tail of the distribution encodes the stochastic spread in the transition
times. A broader tail corresponds to stronger kinetic fluctuations, whereas a more
rapidly decaying tail signals a narrower distribution of first passage events.
With the time distributions, we can calculated the mean first passage time and its fluctuation. The mean first passage time is defined by
\begin{eqnarray}\label{MFPTcal}
\tau =\langle t \rangle=\int_{0}^{+\infty} t F_p(t) dt\;  = \int_{0}^{+\infty}\Sigma(t) dt\;
\end{eqnarray}
\noindent
In principle, we can also calculate the $n$-th moment of time distribution function of first passage time by the relation
\begin{eqnarray}\label{nthmoment}
\langle t^n \rangle=\int_{0}^{+\infty} t^n F_P(t) dt\;.
\end{eqnarray}
\noindent 
Hence second order moment of time distribution can be given as 
\begin{eqnarray}\label{nthmoment}
m_2 =\langle t^2 \rangle=\int_{0}^{+\infty} t^2 F_P(t) dt\; =\int_{0}^{+\infty} 2t \Sigma(t) dt\;.
\end{eqnarray}
\noindent
In principle this procedure can be carried out directly. However, the required integrals
extend to $\tilde{\delta}\to\infty$, which necessitates the introduction of an artificial
upper cutoff and a careful analysis of cutoff dependence. This significantly complicates
the numerical implementation without adding physical insight. To bypass these technical
issues, in the next subsection, we instead adopt the backward Fokker-Planck formulation, in which the explicit
time dependence has already been integrated out. This approach provides direct access to
the relevant first passage observables and is numerically more stable for our purposes.\\

\noindent
At last, we will discuss the first passage kinetic process from high entropy CFT state  to low entropy CFT state. This time the time distribution of first passage process is given by~\cite{Li:2020nsy}
\begin{eqnarray}
F_p(t)=\left.D\frac{\partial}{\partial \tilde{\delta}}\rho(\tilde{\delta},t)\right|_{\tilde{\delta}=\tilde{\delta}_m}\;,
\end{eqnarray}
where we should impose the reflecting boundary condition at $\tilde{\delta}=+\infty$ and the absorbing boundary condition at $\tilde{\delta}=\tilde{\delta}_m$ (transition state at the free energy barrier top). For numerical stability we put the reflecting boundary condition at $\tilde{\delta} = 22$ (way outside the free energy basin)  By numerically solving Fokker-Planck equation, we can also get the time distributions of first passage process from the high entropy  to the low entropy CFT state.\\

\begin{figure}[!htbp]
    \centering
    \includegraphics[width=.6\linewidth]{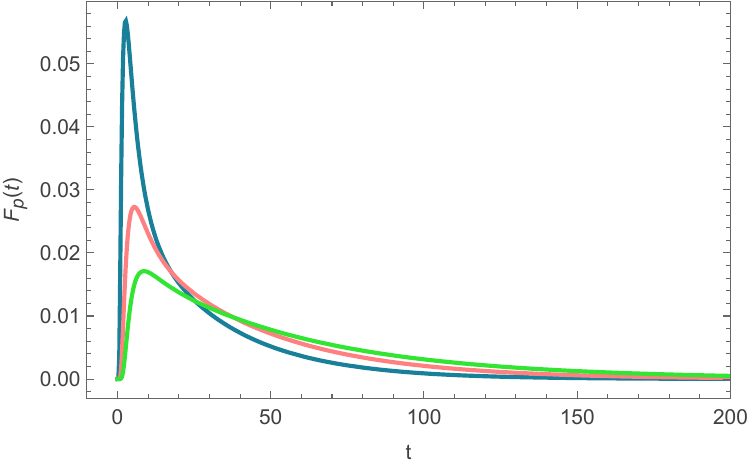}
    \caption{The distributions of first passage time $F_p(t)$ from high entropy CFT state to low entropy CFT state transition at different temperatures. The blue, red and green curves represents $T = 0.46, 0.47$ and $0.48$ respectively, and here for all curves we fix $\tilde{Q}=.1,C=1,R=1$. The initial distribution is Gaussian wave pocket located at the high entropy CFT state. }
    \label{figFpcanhtl1}
\end{figure}

\noindent
Figure~\ref{figFpcanhtl1} shows the first passage time distributions for transitions from the
high-entropy to the low-entropy CFT state at different temperatures. As the temperature increases,
the free-energy barrier opposing this transition grows, as already seen in
Fig.~\ref{figbarrierheight1}. Consequently, the system requires a much longer time, on average, to
complete the transition from the high-entropy to the low-entropy state. Each distribution exhibits a
single peak followed by an exponential tail, indicating activated barrier-crossing dynamics. With
increasing temperature the peak broadens and the tail becomes longer, signaling an enhancement of
kinetic fluctuations.
\FloatBarrier
\noindent
We should mention that, in~\cite{Li:2020khm}, the authors solved the forward Fokker--Planck equation
of the form \eqref{ForwardFPequation} for the Schwarzschild--AdS black hole and
employed a Laplace transform of the probability density
$\rho(\tilde{\delta},t)$ to obtain exact analytical expressions for the mean
first-passage time, higher moments of the passage-time distribution, and their
associated fluctuations. In contrast, in~\cite{Li:2020nsy}, the  focus was on
Reissner--Nordstr\"om--AdS black holes, and solving the forward Fokker-Planck
equation numerically, with the explicit goal of tracking the full time evolution
of the probability density $\rho(\tilde{\delta},t)$.\\

\noindent
Our perspective in the present work is different. Here we are primarily
interested in understanding how the stochastic dynamics of the system responds
to variations of intrinsic CFT parameters, in particular the electric charge
$\tilde{Q}$ and the central charge $C$, across different thermodynamic
ensembles. As briefly mentioned above, our focus in the following section is not on the detailed time evolution of the
probability distribution itself, beyond what has been found here using the
Fokker-Planck equation \eqref{ForwardFPequation}. Instead, the most efficient and physically transparent approach to proceed for our purposes here
is to work with the corresponding backward Fokker--Planck equation, which allows
direct access to first-passage observables without requiring the full solution
for $\rho(\tilde{\delta},t)$. This strategy significantly reduces the numerical
complexity while retaining all the information relevant for the study of
transition dynamics on the off-shell free-energy landscape.

\subsection{Backward Fokker--Planck formulation }
\label{subsec:backwardFP_adjoint}
\noindent
In the forward formulation, the probability density
$\rho(\tilde{\delta},t)$ evolves on the off-shell free-energy landscape
$F(\tilde{\delta})$ (for fixed $R$, $C$ and $\tilde Q$) according to
eq.~\eqref{ForwardFPequation}. While this description is natural for
studying the full time evolution of the probability distribution, it is
not the most convenient framework for computing first-passage
observables. For such questions, it is advantageous to work in the
backward (Kolmogorov) formulation, where one directly solves for the
moments of the first-passage time as functions of the \emph{initial}
value of the order parameter. In the backward formulation, the explicit time evolution of the
probability density is replaced by differential equations for stochastic
observables associated with the trajectories. Physically, while the
forward equation describes how an ensemble of CFT states evolves in
time, the backward equation determines how quantities such as transition
times depend on the initial configuration of the system on the
free-energy landscape.\\

\noindent
The quantity of primary interest is the mean first-passage time (MFPT),
denoted by $\tau(\tilde{\delta})$, which represents the average time
required for a stochastic trajectory starting at $\tilde{\delta}$ to
reach the absorbing point $\tilde{\delta}_m$ for the first time. For
constant diffusion coefficient $D=T$, the backward operator takes the
form
\begin{equation}
\mathcal{L}^\dagger
=
T\left(
\partial_{\tilde{\delta}}^2
-
\beta F'(\tilde{\delta})\,\partial_{\tilde{\delta}}
\right).
\label{Backward_operator_main}
\end{equation}
The MFPT satisfies the backward equation
\begin{equation}
\mathcal{L}^\dagger \tau(\tilde{\delta})=-1,
\label{MFPT_equation_main}
\end{equation}
together with the boundary conditions
\begin{equation}
\tau'(\tilde{\delta}_L)=0,
\qquad
\tau(\tilde{\delta}_m)=0,
\label{MFPT_BC_main}
\end{equation}
corresponding to reflecting and absorbing boundaries, respectively.
Similarly, the second moment of the first-passage time,
$m_2(\tilde{\delta})$, satisfies
\begin{equation}
\mathcal{L}^\dagger m_2(\tilde{\delta})
=
-2\tau(\tilde{\delta}),
\label{M2_equation_main}
\end{equation}
with boundary conditions
\begin{equation}
m_2'(\tilde{\delta}_L)=0,
\qquad
m_2(\tilde{\delta}_m)=0.
\label{M2_BC_main}
\end{equation}
From $\tau(\tilde{\delta})$ and $m_2(\tilde{\delta})$, one can compute
the variance and relative fluctuations of the first-passage time. The
detailed derivation of the backward operator and the associated boundary
conditions are relegated to Appendix~A.\\

\begin{figure}
  \centering
  \includegraphics[width=7.4cm]{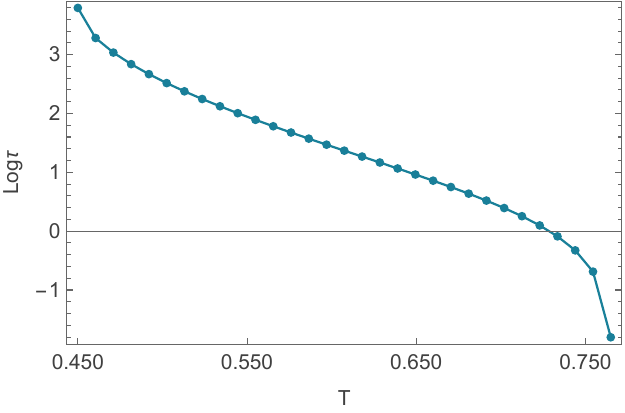}
   \hspace{0.2cm}
  \includegraphics[width=7.4cm]{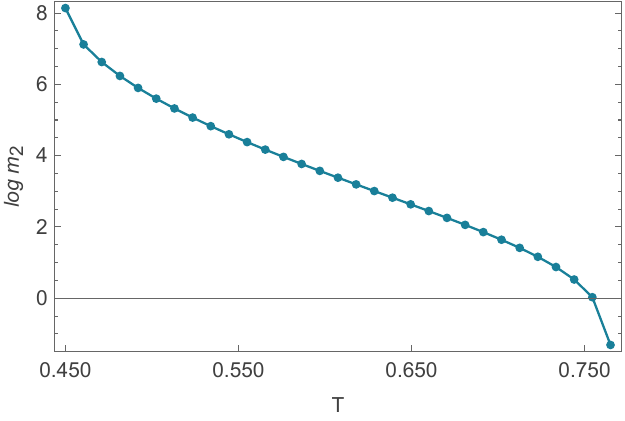}
  \caption{The left panel is the plot of mean first passage time $\tau$ for low to high entropy CFT state transition as a function of CFT temperature $T$ and the right is the second order moment of the first passage time distribution $m_2$ for low to high entropy CFT  state transition. The initial distribution is Gaussian wave pocket located at the low entropy CFT state. For both curves $ \tilde{Q} = 0.1, R = 1, C = 1$ }
   \label{figlogtaum1}
\end{figure}

\noindent
In Fig.~\ref{figlogtaum1} we plot the mean first passage time $\tau$ for transitions from the
low-entropy to the high-entropy CFT state, together with the second moment $m_{2}$ of the
first-passage-time distribution, as functions of the CFT temperature $T$. The vertical axis is shown
on a logarithmic scale. We find that the mean first passage time decreases monotonically with
increasing temperature. This can be understood as follows. As $T$ increases, the free-energy
barrier separating the low- and high-entropy phases through the intermediate saddle point is
progressively lowered, which directly shortens the characteristic transition time. In addition, higher
temperatures enhance thermal diffusion on the free-energy landscape, further facilitating barrier
crossing. Both effects act in the same direction, leading to a rapid suppression of $\tau$ at elevated
temperatures. Note that the reflecting boundary condition is imposed at $\tilde{\delta}=0$ because the probability will not leak out there, and the absorbing boundary condition is imposed at $\tilde{\delta}=\tilde{\delta}_m$ (transition state at free energy barrier top) where the perfect absorber is placed. As the free energy is divergent at $\tilde{\delta}=0$, the reflecting boundary condition is imposed at $\tilde{\delta}=0.1$ here.\\

\begin{figure}
  \centering
  \includegraphics[width=7.2cm]{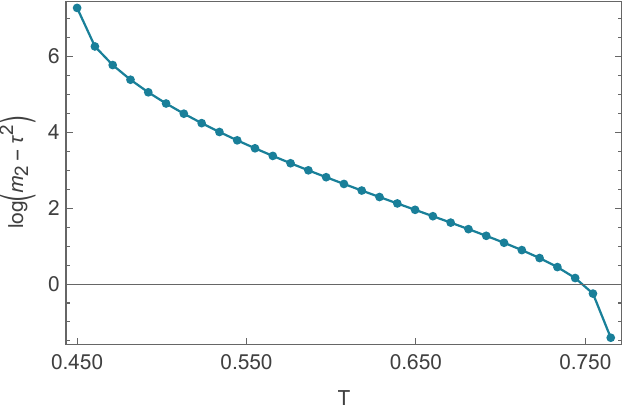}
   \hspace{0.2cm}
  \includegraphics[width=7.5cm]{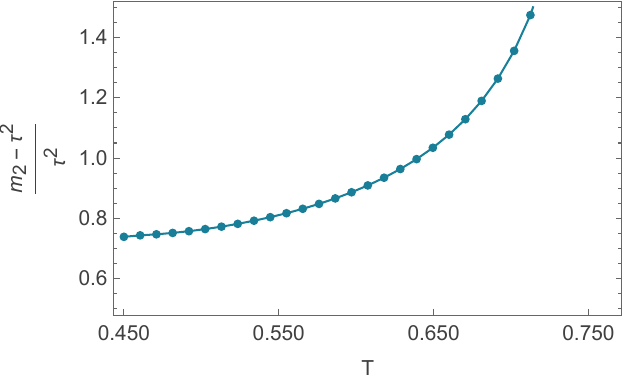}
  \caption{The left panel is the plot of fluctuation $m_{2}-\tau^2$ for low to high entropy CFT state transition as a function of CFT temperature $T$ and the right is the relative fluctuation  ${(m_2 - \tau^2)}/{\tau^2}$ for low to high entropy CFT  state transition. The initial distribution is Gaussian wave pocket located at the low entropy CFT state. For both curves $ \tilde{Q} = 0.1, R = 1, C = 1$ }
   \label{figFluclth1}
\end{figure}

\begin{figure}
  \centering
  \includegraphics[width=7.4cm]{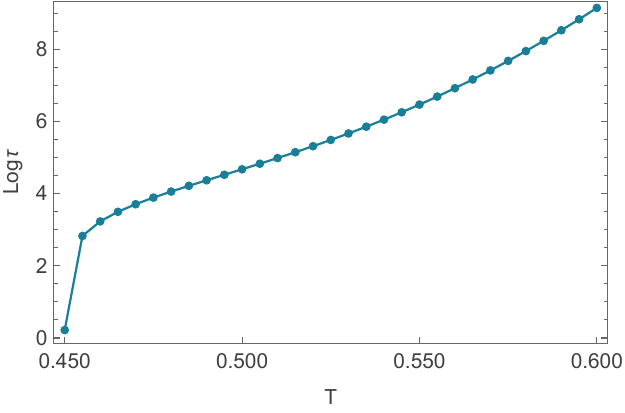}
   \hspace{0.2cm}
  \includegraphics[width=7.4cm]{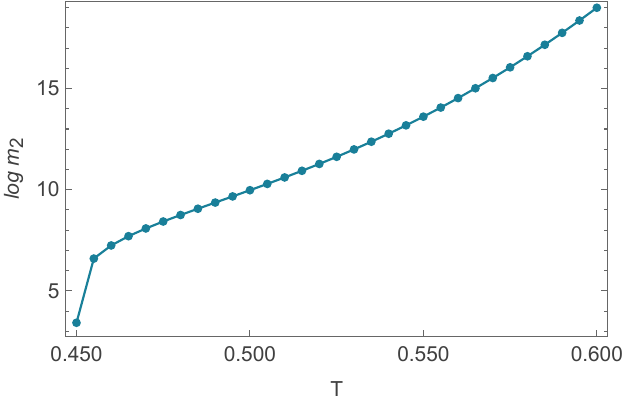}
  \caption{The left panel is the plot of mean first passage time $\tau$ for high to low entropy CFT state transition as a function of CFT temperature $T$ and the right is the second order moment of the first passage time distribution $m_2$ for high to low entropy CFT  state transition. The initial distribution is Gaussian wave pocket located at the high entropy CFT state. For both curves $ \tilde{Q} = 0.1, R = 1, C = 1$ }
   \label{figlogtaumhtl1}
\end{figure}

\noindent
The fluctuations and relative fluctuations of first passage time are depicted in Fig.\ref{figFluclth1}.
It can be seen that the fluctuations will decrease when the temperature increases. This behavior is consistent with the distributions of first passage time displayed in Fig.\ref{figFpcan1}, where the peak becomes sharper at higher temperatures. The relative fluctuations monotonically increase with increase of temperature. This shows higher temperature thermal fluctuations are very dominant with respect to free energy barrier height. Not surprisingly, the relative fluctuations attain the maximum at the temperature where the mean first time is at its minimum.
\noindent
In Fig.~\ref{figlogtaumhtl1} we show the mean first passage time $\tau$ and the second moment $m_2$
for transitions from the high-entropy to the low-entropy CFT state as functions of the CFT temperature
$T$. The mean first passage time increases monotonically with temperature and grows rapidly in the
high-$T$ regime, indicating a strong suppression of the transition. This behavior reflects the
increase of the free-energy barrier separating the two states, which dominates the kinetics of the
high-to-low entropy transition.\\

\begin{figure}
  \centering
  \includegraphics[width=7.2cm]{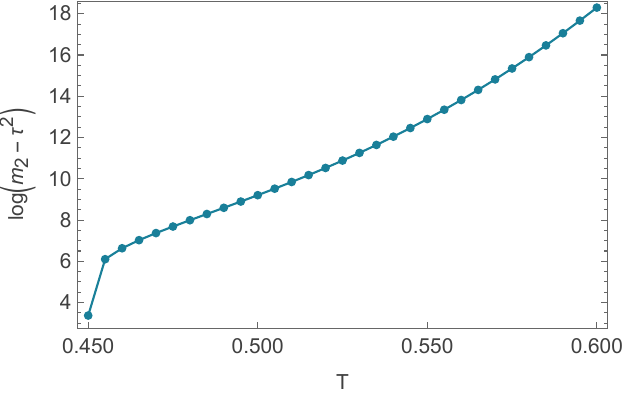}
   \hspace{0.2cm}
  \includegraphics[width=7.5cm]{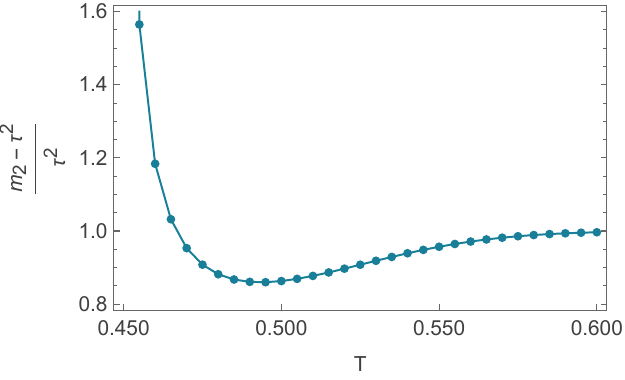}
  \caption{The left panel is the plot of fluctuation $m_{2}-\tau^2$ for high to low entropy CFT state transition as a function of CFT temperature $T$ and the right is the relative fluctuation  ${(m_2 - \tau^2)}/{\tau^2}$ for high to low entropy CFT  state transition. The initial distribution is Gaussian wave pocket located at the high entropy CFT state. For both curves $ \tilde{Q} = 0.1, R = 1, C = 1$ }
   \label{figFluchtl1}
\end{figure}

\noindent
The fluctuations of the first-passage time for transitions from the high-entropy to the
low-entropy CFT state are shown in the left panel of Fig.~\ref{figFluchtl1}.
The fluctuations increase with temperature, in agreement with the corresponding
first-passage time distributions in Fig.~\ref{figFpcanhtl1}, where the peak broadens at
higher temperatures.
The relative fluctuations are displayed in the right panel of Fig.~\ref{figFluchtl1}.
As expected, they attain a maximum at the temperature where the mean first-passage time
reaches its minimum. This behavior can be interpreted as follows.
At low temperatures, the free-energy barrier height is small compared to the thermal scale,
so stochastic fluctuations play a dominant role in the transition dynamics.
As a result, thermal noise has a stronger impact on the kinetics than the barrier itself,
leading to large relative fluctuations.
With increasing temperature, the relative fluctuations decrease rapidly from their maximum,
reach a minimum, and then increase slowly at higher temperatures.
\subsubsection{Effect of variation of $\tilde{Q}$}
\begin{figure}[htbp]
  \centering
  \includegraphics[width=7.2cm]{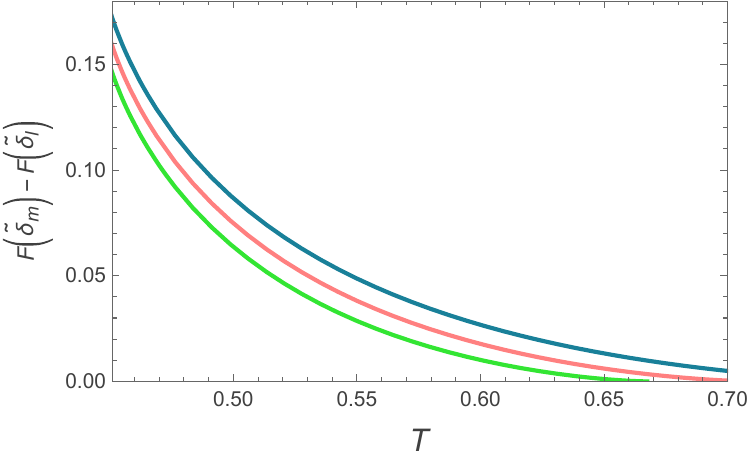}
   \hspace{0.2cm}
  \includegraphics[width=7.3cm]{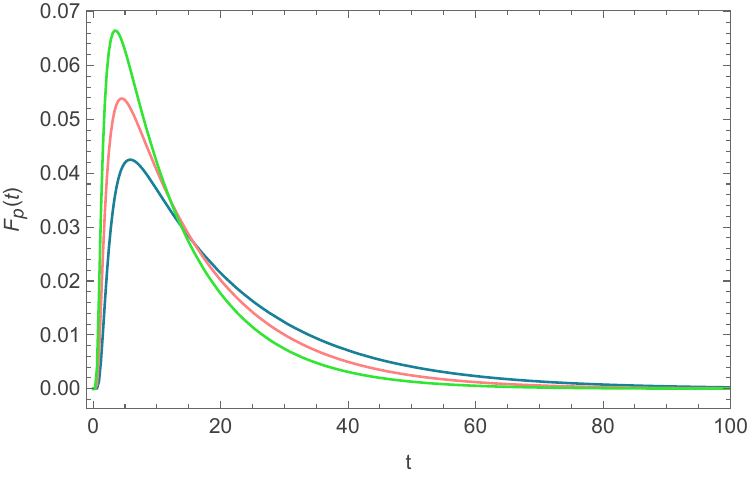}
  \caption{\textit{Left}: Free-energy barrier height 
$F(\tilde{\delta}_m)-F(\tilde{\delta}_l)$ as a function of the CFT temperature $T$
for different values of $\tilde Q$. 
\textit{Right}: Distribution of first passage time $F_p(t)$ as a function of time $t$
for the same set of parameters. In both panels the transition considered is from
the low-entropy to the high-entropy CFT state. We fix $C=1$ and $R=1$.
The blue, red and green curves correspond to $\tilde Q=0.1$, $0.12$ and $0.14$,
respectively.}
\label{figbarheightcanQlth}
\end{figure}

\noindent
We now examine the effect of varying the CFT electric charge $\tilde Q$, while
keeping the boundary curvature radius $R$ and the central charge $C$ fixed.
The left panel of Fig.~\ref{figbarheightcanQlth} shows the corresponding barrier
heights for the transition from the low-entropy to the high-entropy CFT state
for different values of $\tilde Q$, while the right panel displays the associated
first passage time distributions for the same transition. As $\tilde Q$ increases,
the barrier height decreases. Consequently, the mean first passage time $\tau$
is expected to decrease. This expectation is consistent with the behavior of the
distribution $F_p(t)$: with increasing $\tilde Q$ the peak of the distribution
becomes sharper and shifts toward smaller times. In addition, the width of the
distribution decreases, indicating that the fluctuations in the first passage
time are also reduced. These point will be further confirmed from further analysis below.\\

\begin{figure}[htbp]
  \centering
  \includegraphics[width=7.2cm]{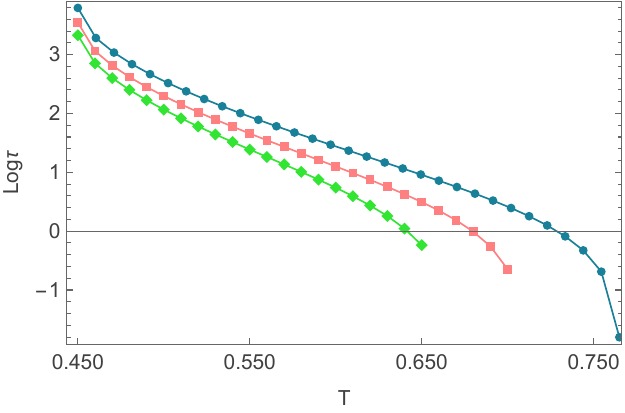}
   \hspace{0.2cm}
  \includegraphics[width=7.3cm]{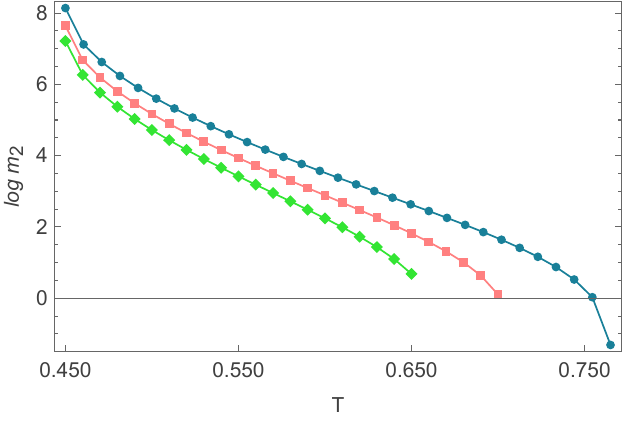}
  \caption{\textit{Left}: Mean first passage time $\tau$ for the transition from the
low-entropy to the high-entropy CFT state as a function of the CFT temperature $T$.
\textit{Right}: the second order moment of the first passage time distribution $m_2$ for the same transition.
 For all the curves, we fix $R=1$ and $C=1$.
The blue, red and green curves correspond to $\tilde Q=0.1$, $0.12$ and $0.14$,
respectively.}
\label{Figlogtau_m_Qlth}
\end{figure}

\noindent
In Fig.~\ref{Figlogtau_m_Qlth} we present the mean first passage time $\tau$ and the
second moment of the first passage time distribution $m_2$ for the transition
from the low-entropy to the high-entropy CFT state at different values of
$\tilde Q$. As anticipated from the behavior of the barrier heights shown in
Fig.~\ref{figbarheightcanQlth}, both $\tau$ and $m_2$ decrease with increasing
$\tilde Q$.\\

\begin{figure}[htbp]
  \centering
  \includegraphics[width=7.2cm]{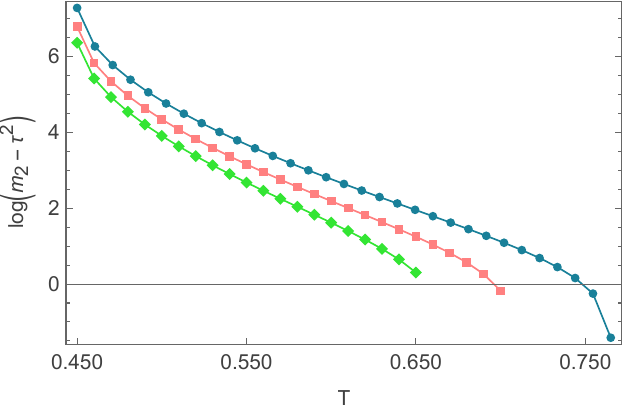}
   \hspace{0.2cm}
  \includegraphics[width=7.3cm]{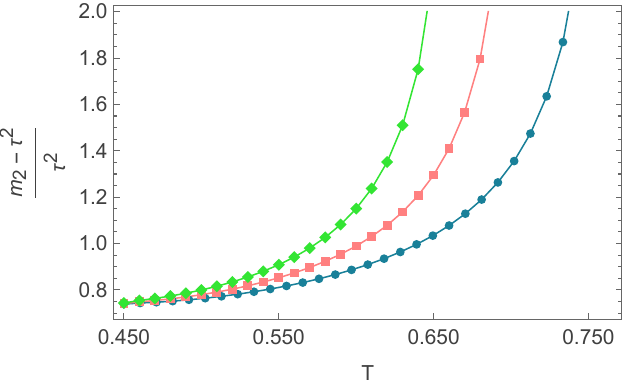}
  \caption{\textit{Left}: Fluctuation $m_{2}-\tau^2$ for the transition from the
low-entropy to the high-entropy CFT state as a function of the CFT temperature $T$.
\textit{Right}: relative fluctuation ${(m_2 - \tau^2)}/{\tau^2}$ for the same transition.
 For all the curves, we fix $R=1$ and $C=1$.
The blue, red and green curves correspond to $\tilde Q=0.1$, $0.12$ and $0.14$,
respectively.}
\label{fig_relativevar_var_can_Q}
\end{figure}

\noindent
In Fig.~\ref{fig_relativevar_var_can_Q} we present the fluctuation
$m_{2}-\tau^2$ and the relative fluctuation ${(m_2-\tau^2)}/{\tau^2}$
for the transition from the low-entropy to the high-entropy CFT state
at different values of $\tilde Q$. Consistent with the behavior of the
barrier heights shown in Fig.~\ref{figbarheightcanQlth}, the fluctuation
$m_{2}-\tau^2$ decreases as $\tilde Q$ increases. However, the relative
fluctuation ${(m_2-\tau^2)}/{\tau^2}$ shows the opposite trend and
increases with $\tilde Q$. This indicates that with increasing $\tilde Q$, although the absolute
fluctuations become smaller, the thermal fluctuations relative to the
characteristic transition time increases significantly and begin to
dominate over the effect associated with the barrier height.\\

\begin{figure}[htbp]
  \centering
  \includegraphics[width=7.2cm]{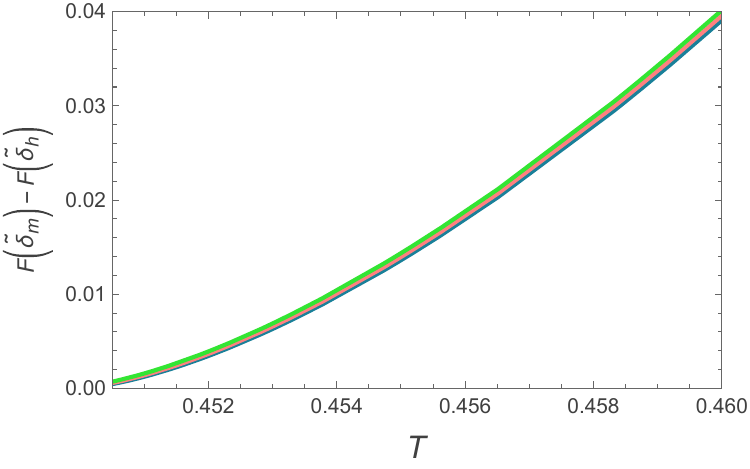}
   \hspace{0.2cm}
  \includegraphics[width=7.3cm]{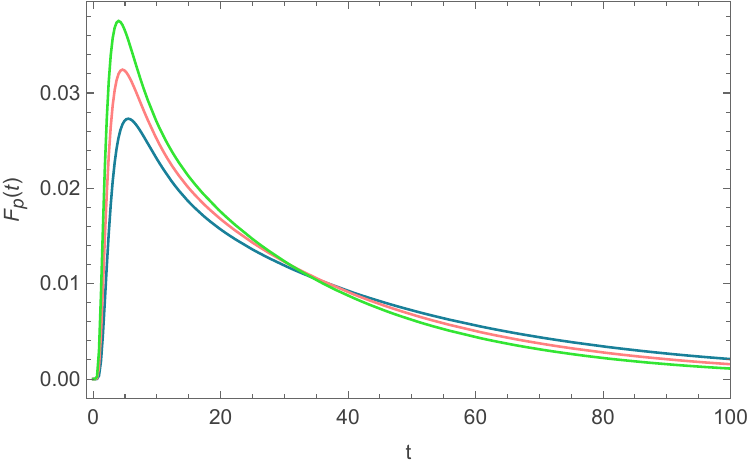}
  \caption{\textit{Left}: Free-energy barrier height 
$F(\tilde{\delta}_m)-F(\tilde{\delta}_h)$ as a function of the CFT temperature $T$
for different values of $\tilde Q$. 
\textit{Right}: Distribution of first passage time $F_p(t)$ as a function of time $t$
for the same set of parameters. In both panels the transition considered is from
the high-entropy to the low-entropy CFT state. We fix $C=1$ and $R=1$.
The blue, red and green curves correspond to $\tilde Q=0.1$, $0.12$ and $0.14$,
respectively.}
\label{fig_barheight_canQhtl}
\end{figure}

\noindent
We now turn to the reverse process, namely the transition from the high-entropy
to the low-entropy CFT state. The left panel of Fig.~\ref{fig_barheight_canQhtl}
shows the corresponding barrier heights $F(\tilde{\delta}_m)-F(\tilde{\delta}_h)$
for this transition at different values of $\tilde Q$, while the right panel
displays the associated first passage time distributions $F_p(t)$. We observe
that the barrier height changes very little with increasing $\tilde Q$ and in
fact shows a slight increase. Nevertheless, the distributions $F_p(t)$ become
sharper, the peak shifts towards smaller times, and the spread decreases as
$\tilde Q$ increases. Although the barrier height shows a slight increase with $\tilde Q$, the
corresponding first passage time distributions shift towards smaller times.
This indicates that the kinetics is not determined solely by the barrier
height but also by changes in the shape of the free-energy landscape, which
affects the effective dynamics.

\begin{figure}[htbp]
  \centering
  \includegraphics[width=7.1cm]{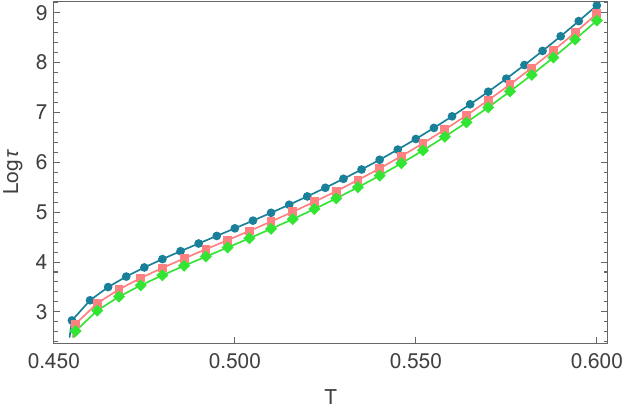}
   \hspace{0.2cm}
  \includegraphics[width=7.3cm]{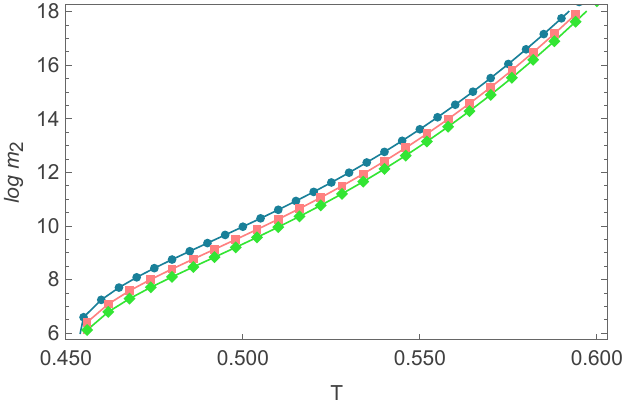}
  \caption{\textit{Left}: Mean first passage time $\tau$ for the transition from the
high-entropy to the low-entropy CFT state as a function of the CFT temperature $T$.
\textit{Right}: the second order moment of the first passage time distribution $m_2$ for the same transition.
 For all the curves, we fix $R=1$ and $C=1$.
The blue, red and green curves correspond to $\tilde Q=0.1$, $0.12$ and $0.14$,
respectively.}
\label{FiglogtauQhtl}
\end{figure}

\noindent
In Fig.~\ref{FiglogtauQhtl} we present the mean first passage time $\tau$ and the
second moment of the first passage time distribution $m_2$ for the transition
from the high-entropy to the low-entropy CFT state at different values of
$\tilde Q$. As anticipated from the behavior of the distribution function $F_p(t)$ shown in
Fig.~\ref{fig_barheight_canQhtl}, both $\tau$ and $m_2$ decrease with increasing$\tilde Q$.\\

\begin{figure}[!htbp]
    \centering
    \includegraphics[width=.5\linewidth]{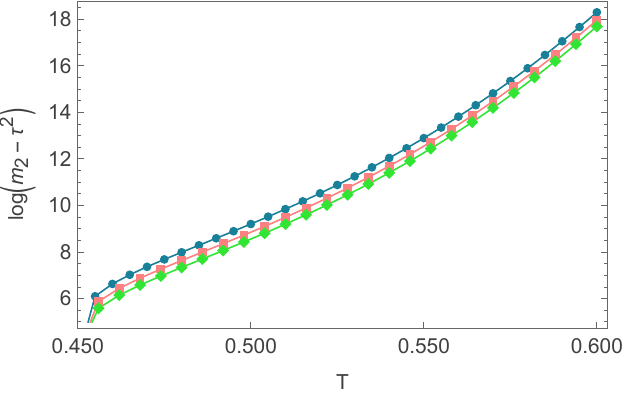}
    \caption{ Fluctuation $m_{2}-\tau^2$ for the transition from the high-entropy to the low-entropy CFT state as a function of the CFT temperature $T$.For all curves we fix $R=1$ and $C=1$.
The blue, red and green curves correspond to $\tilde Q=0.1$, $0.12$ and $0.14$,
respectively.}
    \label{fig_var_can_Q_htl}
\end{figure}

\noindent
In Fig.~\ref{fig_var_can_Q_htl} we present the fluctuation
$m_{2}-\tau^2$ for the transition from the high-entropy to the low-entropy
CFT state at different values of $\tilde Q$. Consistent with the behavior
of the distribution function $F_p(t)$ shown in
Fig.~\ref{fig_barheight_canQhtl}, the fluctuation $m_{2}-\tau^2$ decreases
as $\tilde Q$ increases. Since the barrier height changes only very
slightly with $\tilde Q$, the relative fluctuation also shows negligible
variation. For this reason we do not display it separately.
\FloatBarrier

\subsubsection{Effect of variation of $C$}
\begin{figure}[htbp]
  \centering
  \includegraphics[width=7.2cm]{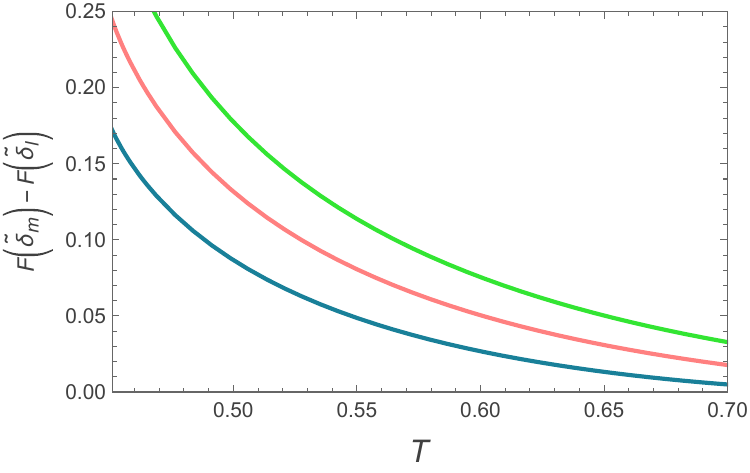}
   \hspace{0.2cm}
  \includegraphics[width=7.3cm]{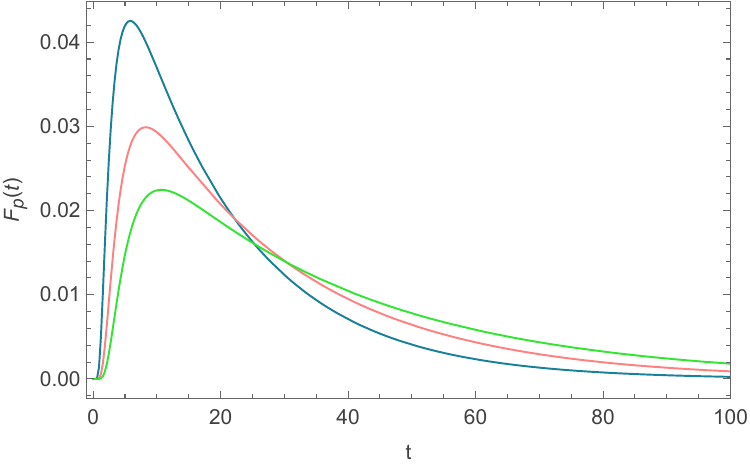}
  \caption{\textit{Left}: Free-energy barrier height 
$F(\tilde{\delta}_m)-F(\tilde{\delta}_l)$ as a function of the CFT temperature $T$
for different values of $C$. 
\textit{Right}: Distribution of first passage time $F_p(t)$ as a function of time $t$
for the same set of parameters. In both panels the transition considered is from
the low-entropy to the high-entropy CFT state. We fix $Q=0.1$ and $R=1$.
The blue, red and green curves correspond to $C=1$, $1.3$ and $1.6$,
respectively.}
\label{fig_barheight_Fp_can_C_lth}
\end{figure}

\noindent
We now examine the effect of varying the central charge $C$, while
keeping the boundary curvature radius $R$ and the CFT electric charge
$\tilde Q$ fixed. The left panel of Fig.~\ref{fig_barheight_Fp_can_C_lth}
shows the barrier heights for the transition from the low-entropy to
the high-entropy CFT state at different values of $C$, while the
right panel shows the corresponding first passage time distributions.
Unlike the case where $\tilde Q$ is varied, here the barrier height
increases as $C$ increases. Consequently, the transition becomes
slower. This is also reflected in the distribution $F_p(t)$: the peak
moves towards larger times and the distribution becomes wider as $C$
increases. This suggests that both the mean first passage time $\tau$
and the associated fluctuations increase with $C$. These features
will be confirmed explicitly from the behavior of $\tau$ and the
fluctuations studied below.\\

\begin{figure}[htbp]
  \centering
  \includegraphics[width=7.2cm]{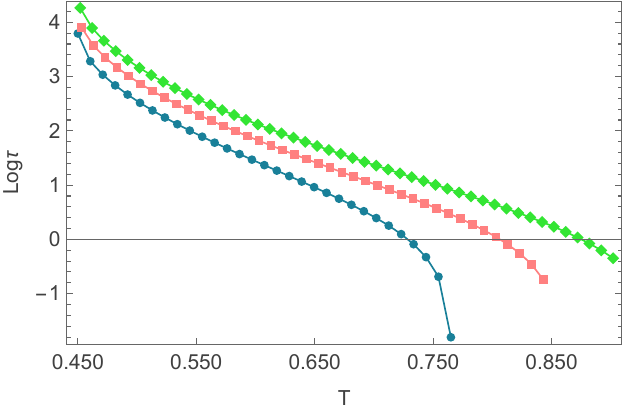}
   \hspace{0.2cm}
  \includegraphics[width=7.3cm]{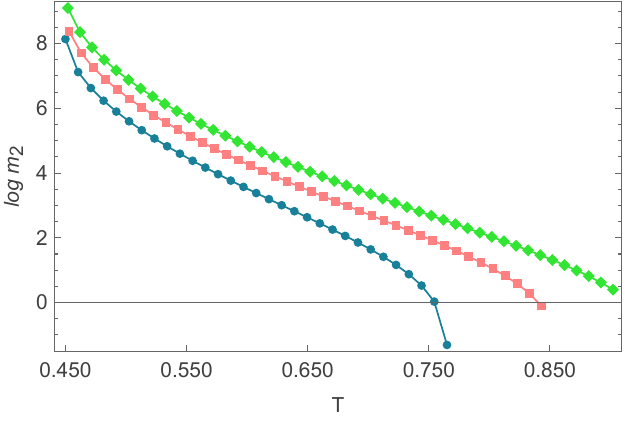}
  \caption{\textit{Left}: Mean first passage time $\tau$ for the transition from the
low-entropy to the high-entropy CFT state as a function of the CFT temperature $T$.
\textit{Right}: the second order moment of the first passage time distribution $m_2$ for the same transition.
 For all the curves, we fix $R=1$ and $\tilde{Q}=0.1$.
The blue, red and green curves correspond to $C=1$, $1.3$ and $1.6$,
respectively.}
\label{Figlogtau_m_C_lth}
\end{figure}

\noindent
In Fig.~\ref{Figlogtau_m_C_lth} we present the mean first passage time $\tau$ and the
second moment of the first passage time distribution $m_2$ for the transition
from the low-entropy to the high-entropy CFT state at different values of
$C$. As anticipated from the behavior of the barrier heights and distribution of first passage time shown in
Fig.~\ref{fig_barheight_Fp_can_C_lth}, both $\tau$ and $m_2$ increase with increasing $C$.\\

\begin{figure}[htbp]
  \centering
  \includegraphics[width=7.2cm]{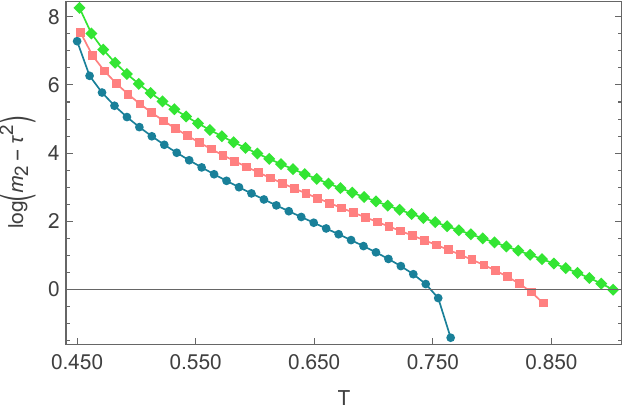}
   \hspace{0.2cm}
  \includegraphics[width=7.3cm]{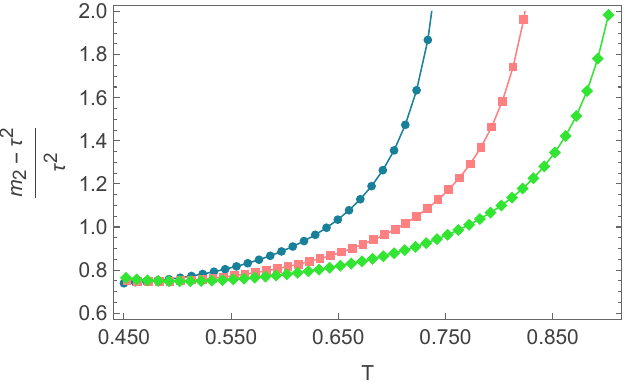}
  \caption{\textit{Left}: Fluctuation $m_{2}-\tau^2$ for the transition from the
low-entropy to the high-entropy CFT state as a function of the CFT temperature $T$.
\textit{Right}: relative fluctuation ${(m_2 - \tau^2)}/{\tau^2}$ for the same transition.
 For all the curves, we fix $R=1$ and $\tilde{Q}=0.1$.
The blue, red and green curves correspond to $C=1$, $1.3$ and $1.6$,
respectively.}
\label{fig_relativevar_var_can_C_lth}
\end{figure}

\noindent
In Fig.~\ref{fig_relativevar_var_can_C_lth} we present the fluctuation
$m_{2}-\tau^2$ and the relative fluctuation ${(m_2-\tau^2)}/{\tau^2}$
for the transition from the low-entropy to the high-entropy CFT state
at different values of $C$. Consistent with the behavior of the
barrier heights and the first passage time distributions shown in
Fig.~\ref{fig_barheight_Fp_can_C_lth}, the fluctuation $m_{2}-\tau^2$
increases with increasing $C$. Interestingly, the relative fluctuation
${(m_2-\tau^2)}/{\tau^2}$ shows the opposite trend and decreases as $C$
increases. This indicates that, at larger $C$, factors other than the
barrier height and thermal diffusion may play a role in determining the
relative magnitude of the fluctuations. A more detailed investigation
of this behavior is left for future work.\\

\begin{figure}[htbp]
  \centering
  \includegraphics[width=7.2cm]{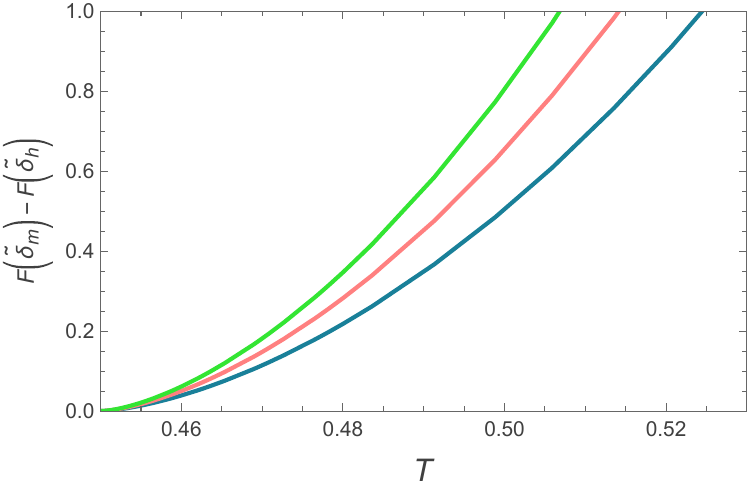}
   \hspace{0.2cm}
  \includegraphics[width=7.3cm]{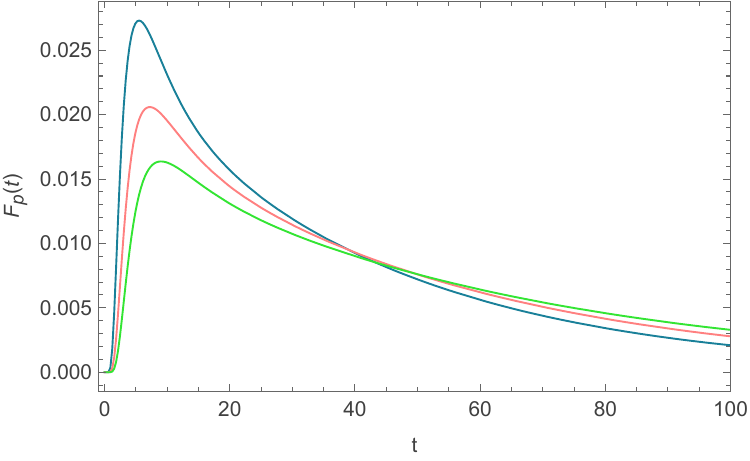}
  \caption{\textit{Left}: Free-energy barrier height 
$F(\tilde{\delta}_m)-F(\tilde{\delta}_h)$ as a function of the CFT temperature $T$
for different values of $C$. 
\textit{Right}: Distribution of first passage time $F_p(t)$ as a function of time $t$
for the same set of parameters. In both panels the transition considered is from
the high-entropy to the low-entropy CFT state. For all the curves we fix $R=1$ and $\tilde{Q}=0.1$.
The blue, red and green curves correspond to $C=1$, $1.3$ and $1.6$,
respectively}
\label{fig_barheight_can_C_htl}
\end{figure}

\noindent
We now turn to the reverse process, namely the transition from the
high-entropy to the low-entropy CFT state. The left panel of
Fig.~\ref{fig_barheight_can_C_htl} shows the corresponding barrier
heights $F(\tilde{\delta}_m)-F(\tilde{\delta}_h)$ for this transition
at different values of $C$, while the right panel displays the
associated first passage time distributions $F_p(t)$. Similar to the
transition from the low-entropy to the high-entropy CFT state shown in
Fig.~\ref{fig_barheight_Fp_can_C_lth}, the barrier height increases as
$C$ increases. Correspondingly, the distributions $F_p(t)$ become
broader: the peak shifts towards larger times and the spread of the
distribution increases with $C$. This indicates that both the mean
first passage time $\tau$ and the associated fluctuations increase
with increasing $C$. These features will be confirmed explicitly by
the plots of $\tau$ and the fluctuations presented below.\\

\begin{figure}[htbp]
  \centering
  \includegraphics[width=7.2cm]{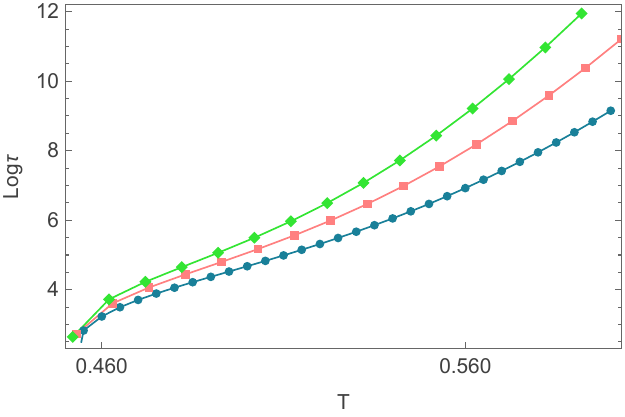}
   \hspace{0.2cm}
  \includegraphics[width=7.3cm]{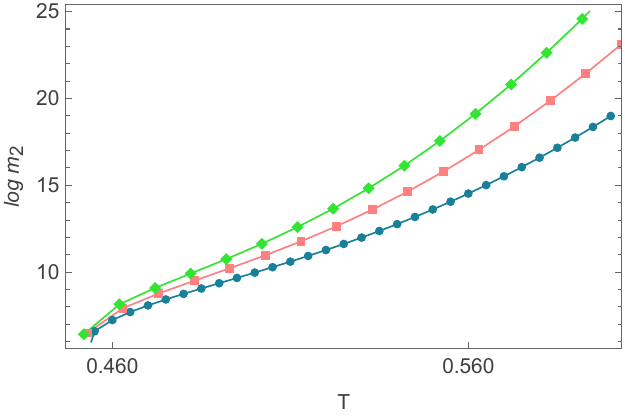}
  \caption{\textit{Left}: Mean first passage time $\tau$ for the transition from the
high-entropy to the low-entropy CFT state as a function of the CFT temperature $T$.
\textit{Right}: the second order moment of the first passage time distribution $m_2$ for the same transition.
 For all curves we fix $R=1$ and $\tilde{Q}=0.1$.
The blue, red and green curves correspond to $C=1$, $1.3$ and $1.6$,
respectively.}
\label{Figlogtau_m_C_htl}
\end{figure}

\noindent
In Fig.~\ref{Figlogtau_m_C_htl}, one notes the the mean first passage time $\tau$ and the
second moment of the first passage time distribution $m_2$ for the transition
from the high-entropy to the low-entropy CFT state at different values of
$C$. As anticipated from the behavior of the barrier heights and distribution of first passage time shown in
Fig.~\ref{fig_barheight_can_C_htl}, both $\tau$ and $m_2$ increases with increasing $C$.\\

\begin{figure}[htbp]
  \centering
  \includegraphics[width=7.2cm]{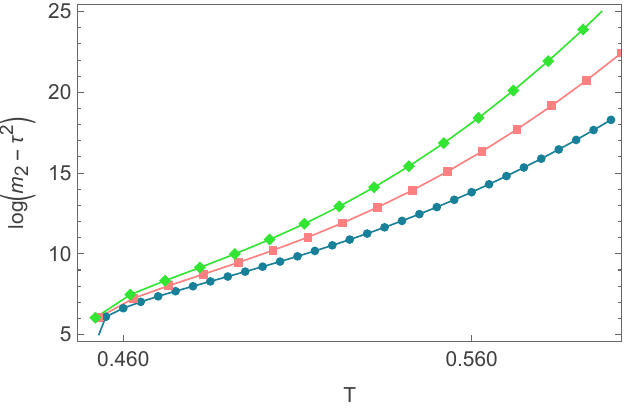}
   \hspace{0.2cm}
  \includegraphics[width=7.3cm]{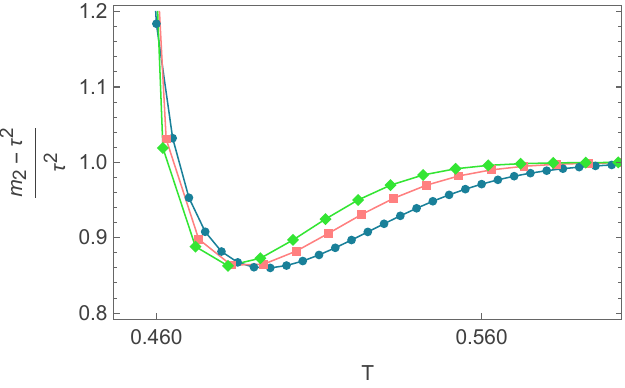}
  \caption{\textit{Left}: Fluctuation $m_{2}-\tau^2$ for the transition from the
low-entropy to the high-entropy CFT state as a function of the CFT temperature $T$.
\textit{Right}: relative fluctuation ${(m_2 - \tau^2)}/{\tau^2}$ for the same transition.
 For all the curves, we fix $R=1$ and $\tilde{Q}=0.1$.
The blue, red and green curves correspond to $C=1$, $1.3$ and $1.6$,
respectively.}
\label{fig_relativevar_var_can_C_htl}
\end{figure}

\noindent
Fig.~\ref{fig_relativevar_var_can_C_htl} shows the fluctuation
$m_{2}-\tau^2$ and the relative fluctuation ${(m_2-\tau^2)}/{\tau^2}$
for the transition from the high-entropy to the low-entropy CFT state
at different values of $C$. As expected from the behavior of the first
passage time distributions shown in Fig.~\ref{fig_barheight_can_C_htl},
the fluctuation $m_{2}-\tau^2$ increases with increasing $C$. The
relative fluctuation ${(m_2-\tau^2)}/{\tau^2}$, however, shows a
nontrivial behavior: it decreases as $C$ increases at lower temperatures,
while at higher temperatures it increases as $C$ increases. The
behavior at higher temperatures is expected, since the barrier height
grows with increasing $C$, and larger fluctuations. The decrease of the relative fluctuation with $C$
at lower temperatures, however, is less clear and may indicate the
presence of additional effects beyond the simple barrier–controlled
picture. It would be nice to explore and understand these issues in future.\\
\FloatBarrier

\section{Fokker-Planck Kinetics in the fixed $(\tilde\Phi,{\cal V},C)$ ensemble}
\label{subsec:kinetics_grand}


\noindent
In this section, we discuss the stochastic dynamics in the  ensemble at fixed
$(\tilde{\Phi},{\cal V},C)$, where the natural order parameter is the CFT charge
$\tilde{Q}$, and at fixed $\tilde{\Phi}$, the off-shell free energy $W$ is given as
\be
W \equiv E-TS-\tilde{\Phi}\tilde{Q}=W(\tilde{Q};T,\tilde{\Phi},R,C)\,.
\ee
The stationary points of this free energy reproduce the equilibrium branches discussed in
Sec.~\ref{subsec:phase_diagrams_grand}. In the regime $\tilde{\Phi}<\tilde{\Phi}_{\rm
crit}$ and $T>T_{\min}$, there are three different phases: a local minima, i.e., the high entropy CFT state, the CFT state with $W =0, \tilde{Q} = 0$ dual to thermal AdS, and finally the local maxima low entropy state which acts as barrier top. This is precisely the setting in which a first-passage analysis is natural. Following the the detailed analysis in the previous section, we will try to be brief with the set up here and model thermally activated switching on the off-shell
landscape as a one-dimensional diffusion process in the collective coordinate $\tilde{Q}$.
The probability density $\rho(\tilde{Q},t)$ evolves under thermal noise on the effective
potential $W(\tilde{Q})$.

\subsection{Fokker--Planck equation and first-passage observables}
\label{subsubsec:FP_forward_grand}

\noindent
The forward Fokker-Planck equation governing the probabilistic evolution in this ensemble, with the order
parameter $\tilde{Q}$ is
\begin{eqnarray}
\label{eq:FP_forward_grand}
\frac{\partial \rho(\tilde{Q},t)}{\partial t}
=
D\,\frac{\partial}{\partial \tilde{Q}}
\left\{
e^{-\beta W(\tilde{Q})}\,
\frac{\partial}{\partial \tilde{Q}}
\Big[
e^{\beta W(\tilde{Q})}\,\rho(\tilde{Q},t)
\Big]
\right\}\,,
\end{eqnarray}
where $\beta\equiv 1/T$ and $D$ is a diffusion coefficient.  In what follows, the forward
Fokker-Planck equation plays a mainly auxiliary role: we introduce it to fix conventions and to define the
appropriate boundary conditions for first-passage problems. Depending on the transition of interest, one imposes reflecting or absorbing boundary
conditions at the endpoints of the chosen computational domain.  At a generic boundary
$\tilde{Q}=\tilde{Q}_0$, the reflecting condition enforces vanishing probability current,
\begin{eqnarray}
\label{eq:bc_reflect_grand}
\left.
e^{-\beta W(\tilde{Q})}\frac{\partial}{\partial \tilde{Q}}
\Big[e^{\beta W(\tilde{Q})}\rho(\tilde{Q},t)\Big]
\right|_{\tilde{Q}=\tilde{Q}_0}
=0\,,
\end{eqnarray}
equivalently
\begin{eqnarray}
\label{eq:bc_reflect_grand_alt}
\left.\Big(\beta W'(\tilde{Q})\,\rho(\tilde{Q},t)+\partial_{\tilde{Q}}\rho(\tilde{Q},t)\Big)\right|_{\tilde{Q}=\tilde{Q}_0}=0\,.
\end{eqnarray}
An absorbing boundary at $\tilde{Q}=\tilde{Q}_0$ implements a perfect sink,
\begin{eqnarray}
\label{eq:bc_absorb_grand}
\rho(\tilde{Q}_0,t)=0\,.
\end{eqnarray}
\noindent
For first passage from $W = 0$ CFT state to high entropy CFT state (and vice versa), we place the absorbing boundary at the barrier top $\tilde{Q}=\tilde{Q}_l$ (low entropy state), while imposing reflecting boundary
conditions on the opposite side to prevent probability leakage out of the physical domain.
In practice, if the landscape becomes singular or numerically stiff near an endpoint, we
implement the reflecting boundary at a small cutoff rather than
at the formal endpoint. This does not affect the first-passage observables provided the cutoff
lies deep inside the corresponding basin.\\

\noindent
As our primary interest is in mean first passage times and their fluctuations as functions of the
thermodynamic control parameters, we work with the
backward (Kolmogorov) formulation, which yields boundary-value problems for the moments
of the first-passage time directly, without evolving the full distribution in time. Let $\tau(\tilde{Q})$ denote the mean first passage time to reach an absorbing boundary,
given an initial condition at $\tilde{Q}$. The backward equation associated with
\eqref{eq:FP_forward_grand} takes the form
\be
\label{eq:backward_operator_grand}
{\cal L}^{\dagger} f(\tilde{Q})
=
D\,e^{\beta W(\tilde{Q})}\frac{\partial}{\partial \tilde{Q}}
\left(
e^{-\beta W(\tilde{Q})}\frac{\partial f(\tilde{Q})}{\partial \tilde{Q}}
\right),
\ee
and the MFPT satisfies
\be
\label{eq:MFPT_grand}
{\cal L}^{\dagger}\tau(\tilde{Q})=-1\,.
\ee
Similarly, the second moment $m_2(\tilde{Q})\equiv \langle t^2\rangle$ obeys
\be
\label{eq:m2_grand}
{\cal L}^{\dagger}m_2(\tilde{Q})=-2\,\tau(\tilde{Q})\,.
\ee
Once $\tau$ and $m_2$ are known, the variance and relative fluctuation are obvious.
The boundary conditions mirror the first-passage setup used in the last section, which we spell out below in the current notations, just for completeness.
For a transition out of a given basin, we impose
\begin{itemize}
\item \emph{Absorbing boundary at the transition state} (barrier top):
\be
\label{eq:bc_absorb_MFPT_grand}
\tau(\tilde{Q}_l)=0,\qquad m_2(\tilde{Q}_l)=0\,.
\ee
\item \emph{Reflecting boundary condition} (no probability flux through the boundary):
\be
\label{eq:bc_reflect_MFPT_grand}
\left.\frac{\partial \tau(\tilde{Q})}{\partial \tilde{Q}}\right|_{\tilde{Q}=\tilde{Q}_{\rm 0}}=0,
\qquad
\left.\frac{\partial m_2(\tilde{Q})}{\partial \tilde{Q}}\right|_{\tilde{Q}=\tilde{Q}_{\rm 0}}=0\,.
\ee
\end{itemize}
Here $\tilde{Q}_{\rm 0}$ denotes a reflecting cutoff chosen on the side opposite to the
absorber.\\

\begin{figure}[!htbp]
    \centering
    \includegraphics[width=.5\linewidth]{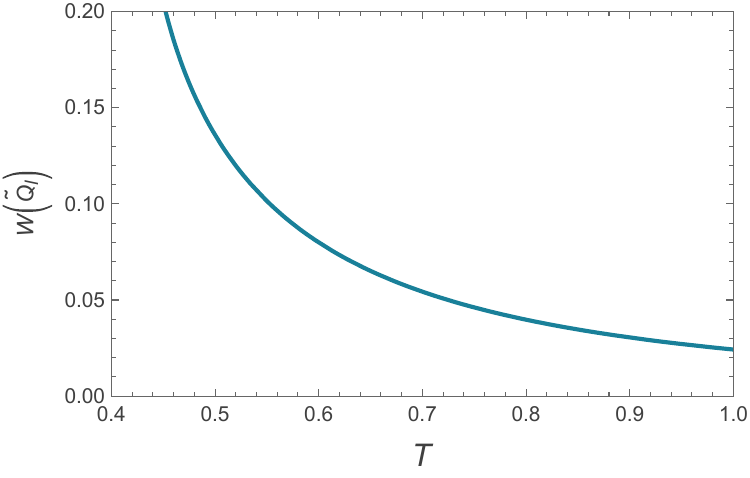}
    \caption{ barrier height
$W(\tilde{Q}_l)$
between the low entropy and thermal CFT states. Here  we fix $C=1$, $R=1$ and $\tilde{\Phi} = 0.2\tilde{\Phi}_c $}
    \label{fig_bar_height_grand_VtL}
\end{figure}

\noindent
In the numerics we initialize the process at the location of the relevant equilibrium branch.
Concretely, the equilibrium values $\tilde{Q}_{l,h}$ are obtained by solving
$\partial_{\tilde{Q}}W(\tilde{Q})=0$.
Before turning to the kinetic analysis, we first examine the barrier heights that control the stochastic dynamics. In the grand canonical ensemble the thermal CFT state lies at $\tilde{Q}=0$ with free energy $W=0$. The barrier separating the thermal CFT state from the high-entropy CFT state is therefore set by the free energy of the intermediate low-entropy state, namely $W(\tilde{Q}_l)$. Figure~\ref{fig_bar_height_grand_VtL} shows this barrier height as a function of the CFT temperature $T$. We observe that $W(\tilde{Q}_l)$ decreases monotonically with increasing temperature.\\

\begin{figure}
  \centering
  \includegraphics[width=7.4cm]{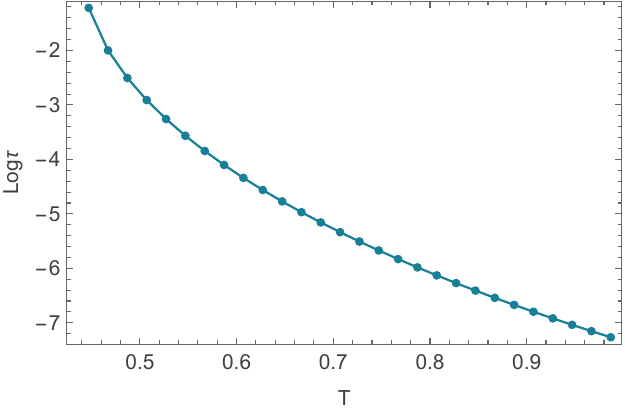}
   \hspace{0.2cm}
  \includegraphics[width=7.4cm]{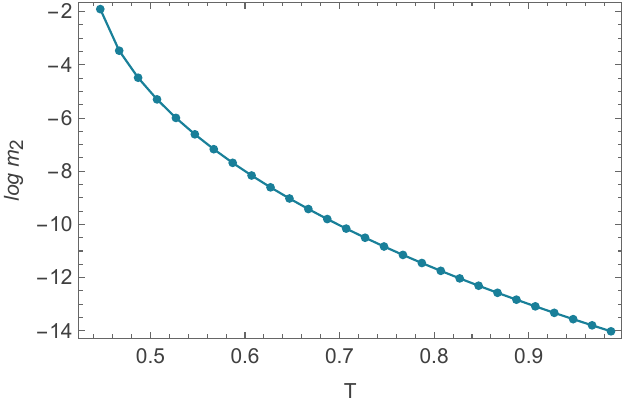}
  \caption{The left panel is the plot of mean first passage time $\tau$ for thermal CFT to high entropy CFT state transition as a function of CFT temperature $T$ and the right is the second order moment of the first passage time distribution $m_2$ for thermal CFT to high entropy CFT  state transition.  For both curves $ \tilde{\Phi} = 0.2\tilde{\Phi}_{c}, R = 1, C = 1$ }
   \label{figlogtaum_grand_VtH}
\end{figure}

\noindent
Fig.~\ref{figlogtaum_grand_VtH} shows the mean first passage time $\tau$ for the transition
from the thermal CFT state to the high-entropy CFT state, together with the second moment
$m_{2}$ of the first-passage-time distribution, as functions of the CFT temperature $T$.
The vertical axis is shown on a logarithmic scale. We observe that the mean first passage time
decreases monotonically with increasing temperature. This behavior can be understood from
the corresponding free-energy barrier. As the temperature increases, the barrier height
$W(\tilde{Q}_{l})$  decreases. As a result, the characteristic transition time becomes
shorter. In addition, higher temperatures enhance thermal diffusion on the free-energy landscape,
which further facilitates barrier crossing.

\noindent
For this transition the reflecting boundary condition is imposed at $\tilde{Q}=0$, corresponding
to the thermal CFT state.
The absorbing boundary condition is imposed at $\tilde{Q}=\tilde{Q}_{l}$, which corresponds
to the transition state located at the free-energy barrier top.\\

\begin{figure}
  \centering
  \includegraphics[width=7.2cm]{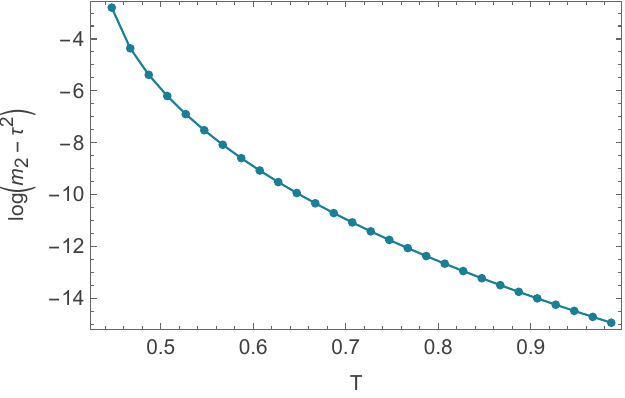}
   \hspace{0.2cm}
  \includegraphics[width=7.5cm]{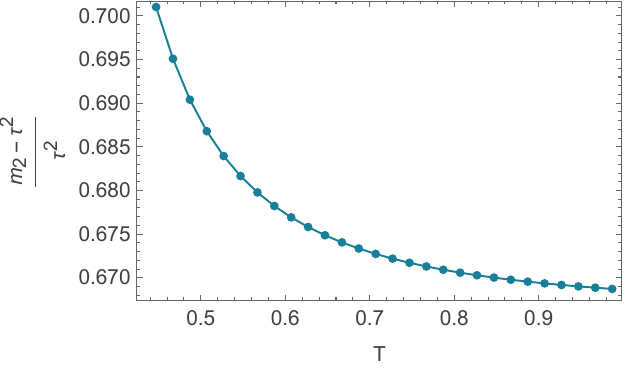}
  \caption{The left panel is the plot of fluctuation $m_{2}-\tau^2$ for thermal CFT to high entropy CFT state transition as a function of CFT temperature $T$ and the right is the relative fluctuation  ${(m_2 - \tau^2)}/{\tau^2}$ for thermal CFT to high entropy CFT  state transition. For both curves $  \tilde{\Phi} = 0.2\tilde{\Phi}_{c}, R = 1, C = 1$ }
   \label{figFlucgrand_VtH}
\end{figure}

\noindent
The second-order fluctuation $m_{2}-\tau^{2}$ and the relative fluctuation
$(m_{2}-\tau^{2})/\tau^{2}$ of the first passage time are shown in
Fig.~\ref{figFlucgrand_VtH}. The vertical axis in the left panel is displayed
on a logarithmic scale. We observe that the fluctuation is largest near the
minimal temperature $T_{\min}$ and decreases as the temperature increases.
At sufficiently high temperatures the fluctuation becomes very small. The
relative fluctuation also decreases monotonically with temperature, although
the overall variation across the temperature range is very small. This behavior
is closely related to the structure of the free-energy landscape. In
particular, the barrier height governing the transition does not change
significantly with temperature, leading to only mild variations in the
relative fluctuations. Overall, both the mean transition time
and its fluctuations are controlled by the barrier structure of the
off-shell free-energy landscape. The relatively small magnitude of the
relative fluctuations indicates that thermal fluctuations have only a
limited impact on the kinetics of the transition from the thermal CFT state
to the high-entropy CFT state.\\

\begin{figure}[!htbp]
    \centering
    \includegraphics[width=.5\linewidth]{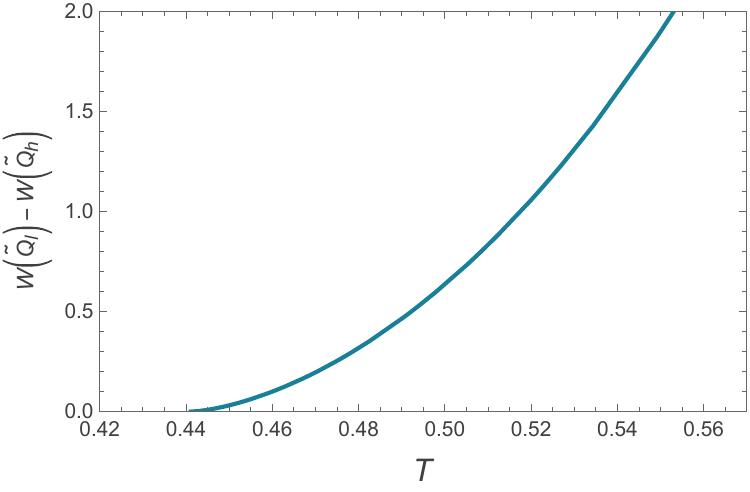}
    \caption{ barrier height
$W(\tilde{Q}_l)$-$W(\tilde{Q}_h)$
between the low entropy and high entropy CFT states. Here  we fix $C=1$, $R=1$ and $\tilde{\Phi} = 0.2\tilde{\Phi}_c $}
    \label{fig_bar_height_grand_LtV}
\end{figure}

 \noindent
We now turn to the reverse process, namely the transition from the high-entropy CFT state to the thermal CFT state. In this case the system must cross the intermediate low-entropy CFT state, which acts as the barrier separating the two phases. The corresponding barrier height is therefore given by $W(\tilde{Q}_l)-W(\tilde{Q}_h)$. Figure~\ref{fig_bar_height_grand_LtV} shows this barrier height as a function of the CFT temperature $T$. We observe that it increases monotonically with increasing temperature.\\

\begin{figure}
  \centering
  \includegraphics[width=7.4cm]{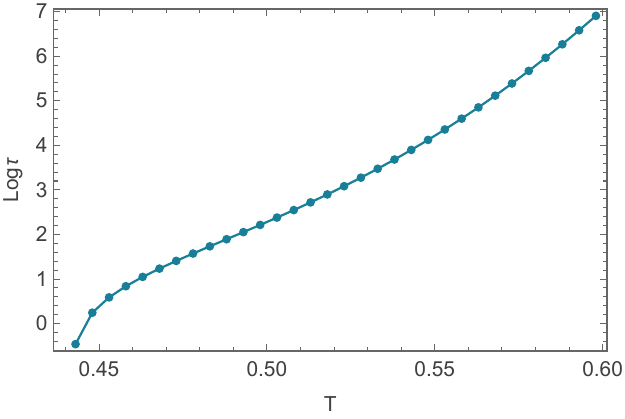}
   \hspace{0.2cm}
  \includegraphics[width=7.4cm]{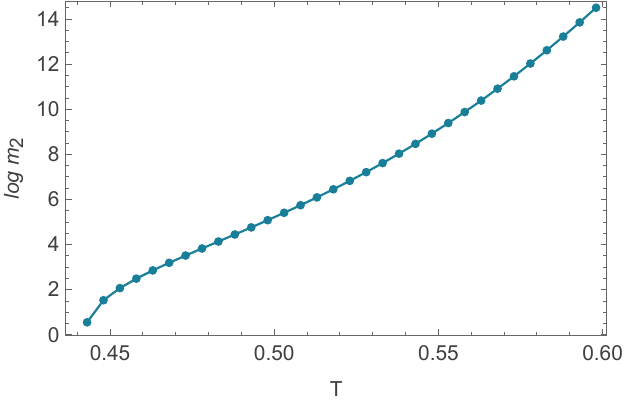}
  \caption{The left panel is the plot of mean first passage time $\tau$ for high entropy to thermal CFT state transition as a function of CFT temperature $T$ and the right is the second order moment of the first passage time distribution $m_2$ for high entropy  to thermal CFT  state transition.For both curves $ \tilde{\Phi} = 0.2\tilde{\Phi}_{c}, R = 1, C = 1$ }
   \label{figlogtaum_grand_HtV}
\end{figure}

\noindent
The behavior of the mean first passage time $\tau$ and its second moment $m_{2}$
for the transition from the high-entropy CFT state to the thermal CFT state are
shown in Fig.~\ref{figlogtaum_grand_HtV}. In this case, both $\tau$ and $m_{2}$
increase with increasing temperature. This indicates that the escape of the
high-entropy CFT state to the thermal CFT state becomes progressively more
difficult at higher temperatures. The reason can be understood from the
corresponding free-energy barrier. As shown in figure \ref{fig_bar_height_grand_LtV}, the barrier height
$W(\tilde{Q}_{l})-W(\tilde{Q}_{h})$ increases with temperature, which suppresses
the probability of crossing the barrier under thermal fluctuations. As a
result, the high-entropy CFT state becomes increasingly stable as the
temperature rises, in agreement with the behavior expected from the
free-energy landscape.\\

\begin{figure}
  \centering
  \includegraphics[width=7.2cm]{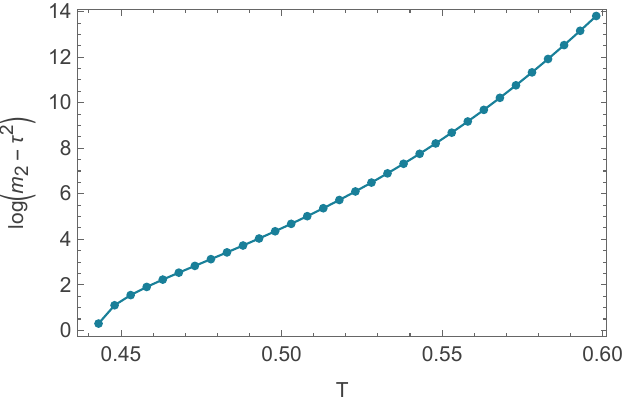}
   \hspace{0.2cm}
  \includegraphics[width=7.5cm]{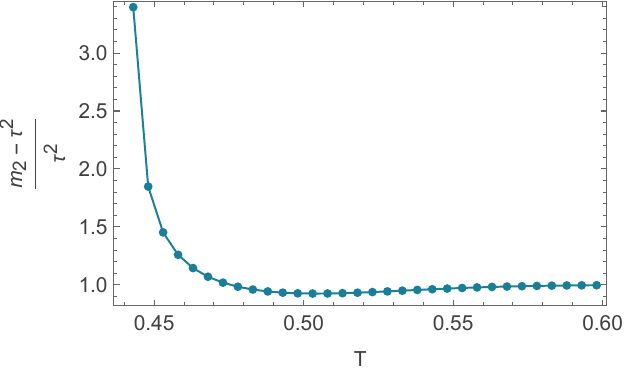}
  \caption{The left panel is the plot of fluctuation $m_{2}-\tau^2$ for high entropy CFT to thermal CFT state transition as a function of CFT temperature $T$ and the right is the relative fluctuation  ${(m_2 - \tau^2)}/{\tau^2}$ for high entropy CFT to thermal CFT  state transition. For both curves $ \tilde{\Phi} = 0.2\tilde{\Phi}_{c}, R = 1, C = 1$ }
   \label{figFlucgrand_HtV}
\end{figure}

\noindent
The variance $m_{2}-\tau^{2}$ and the relative variance
$(m_{2}-\tau^{2})/\tau^{2}$ for the transition from the high-entropy CFT
state to the thermal CFT state are shown in
Fig.~\ref{figFlucgrand_HtV}. The variance increases with increasing
temperature, indicating that the fluctuations in the first passage time
become larger as the temperature rises. This behavior follows from the
corresponding free-energy barrier. As discussed earlier, the barrier
height $W(\tilde{Q}_{l})-W(\tilde{Q}_{h})$ increases with temperature,
which suppresses the probability of barrier crossing and leads to
larger fluctuations in the transition time. The relative fluctuation, however, shows a different behavior. It decreases rapidly at lower temperatures, reaches a minimum, and then approaches an approximately constant value at higher temperatures. This behavior can be understood from the interplay between the free-energy barrier and thermal fluctuations. At low temperatures the barrier height is small compared to the thermal fluctuations, leading to relatively large fluctuations in the first passage time. As the temperature increases, the barrier height grows and begins to dominate the dynamics, which suppresses the relative fluctuation. At sufficiently high temperatures the relative fluctuation varies only weakly with temperature and approaches a nearly constant value.\\
\FloatBarrier
\noindent
Overall, the stochastic dynamics in the this ensemble is largely governed by the structure of the off-shell free-energy landscape. In particular, the temperature dependence of the barrier heights plays the dominant role in determining both the mean first passage time and its fluctuations. When the barrier decreases, the transition becomes easier and the characteristic timescale is reduced, whereas an increasing barrier suppresses the transition and stabilizes the corresponding phase. The behavior of the variance and relative variance further reflects this barrier-controlled dynamics. These results are therefore consistent with the thermodynamic picture obtained from the off-shell phase diagrams.

\subsubsection{Effect of variation of $C$ }
\begin{figure}[!htbp]
    \centering
    \includegraphics[width=.5\linewidth]{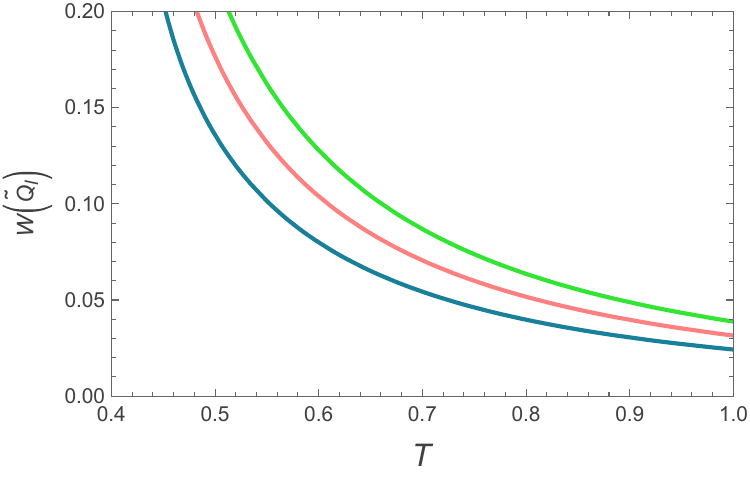}
    \caption{ barrier height
$W(\tilde{Q}_l)$
between the low entropy and thermal CFT states. Here for all curves we fix  $R=1$ and $\tilde{\Phi} = 0.2\tilde{\Phi}_c $. The red, blue and green curves corresponds to $C =1,1.3$ and $1.6$ respectively.}
    \label{fig_barheight_grand_C_VtL}
\end{figure}

\noindent
We now examine the effect of varying the central charge $C$ on the kinetics. As the stochastic dynamics is governed by the
free-energy barrier, we first consider the corresponding barrier height.
Figure~\ref{fig_barheight_grand_C_VtL} shows the barrier height $W(\tilde{Q}_l)$
between the thermal CFT state and the low-entropy CFT state for different
values of $C$. The barrier height increases  with increasing $C$.\\

\begin{figure}
  \centering
  \includegraphics[width=7.4cm]{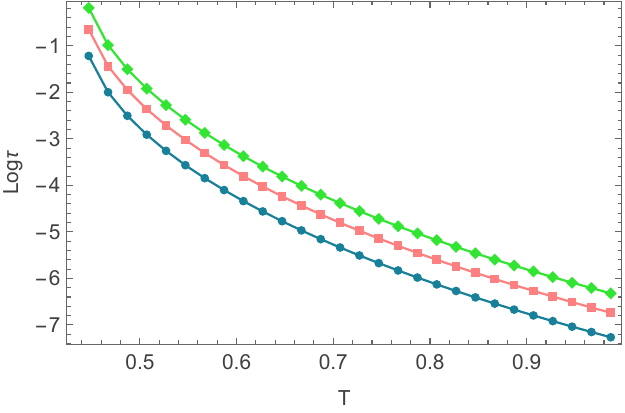}
   \hspace{0.2cm}
  \includegraphics[width=7.4cm]{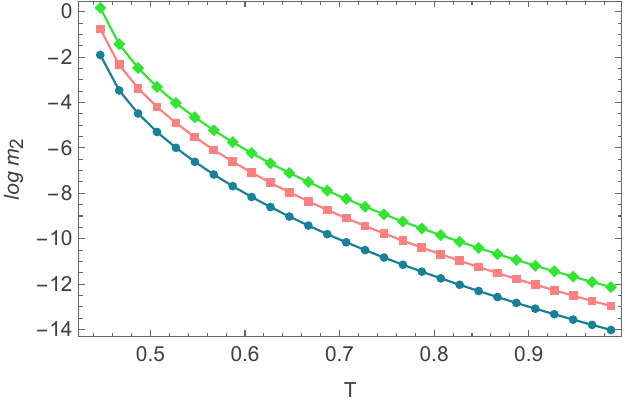}
 \caption{\textit{Left}: Mean first passage time $\tau$ for the transition from the thermal CFT state to the high-entropy CFT state as a function of the CFT temperature $T$. 
\textit{Right}: Second moment of the first-passage-time distribution $m_2$ for the same transition.
For all the curves, we fix $\tilde{\Phi}=0.2\,\tilde{\Phi}_{c}$ and $R=1$. 
The blue, red and green curves correspond to $C=1$, $1.3$ and $1.6$, respectively.}
   \label{figlogtaum_grand_C_VtH}
\end{figure}

\noindent
In Fig.~\ref{figlogtaum_grand_C_VtH} we show the mean first passage time $\tau$ for the transition
from the thermal CFT state to the high-entropy CFT state, together with the second moment
$m_{2}$ of the first-passage-time distribution, as functions of the CFT temperature $T$
for different values of $C$. The vertical axis is shown on a logarithmic scale. We observe
that the mean first passage time increases with increasing $C$, which is consistent with
the behavior of the barrier height, since the barrier height increases as $C$ increases.\\

\begin{figure}
  \centering
  \includegraphics[width=7.2cm]{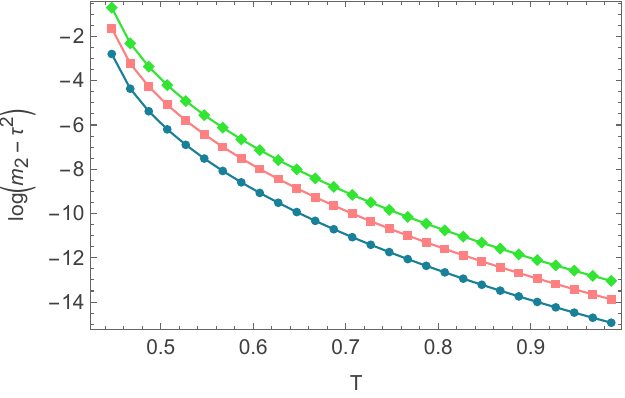}
   \hspace{0.2cm}
  \includegraphics[width=7.5cm]{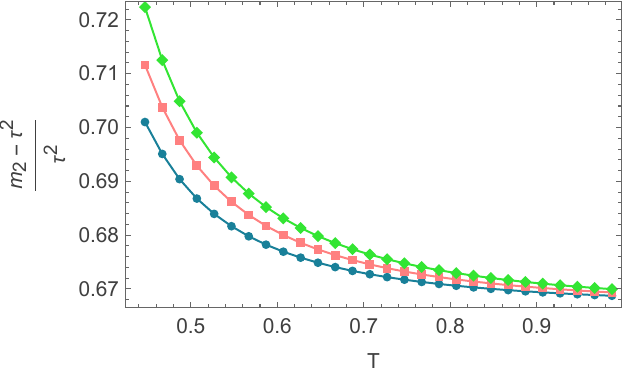}
  \caption{\textit{Left}: Fluctuation $m_{2}-\tau^{2}$ for the transition from the thermal CFT state to the high-entropy CFT state as a function of the CFT temperature $T$. 
\textit{Right}: Relative fluctuation $(m_{2}-\tau^{2})/\tau^{2}$ for the same transition.
For all curves we fix $\tilde{\Phi}=0.2\,\tilde{\Phi}_{c}$ and $R=1$.
The blue, red and green curves correspond to $C=1$, $1.3$ and $1.6$, respectively.}
   \label{figFlucgrand_C_VtH}
\end{figure}

\noindent
The second-order fluctuation $m_{2}-\tau^{2}$ and the relative fluctuation
$(m_{2}-\tau^{2})/\tau^{2}$ of the first passage time for the transition from the
thermal CFT state to the high-entropy CFT state are shown in
Fig.~\ref{figFlucgrand_C_VtH}. The vertical axis in the left panel is displayed
on a logarithmic scale. We observe that both the fluctuation and the relative
fluctuation increase with increasing $C$, consistent with the behavior of the
barrier heights shown in Fig.~\ref{fig_barheight_grand_C_VtL}. This indicates
that, for the transition from the thermal CFT state to the high-entropy CFT
state in the grand canonical ensemble, the increase in the barrier height with
$C$ plays a dominant role compared to thermal fluctuations.\\

\begin{figure}[!htbp]
    \centering
    \includegraphics[width=.5\linewidth]{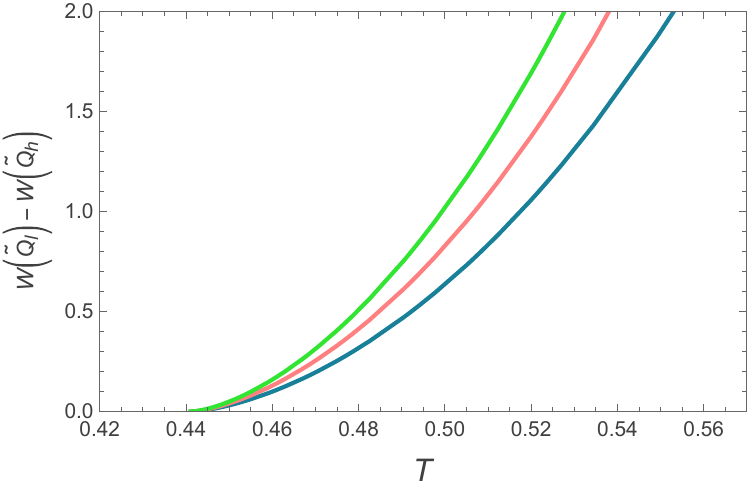}
    \caption{ barrier height
$W(\tilde{Q}_l)-W(\tilde{Q}_h)$
between the low entropy and high entropy CFT states. Here for all curves we fix  $R=1$ and $\tilde{\Phi} = 0.2\tilde{\Phi}_c $. The red, blue and green curves corresponds to $C =1,1.3$ and $1.6$ respectively.}
    \label{fig_barheight_grand_C_LtV}
\end{figure}

\noindent
The reverse process corresponds to the transition from the high-entropy
CFT state to the thermal CFT state. We first examine the corresponding
barrier height. Figure~\ref{fig_barheight_grand_C_LtV} shows the barrier
height $W(\tilde{Q}_l)-W(\tilde{Q}_h)$ between the low-entropy and the
high-entropy CFT states for different values of $C$. We observe that the
barrier height increases with increasing $C$.\\

\begin{figure}
  \centering
  \includegraphics[width=7.4cm]{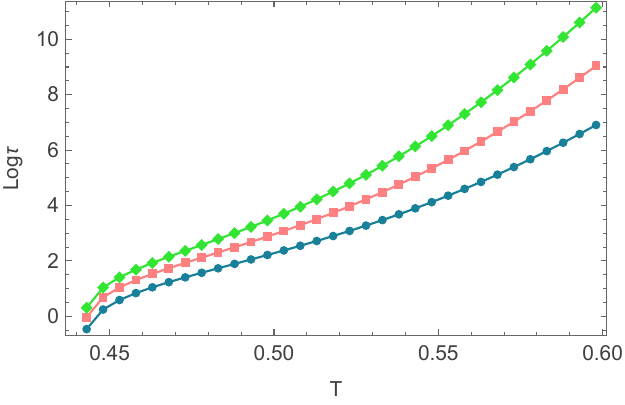}
   \hspace{0.2cm}
  \includegraphics[width=7.4cm]{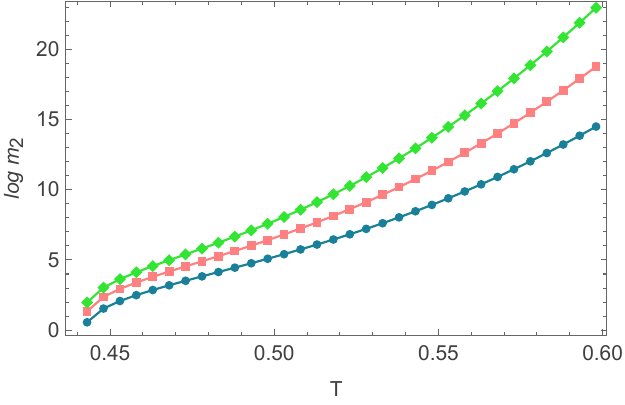}
 \caption{\textit{Left}: Mean first passage time $\tau$ for the transition from the high entropy CFT state to the thermal CFT state as a function of the CFT temperature $T$. 
\textit{Right}: Second moment of the first-passage-time distribution $m_2$ for the same transition.
For all curves we fix $\tilde{\Phi}=0.2\,\tilde{\Phi}_{c}$ and $R=1$. 
The blue, red and green curves correspond to $C=1$, $1.3$ and $1.6$, respectively.}
   \label{figlogtaum_grand_C_HtV}
\end{figure}

\noindent
Fig.~\ref{figlogtaum_grand_C_HtV} captures the mean first passage time $\tau$ for the transition
from the high-entropy CFT state to thermal CFT state together with the second moment
$m_{2}$ of the first-passage-time distribution, as functions of the CFT temperature $T$
for different values of $C$. The vertical axis is shown on a logarithmic scale. We observe
that the mean first passage time increases with increasing $C$, which is consistent with
the behavior of the barrier height figure \ref{fig_barheight_grand_C_LtV}, since the barrier height increases as $C$ increases.\\

\begin{figure}[!htbp]
    \centering
    \includegraphics[width=.5\linewidth]{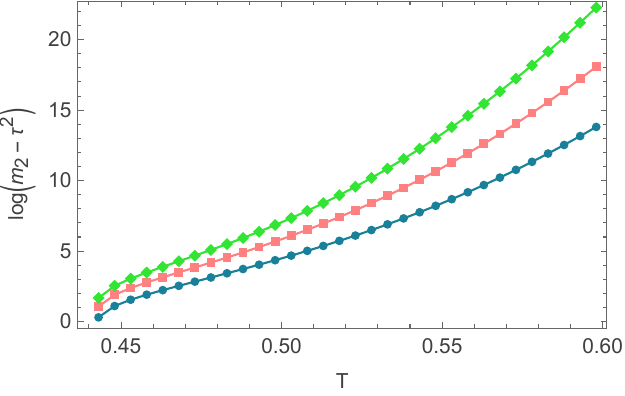}
    \caption{Fluctuation $m_{2}-\tau^{2}$ for the transition from the high entropy CFT state to thermal CFT state as a function of the CFT temperature $T$. For all the curves, we fix $\tilde{\Phi}=0.2\,\tilde{\Phi}_{c}$ and $R=1$.
The blue, red and green curves correspond to $C=1$, $1.3$ and $1.6$, respectively.}
    \label{figFlucgrand_C_HtV}
\end{figure}

\noindent
Finally, we present the fluctuation $m_{2}-\tau^{2}$ of the first passage time
for the transition from the high-entropy CFT state to the thermal CFT state in
Fig.~\ref{figFlucgrand_C_HtV}. We observe that the fluctuation increases with
increasing $C$, consistent with the behavior of the barrier heights shown in
Fig.~\ref{fig_barheight_grand_C_LtV}. We have also computed the relative
fluctuation $(m_{2}-\tau^{2})/\tau^{2}$ for different values of $C$, but do not
find any noticeable variation and therefore do not display it here. This
suggests that, for the transition from the high-entropy CFT state to the
thermal CFT state, the increase in the barrier
height with $C$ plays a more dominant role than thermal fluctuations.\\

\FloatBarrier

\section{Remarks} \label{sec:conclusion}
\noindent
In this work, we studied the thermodynamics and phase structure of holographic conformal field theories dual to spherically symmetric charged AdS black holes using an off-shell free energy framework. We considered three ensembles of the dual CFT with fixed: $(\tilde Q,{\cal V},C)$, $(\tilde \Phi,{\cal V},C)$, and  $(\tilde Q,{\cal V},\mu)$. We presented a detailed construction of off-shell free energy and identified the appropriate order parameters for each of the three cases. Following this, we discussed the phase structure in different parameter regimes and compared it with the on-shell analysis performed earlier in~\cite{Cong:2021jgb}).  An important advantage  of the off-shell phase structure is that some of the thermodynamic quantities can be taken to be control parameters, giving a handle on exploring dynamics close to equilibrium. Taking advantage of this set up,  we studied the transitions between competing states using a stochastic description on the various phases given by off-shell free energy. This is described by an ensemble-dependent Fokker–Planck equation, allowing us to compute the first-passage-time distribution, including the mean first passage time and its fluctuations over a range of temperatures. The stochastic analysis was presented wherever applicable, i.e.,  to the fixed $(\tilde Q,{\cal V},C)$ and $(\tilde \Phi,{\cal V},C)$ ensembles.  We also examined how the phase structure and the associated kinetics depend on the electric charge $\tilde Q$ and the central charge $C$.
For both the fixed $(\tilde \Phi,{\cal V},C)$) and  $(\tilde Q,{\cal V},C)$) ensembles, the phase transitions are first order, and the stochastic dynamics is governed by barrier crossing on the free-energy landscape. In contrast, the new ensemble with fixed $(\tilde Q,{\cal V},\mu)$ exhibits a zeroth-order phase transition, where the free energy is discontinuous and no barrier structure is present. As a result, the stochastic framework used in the previous two ensembles cannot be applied directly in this case. Nevertheless, for any future study of stochastic or statistical properties in this ensemble, the off-shell free energy in eq.~\eqref{Unique_Offshell} provides a natural candidate for an effective potential. The  results on the kinetics using the Fokker-Planck equation are summarized in Table \ref{tab:stochastic_summary}. We leave a detailed analysis of such dynamics for future work.

\begin{table}[t]
\centering
\small
\renewcommand{\arraystretch}{1.35}
\setlength{\tabcolsep}{5pt}

\begin{tabular}{|p{1.7cm}|p{2.0cm}|c|c|c|c|}
\hline
\textbf{Ensemble} & \textbf{Transition} & \textbf{Parameter}
& \textbf{MFPT} & \textbf{Relative fluctuation} & \textbf{Figure} \\
 &  & 
& &  & \\
\hline

\multirow{6}{=}{{\bf Fixed $(\tilde Q,\mathcal V,C)$}}
& \multirow{3}{=}{Low $\rightarrow$ High}
& $T$ & Decreases & Increases
& Figs.~\ref{figlogtaum1} and \ref{figFluclth1} \\
\cline{3-6}
& & $\tilde Q$ & Decreases & Increases
& Figs.~\ref{Figlogtau_m_Qlth} and \ref{fig_relativevar_var_can_Q} \\
\cline{3-6}
& & $C$ & Increases & Decreases
& Figs.~\ref{Figlogtau_m_C_lth} and \ref{fig_relativevar_var_can_C_lth} \\
\cline{2-6}

& \multirow{3}{=}{High $\rightarrow$ Low}
& $T$ & Increases & Max $\rightarrow$ Min $\rightarrow$ Saturation
& Figs.~\ref{figlogtaumhtl1} and \ref{figFluchtl1} \\
\cline{3-6}
& & $\tilde Q$ & Decreases & Nearly unchanged
& Fig.~\ref{FiglogtauQhtl} \\

\cline{3-6}
& & $C$ & Increases & $\downarrow$ at low $T$, $\uparrow$ at high $T$
& Figs.~\ref{Figlogtau_m_C_htl} and \ref{fig_relativevar_var_can_C_htl} \\
\hline

\multirow{4}{=}{{\bf Fixed $(\tilde\Phi,\mathcal V,C)$}}
& \multirow{2}{=}{Thermal $\rightarrow$ High}
& $T$ & Decreases & Decreases
& Figs.~\ref{figlogtaum_grand_VtH} and \ref{figFlucgrand_VtH} \\
\cline{3-6}
& & $C$ & Increases & Increases
& Figs.~\ref{figlogtaum_grand_C_VtH} and \ref{figFlucgrand_C_VtH} \\
\cline{2-6}

& \multirow{2}{=}{High $\rightarrow$ Thermal}
& $T$ & Increases & Max $\rightarrow$ Min $\rightarrow$ Saturation
& Figs.~\ref{figlogtaum_grand_HtV} and \ref{figFlucgrand_HtV} \\
\cline{3-6}
& & $C$ & Increases & Nearly unchanged
& Fig.~\ref{figlogtaum_grand_C_HtV}
\\
\hline

\end{tabular}

\caption{Summary of the stochastic first-passage results in the canonical and grand canonical ensembles. Since the second moment $m_2$ and the fluctuation $m_2-\tau^2$ follow the same qualitative trend as the MFPT $\tau$, only the MFPT and the relative fluctuation are listed.}
\label{tab:stochastic_summary}
\end{table}

\section*{Acknowledgements}
We thank Malay K. Bandyopadhyay for helpful discussions. C.B. thanks Renann Lipinski Jusinskas for hospitality at FZU and CEICO, Prague. C.B. gratefully acknowledges ARG-MATRICS grant no. ANRF/ARGM/2025/002280/MTR.

\appendix
\section{The backward Fokker-Planck equation}
\paragraph{Forward operator and current:}
For fixed temperature $T$ we denote $\beta\equiv 1/T$ and write the forward equation \eqref{ForwardFPequation} as
\begin{equation}
\partial_t \rho(\tilde\delta,t)=\mathcal{L}\rho(\tilde\delta,t),
\qquad
\mathcal{L}\rho
\equiv
\partial_{\tilde\delta}\!\left[
D(\tilde\delta)\,e^{-\beta F(\tilde\delta)}\,
\partial_{\tilde\delta}\!\Big(e^{\beta F(\tilde\delta)}\,\rho\Big)
\right],
\label{eq:forward_operator_full}
\end{equation}
so that the continuity form reads
\begin{equation}
\partial_t\rho + \partial_{\tilde\delta}J=0,
\qquad
J(\tilde\delta,t)\equiv
-\,D(\tilde\delta)\,e^{-\beta F(\tilde\delta)}\,
\partial_{\tilde\delta}\!\Big(e^{\beta F(\tilde\delta)}\,\rho(\tilde\delta,t)\Big).
\label{eq:current_def_full}
\end{equation}
In our implementation we take $D(\tilde\delta)=T$ (constant for fixed $T$), but we keep $D(\tilde\delta)$
in the derivation to make the adjoint structure transparent.

\paragraph{Adjointness by explicit integration by parts.}
Consider the standard $L^2$ inner product on the interval $[\tilde\delta_L,\tilde\delta_B]$,
\begin{equation}
\langle f,g\rangle \equiv \int_{\tilde\delta_L}^{\tilde\delta_B} d\tilde\delta\; f(\tilde\delta)\,g(\tilde\delta),
\label{eq:inner_product}
\end{equation}
and let $f(\tilde\delta)$ and $g(\tilde\delta)$ be sufficiently smooth test functions.
Starting from the forward operator \eqref{eq:forward_operator_full}, we compute
\begin{align}
\langle f,\mathcal{L}g\rangle
&=
\int_{\tilde\delta_L}^{\tilde\delta_B} d\tilde\delta\;
f\,
\partial_{\tilde\delta}\!\left[
D\,e^{-\beta F}\,
\partial_{\tilde\delta}\!\Big(e^{\beta F}\,g\Big)
\right]
\nonumber\\
&=
\left[
f\,D\,e^{-\beta F}\,
\partial_{\tilde\delta}\!\Big(e^{\beta F}\,g\Big)
\right]_{\tilde\delta_L}^{\tilde\delta_B}
-
\int_{\tilde\delta_L}^{\tilde\delta_B} d\tilde\delta\;
f'\,D\,e^{-\beta F}\,
\partial_{\tilde\delta}\!\Big(e^{\beta F}\,g\Big),
\label{eq:adjoint_step1}
\end{align}
where primes denote $\partial_{\tilde\delta}$.
We now expand the derivative
\begin{equation}
\partial_{\tilde\delta}\!\Big(e^{\beta F}\,g\Big)
=
e^{\beta F}\Big(g'+\beta F' g\Big),
\label{eq:expand_exp}
\end{equation}
so that $e^{-\beta F}\partial_{\tilde\delta}(e^{\beta F}g)=g'+\beta F' g$.
Inserting this into \eqref{eq:adjoint_step1} yields
\begin{align}
\langle f,\mathcal{L}g\rangle
&=
\left[
f\,D\,(g'+\beta F' g)
\right]_{\tilde\delta_L}^{\tilde\delta_B}
-
\int_{\tilde\delta_L}^{\tilde\delta_B} d\tilde\delta\;
D\,f'\,(g'+\beta F' g)
\nonumber\\
&=
\left[
f\,D\,(g'+\beta F' g)
\right]_{\tilde\delta_L}^{\tilde\delta_B}
-
\int_{\tilde\delta_L}^{\tilde\delta_B} d\tilde\delta\;
D\,f' g'
-
\int_{\tilde\delta_L}^{\tilde\delta_B} d\tilde\delta\;
D\,\beta F'\,f' g.
\label{eq:adjoint_step2}
\end{align}
We integrate by parts once more in the first bulk term:
\begin{equation}
-\int_{\tilde\delta_L}^{\tilde\delta_B} d\tilde\delta\; D\,f' g'
=
-\left[ D\,f' g \right]_{\tilde\delta_L}^{\tilde\delta_B}
+
\int_{\tilde\delta_L}^{\tilde\delta_B} d\tilde\delta\;
(D\,f')' g.
\label{eq:adjoint_step3}
\end{equation}
Combining \eqref{eq:adjoint_step2} and \eqref{eq:adjoint_step3}, we arrive at the adjoint identity
\begin{align}
\langle f,\mathcal{L}g\rangle
&=
\int_{\tilde\delta_L}^{\tilde\delta_B} d\tilde\delta\;
\underbrace{\Big[(D f')' - D\,\beta F' f'\Big]}_{\equiv\;\mathcal{L}^\dagger f}\; g
\nonumber\\
&\quad
+
\left[
f\,D\,(g'+\beta F' g) - D\,f' g
\right]_{\tilde\delta_L}^{\tilde\delta_B}.
\label{eq:adjoint_identity_full}
\end{align}
Therefore the backward (adjoint) operator is
\begin{equation}
\mathcal{L}^\dagger f(\tilde\delta)
=
(D(\tilde\delta)\,f'(\tilde\delta))'
-
D(\tilde\delta)\,\beta F'(\tilde\delta)\,f'(\tilde\delta).
\label{eq:backward_operator_general}
\end{equation}
In the case of constant diffusion coefficient $D(\tilde\delta)=T$ used throughout our numerics,
\eqref{eq:backward_operator_general} reduces to the familiar form
\begin{equation}
\mathcal{L}^\dagger
=
T\left(\partial_{\tilde\delta}^2-\beta F'(\tilde\delta)\,\partial_{\tilde\delta}\right).
\label{eq:backward_operator_constantD}
\end{equation}

\paragraph{Boundary conditions and vanishing of boundary terms.}
For the first-passage problem we work on $\tilde\delta\in[\tilde\delta_L,\tilde\delta_B]$ with
a reflecting boundary at $\tilde\delta=\tilde\delta_L$ and an absorbing boundary at
$\tilde\delta=\tilde\delta_B$ (in our applications $\tilde\delta_B=\tilde\delta_m$, the barrier top).
The forward boundary conditions are
\begin{equation}
J(\tilde\delta_L,t)=0
\quad \Longleftrightarrow\quad
\big(g'+\beta F' g\big)\big|_{\tilde\delta_L}=0,
\qquad
g(\tilde\delta_B,t)=0,
\label{eq:forward_BCs}
\end{equation}
where $g$ stands for the forward density $\rho$.
With these forward conditions, the boundary term in \eqref{eq:adjoint_identity_full} vanishes provided the
backward test function $f$ satisfies
\begin{equation}
f'(\tilde\delta_L)=0,
\qquad
f(\tilde\delta_B)=0,
\label{eq:backward_BCs_full}
\end{equation}
which are precisely the reflecting/absorbing conditions in the backward formulation.

\paragraph{Mean first-passage time.}
Let $\mathcal{T}$ denote the first-passage time to reach the absorbing point $\tilde\delta_B$.
The mean first-passage time (MFPT) for a trajectory starting at $\tilde\delta$ is
\begin{equation}
\tau(\tilde\delta)\equiv \big\langle \mathcal{T}\big\rangle_{\tilde\delta(0)=\tilde\delta}.
\label{eq:mfpt_def_full}
\end{equation}
Let $\rho(\tilde\delta,t\,|\,\tilde\delta_0)$ solve the forward equation with initial condition
$\rho(\tilde\delta,0\,|\,\tilde\delta_0)=\delta(\tilde\delta-\tilde\delta_0)$ and the first-passage
boundary conditions \eqref{eq:forward_BCs}.
Multiplying $\partial_t\rho=\mathcal{L}\rho$ by $\tau(\tilde\delta)$ and integrating over
$[\tilde\delta_L,\tilde\delta_B]$, we obtain
\begin{equation}
\frac{d}{dt}\int_{\tilde\delta_L}^{\tilde\delta_B} d\tilde\delta\;
\tau(\tilde\delta)\,\rho(\tilde\delta,t\,|\,\tilde\delta_0)
=
\int_{\tilde\delta_L}^{\tilde\delta_B} d\tilde\delta\;
\big(\mathcal{L}^\dagger \tau\big)(\tilde\delta)\,\rho(\tilde\delta,t\,|\,\tilde\delta_0),
\label{eq:tau_weighted_evol_full}
\end{equation}
where we used \eqref{eq:adjoint_identity_full} and the boundary conditions
\eqref{eq:forward_BCs}--\eqref{eq:backward_BCs_full} to drop the boundary term.
Choosing $\tau$ such that $\mathcal{L}^\dagger\tau=-1$ makes the right-hand side equal to
$-\int \rho = -\Sigma(\tilde\delta_0,t)$, where $\Sigma$ is the survival probability.
Integrating \eqref{eq:tau_weighted_evol_full} over time then yields the standard relation
$\tau(\tilde\delta_0)=\int_0^\infty dt\,\Sigma(\tilde\delta_0,t)$, confirming that $\tau$ is the MFPT.
Hence $\tau(\tilde\delta)$ is determined by the backward boundary-value problem
\begin{equation}
\mathcal{L}^\dagger\tau(\tilde\delta)=-1,
\qquad
\tau'(\tilde\delta_L)=0,
\qquad
\tau(\tilde\delta_B)=0.
\label{eq:mfpt_backward_BVP}
\end{equation}
For constant $D(\tilde\delta)=T$ this becomes
\begin{equation}
-1
=
T\Big(\tau''(\tilde\delta)-\beta F'(\tilde\delta)\,\tau'(\tilde\delta)\Big).
\label{eq:mfpt_backward_explicit_full}
\end{equation}

\paragraph{Second moment.}
The same construction yields an equation for the second moment
\begin{equation}
m_2(\tilde\delta)\equiv \big\langle \mathcal{T}^2\big\rangle_{\tilde\delta(0)=\tilde\delta}.
\label{eq:m2_def_full}
\end{equation}
Applying \eqref{eq:adjoint_identity_full} with $f=m_2$ and using the recursion for moments of
first-passage times leads to
\begin{equation}
\mathcal{L}^\dagger m_2(\tilde\delta)=-2\,\tau(\tilde\delta),
\qquad
m_2'(\tilde\delta_L)=0,
\qquad
m_2(\tilde\delta_B)=0.
\label{eq:m2_backward_BVP}
\end{equation}
For constant $D(\tilde\delta)=T$ this reduces to
\begin{equation}
-2\,\tau(\tilde\delta)
=
T\Big(m_2''(\tilde\delta)-\beta F'(\tilde\delta)\,m_2'(\tilde\delta)\Big).
\label{eq:m2_backward_explicit_full}
\end{equation}
From $\tau$ and $m_2$ we obtain the variance
$\mathrm{Var}(\mathcal{T})=m_2-\tau^2$ and the relative fluctuation
$(m_2-\tau^2)/\tau^2$ .
Equations \eqref{eq:mfpt_backward_explicit_full} and \eqref{eq:m2_backward_explicit_full}
are linear, second-order boundary-value problems on the interval
$[\tilde\delta_L,\tilde\delta_B]$.  For each fixed temperature $T$ we solve them numerically with
the boundary conditions \eqref{eq:mfpt_backward_BVP} and \eqref{eq:m2_backward_BVP}, which
bypasses the explicit time evolution of the forward equation and directly yields the first-passage
observables entering our figures.

\bibliographystyle{JHEP}
\bibliography{main.bib}


\end{document}